\documentclass[14pt]{article}
\pdfoutput=1
\usepackage{amsmath,amssymb,amsfonts,amsthm,bm,bbm,cancel,wasysym}
\usepackage{epsfig,graphics,graphicx,epstopdf}
\usepackage{array,booktabs,colortbl,colordvi,multirow}
\usepackage{colordvi,color,xcolor}
\usepackage{hyperref}
\usepackage{rotating}

\usepackage{verbatim}
\usepackage{cite}
\usepackage{subfig}
\usepackage{setspace}
\usepackage{url}
\usepackage[percent]{overpic}
\usepackage{slashed}
\usepackage{authblk}
\usepackage{xspace}
\usepackage{fullpage}
\usepackage{hyperref}

\def\ie{{\it i.e.}}
\def\eg{{\it e.g.}}
\def\etc{{\it etc}}

\def\to{\rightarrow}

\newskip\zatskip \zatskip=0pt plus0pt minus0pt
\def\matth{\mathsurround=0pt}
\def\lsim{\mathrel{\mathpalette\atversim<}}
\def\gsim{\mathrel{\mathpalette\atversim>}}
\def\atversim#1#2{\lower0.7ex\vbox{\baselineskip\zatskip\lineskip\zatskip
  \lineskiplimit 0pt\ialign{$\matth#1\hfil##\hfil$\crcr#2\crcr\sim\crcr}}}

\begin{document}


\begin{flushright}
SLAC-PUB-17633\\
\today
\end{flushright}
\vspace*{5mm}

\renewcommand{\thefootnote}{\fnsymbol{footnote}}
\setcounter{footnote}{1}

\begin{center}

{\Large {\bf Portal Matter and Dark Sector Phenomenology at Colliders}}\\

\vspace*{0.75cm}

{\bf Thomas G. Rizzo}~\footnote{rizzo@slac.stanford.edu}

\vspace{0.5cm}

{SLAC National Accelerator Laboratory}\ 
{2575 Sand Hill Rd., Menlo Park, CA, 94025 USA}

\end{center}
\vspace{.5cm}


\begin{abstract}
\noindent  
If dark matter (DM) interacts with the Standard Model (SM) via the kinetic mixing (KM) portal, it necessitates the existence of massive, likely $\gsim 1$ TeV, enabler portal matter (PM) 
particles that carry both dark and SM quantum numbers which will appear in vacuum polarization-like loop graphs. Such heavy states are only directly accessible at high energy colliders and  
apparently lie at mass scales beyond the direct kinematic reach of the ILC, CEPC and FCC-ee. A likely possibility is that these new particles are part of the 'next step' toward a UV-complete 
scenario describing both the SM and dark sector physics. A simple and straightforward example of such a scenario involving a non-abelian dark sector gauge group is employed in this work 
to demonstrate some of the range expected from this new physics. Here we present a broad survey of existing analyses designed to explore the nature of such PM states in a array of 
collider contexts, particularly at the LHC,  and point out some of the future directions where additional work is obviously required in the hunt for new signatures as well as in model building 
directions.
\end{abstract}
\vspace{0.5cm}
\renewcommand{\thefootnote}{\arabic{footnote}}
\setcounter{footnote}{0}
\thispagestyle{empty}
\vfill
\newpage
\setcounter{page}{1}

\section{Introduction and Background}

Portals\cite{vectorportal}, sometimes represented as effective operators of various dimensions, are an efficient way to categorize broad classes of possible interactions between the visible 
states of the Standard Model (SM) and dark matter (DM) or, more generally, the various hidden states of the dark sector. Most of these portals can only function, however, if other new 
`mixed' enabler states also exist. In the case of the well-studied, renormalizable, dimension-4 vector boson/kinetic mixing (KM) portal\cite{vectorportal,KM}, new enabling particles must 
exist which carry both SM and dark charges to generate the required loop-order, 2-point vacuum polarization-like graphs which are responsible for the KM. In the simplest case, where the low 
energy gauge interactions of the dark sector are described by an abelian $U(1)_D$  gauge group (under which all the SM particles are neutral, \ie, have $Q_D=0$) which kinetically mixes with 
the SM hypercharge, $U(1)_Y$, these new states, hereafter referred to as portal matter (PM)\cite{Rizzo:2018vlb,Rueter:2019wdf,Kim:2019oyh,Rueter:2020qhf,Wojcik:2020wgm},  
must carry both hypercharge and also have $Q_D \neq 0$.  Obviously, the KM setup
itself critically relies upon the existence of such new particles, so one would do well to examine the nature of these hybrid PM states and learn what they can tell us more generally 
about DM and the physics of the dark sector. Since these particles carry SM quantum numbers, the only reason for them not to have already been observed is that they must be relatively 
massive, \ie, likely $\gsim 1$ TeV for fermionic PM (with the details depending upon its color charge) 
and/or have atypical decays, as in the case of scalar PM, which makes present and future colliders the natural places to search for, discover and explore 
their properties. As noted, PM fields may be either fermionic, bosonic or more generally be some combination of 
both species and will more than likely also carry other additional SM quantum numbers besides the required hypercharge, \ie, QCD color and/or $SU(2)_L$ weak isospin.

Of course, once we accept that PM states must exist, not only will they lead to obvious new physics when they can appear directly or indirectly in tree-level reactions, they can also lead to new 
loop-induced processes beyond the simple KM they were 'designed' to accomplish; such possibilities\cite{Rizzo:2021lob}, including those in the flavor sector, have been to some extent previously 
discussed\cite{Rueter:2019wdf,Rueter:2020qhf,Wojcik:2020wgm} and in many way are, \eg, similar to the familiar vector-like fermion and multi-Higgs extensions of the SM. 
However, the fact that these fields now carry additional dark charges can make 
important alterations in the resulting phenomenology; the non-collider aspects of this physics are beyond the scope of the present work.

The study of PM can either be addressed from a bottom-up or from a top-down perspective - both from which we can glean important information about their apparent nature. In 
either case, however, at or below the weak scale where the effects of PM are indirect, the relevant physics is relatively straightforward and quite familiar. The PM fields at 1-loop order will 
generate a term in the Lagrangian of the form
\begin{equation}
{\cal L}_{KM}=  \frac{\epsilon}{2c_w} \hat V_{\mu\nu} \hat B^{\mu\nu}\,,  
\end{equation}
where $\hat B^{\mu\nu}$ is the SM hypercharge gauge boson field strength tensor, $\hat V_{\mu\nu}$ is that for the analogous $U(1)_D$ gauge field tensor, \ie, that of the dark photon (DP), 
$c_w=\cos \theta_w$, and $\epsilon$ if the strength of the KM generated by PM loops and is given by
\begin{equation}
\epsilon =c_w \frac{g_D g_Y}{24\pi^2} \sum_i ~\eta_i \frac{Y_i}{2}  N_{c_i}~Q_{D_i}~ ln \frac{m^2_i}{\mu^2}\,,
\end{equation}
with $g_{Y,D}$ being the $U(1)_{Y,D}$ gauge couplings and $m_i(Y_i,Q_{D_i}, N_{c_i})$ are the mass (hypercharge, dark charge, number of colors) of the $i^{th}$ PM field. Here, 
$\eta_i=1(1/2)$ if the PM is a chiral 
fermion(complex scalar) and the hypercharge is normalized so that the electric charge is given by $Q_{em}=T_{3L}+Y/2$ as usual. In UV-complete theories, and as will be the case 
in the setups which we will be discussing below, the sum
\begin{equation}
\sum_i ~\eta_i \frac{Y_i}{2} N_{c_i} Q_{D_i}=0\,,
\end{equation}
so that $\epsilon$ is both finite and, if the PM masses are known, calculable within these frameworks and, for $O(1)$ mass splittings between the PM states, typically lying in the 
experimentally interesting range $\epsilon=10^{-4}-10^{-3}$. One trivially obvious statement that one can make from this is that the set of PM fields {\it must} consist of more than a single 
state for such a cancellation as above to 
occur.  We note that in what follows we will assume that the dark photon, $V$, acquires its mass via the vacuum expectation value(s) of one or more dark Higgs fields and that the masses of 
the DM, DP and the physical dark Higgs fields will all roughly lie in the interesting range below $\sim 1$ GeV.  After applying field redefinitions to remove the KM, \ie, $\hat V \to V$ and 
$\hat B \to B+\frac{\epsilon}{c_w}V$, to leading order in $\epsilon$, and after spontaneous symmetry breaking of both the SM and the dark gauge group,  
the resulting $Z-V$ mass mixing matrix can be easily diagonalized. One then finds that for masses in the range, $m_V^2/m_Z^2<<1$, the DP will couple 
to the SM fields in the expected fashion, as $\simeq e\epsilon Q_{em}$, to leading order in the small parameters, with $e$ being the proton's charge. It is important to note that $\epsilon$ depends 
only upon the {\it ratios} of the masses of the PM states and not on their specific masses so that $\epsilon$ itself provides no obvious guidance as to where to look for the actual PM 
particles. However, as we will see below, in the case of a simple scalar PM scenario at least, one finds that there can be additional model-dependent restrictions on the spectrum of PM masses 
due to the requirements of electroweak symmetry breaking in the SM. 

\begin{figure}
\centerline{\includegraphics[width=5.0in,angle=0]{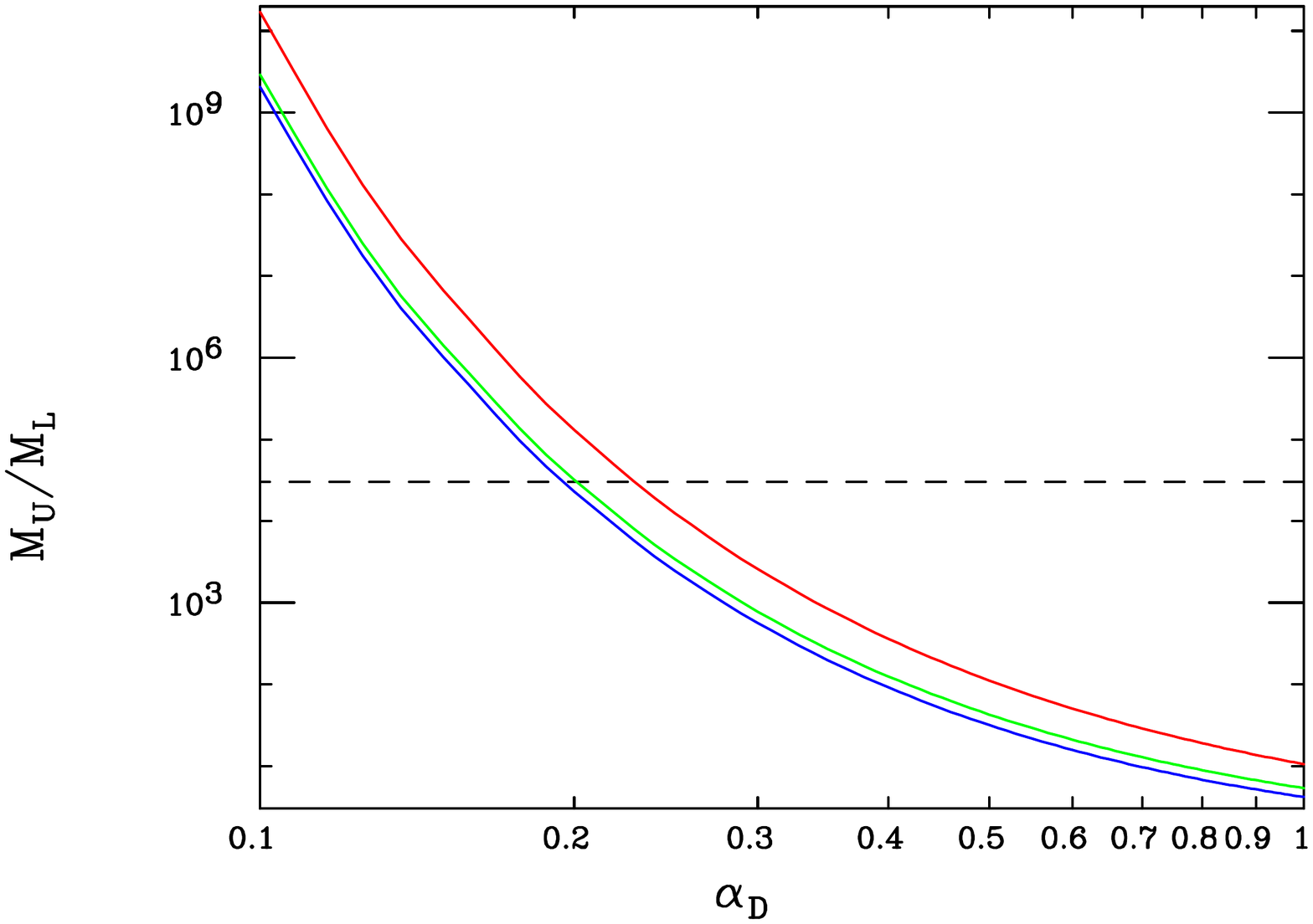}}
\vspace*{-2.3cm}
\centerline{\includegraphics[width=5.0in,angle=0]{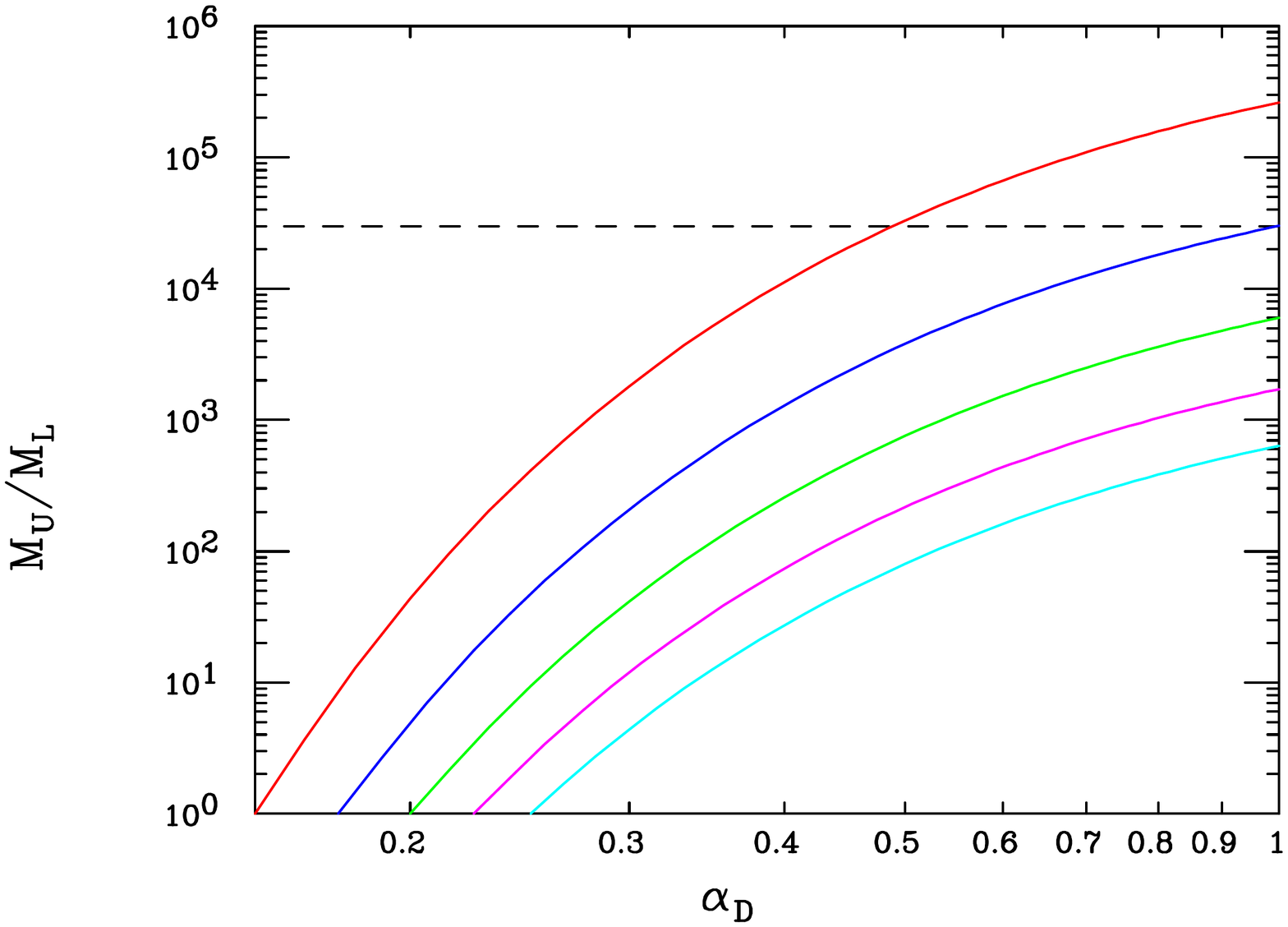}}
\vspace*{-1.30cm}
\caption{The running of $\alpha_D$ in the pseudo-Dirac DM example discussed in the text. The top panel shows the location of the Landau pole, $M_U$,  at the 1-loop(red), 2-loop(blue) and 
3-loop(green) level in RGE running, employing the $\bar{MS}$ scheme, as a function of $\alpha_D(M_L)$ where $M_L$ is the low scale associated with the DM, DP and dark Higgs fields, 
$\simeq 100$ MeV. The dashed line corresponds to $M_U=3$ TeV when $M_L=100$ MeV as a guide for the eye. The lower panel shows the 3-loop running of $\alpha_D$ (on the $x$-axis) 
in this same scenario as a function of $M_U/M_L$. The curves, from top to bottom, are for $\alpha_D(M_L)=0.15,0.175,0.20,0.225$ and 0.25, respectively. 
Here we see that, \eg, if $\alpha_D(M_L=100$ MeV)$\geq 0.175$ then $\alpha_D \geq 1$ for $M_U \geq $ 3 TeV, again indicated by the dashed line. }
\label{fig1}
\end{figure}

However, one reason that we might expect that the PM states, as well as a more complex dark gauge sector, may not be too far away in energy above the weak scale (at least for part of the 
parameter space where the DM coupling is somewhat large) is to consider the running of the dark gauge coupling, $g_D$ or more precisely below, $\alpha_D=g_D^2/4\pi$. As is well-known, 
a $U(1)$ gauge theory is not asymptotically free and eventually will become strongly coupled or even have 
a  Landau pole at some point as the energy scale increases. Here, we will take advantage of this observation noting that in a UV-complete theory one would expect new physics of some form to 
enter before either of these things can happen.  In order to be specific for demonstration purposes, let us consider the case of light fermionic DM, having $Q_D=1$, together with the  DP and dark 
Higgs all lying in roughly the same mass range $\sim 100$ MeV. To be even more specific, and in order to avoid direct detection bounds due to inelastic DM scattering as well as 
the strong constraints from the CMB\cite{Aghanim:2018eyx,Slatyer:2015jla,Liu:2016cnk,Leane:2018kjk} for fermionic DM in this mass range, we consider the scenario where 
the DM is pseudo-Dirac with the relevant mass splitting generated by the same dark Higgs vev that is responsible for the mass of the DP. Assuming that these few fields are the only light 
states, we can then run the value $\alpha_D$ in a known manner from this low energy scale, $M_L$, up to some higher scale, $M_U$ (or until some new physics with $Q_D\neq 0$ enters the 
RGE's) where one reaches a region of strong coupling or even hits a Landau pole{\footnote {Any additional light fields with $Q_D\neq 0$ will only strengthen these arguments below since then 
$\alpha_D$ will only run more quickly.}}. Recall that 
the SM fields will {\it not} enter into this calculation as they all have $Q_D=0$ and so will not couple to the DP to LO in the $\epsilon \to 0$ limit. The results of these simple considerations can be 
found in Fig.~\ref{fig1} from\cite{inprogress}. In this Figure, we can see that if $\alpha_D(M_L) \geq 0.175(0.20)$, a not infrequent assumption made in many phenomenological 
analyses\cite{Alexander:2016aln,Battaglieri:2017aum,Bertone:2018krk,Schuster:2021mlr}, its value 
will become non-perturbative (or even hit a Landau pole\cite{Gockeler:1997dn}) before $M_U\simeq $ a few TeV when running up from the $M_L=100$ MeV scale. Semi-quantitatively, we see 
that these results are not very dependent as to whether they are obtained at the 1-, 2- or 3-loop level as can be gleaned from this Figure.  Although these results are only indicative, they give 
some credence to the likelihood that the PM fields and an associated more complex, non-abelian dark sector gauge structure, \eg, an 
$SU(2)_D$\cite{Rueter:2019wdf,Wojcik:2020wgm,Murgui:2021eqf}, are likely not to be too far 
above the weak scale and may be accessible at the HL-LHC and at other planned future colliders. 

In this White Paper, we will examine some of the generic phenomenological implications of a set of `relatively simple' PM models and their associated dark sectors which will act as guideposts to 
the physics scenarios resulting from the more realist and complex scenarios that may appear in a fully UV-complete KM model. The possibilities here are rather wide ranging and so it 
should be noted that much of this work is still in progress so that only preliminary results are available in some cases at this point. We will, however, rely {\it quite} heavily on the results 
as presented in earlier analyses\cite{Rizzo:2018vlb,Rueter:2019wdf,Rueter:2020qhf,Wojcik:2020wgm} for much that appears below.

The outline of this paper is as follows: In Section 2, we review some basic ideas and mechanics behind the setup for a simple low-scale $U(1)_D$ fermionic PM scenario; here the PM 
fields must be heavy, vector-like copies of the usual SM fermion representations to allow for the PM to decay reasonably quickly 
to agree with cosmological constraints. In Section 3, we consider one 
example of a more UV-complete scenario of PM with an enlarged dark sector gauge group, based on an $E_6$-inspired framework, which embodies the previously discussed simple model 
but with a highly enriched phenomenology. Section 4 contains a broad overview and discussion of some of the $e^+e^-, \gamma\gamma$ and LHC signatures for this scenario, reviewing several 
past studies and pointing out where more work is clearly needed. Section 5 contains a discussion of the phenomenology of an alternative scalar PM model where additional scalar doublets 
carrying dark charges act as as both the dark Higgs fields breaking $U(1)_D$ as well as PM.  Finally, in the last Section we summarize the essential results and conclusions as well as point 
out some possible future directions.

\section{Simple Ideas and Basics}

For simplicity we begin our discussion assuming PM is fermionic and that the relevant gauge group below the mass scale of the PM fields themselves, $\lsim 1$ TeV, is simply 
$G_{SM}\times U(1)_D$. In such a case, the PM fermions must be vector-like, \ie, VLF's, with respect to both gauge groups in order to avoid anomalies, constraints from electroweak 
fits\cite{deBlas:2021wap} and unitarity\cite{Chanowitz:1978uj}, as well as having any significant contribution to the Higgs to $gg/\gamma \gamma$ partial decay widths which would be 
the case of, \eg, a fourth generation of {\it chiral} fermions.  At the electroweak scale or below, 
being vector-like, these PM fermions are allowed to have (apparent) bare mass terms and thus are not generated by their couplings to the SM Higgs. Of course in a more 
UV-complete theory such masses might be generated by, \eg, the Higgs fields that are responsible for the breaking of a larger dark gauge group, $G_{dark}$ down to $U(1)_D$. 
Since the PM fields carry $Q_D\neq 0$ as well as hypercharge, the lightest among them will be stable unless they are allowed to mix with one or more of the lighter SM fermion fields, \ie, 
$f= (\nu,e)_L^T, (u,d)_L^T, e_R,u_R,d_R$ (suppressing the generation index) through, \eg, the vev of a complex isosinglet dark Higgs field, $h_D$, which may or may not also be the dominant 
source of the DP's mass. This implies that the PM fields, $F$, must transform in a vector-like manner\cite{Chen:2017hak} similar to one or more of these SM fermion representations, \ie, 
$F=(N,E')^T,(U',D')^T,E,U,D$ (again dropping any potential generation indices) with the primes only being employed here to help distinguish isosinglet from isodoublet fermions with the 
same electromagnetic and color charges.  Of course, 
one can also imagine dark Higgs fields which can transform as $SU(2)_L$ doublets as long as their vevs are sufficiently suppressed to below the $\lsim 1$ GeV level to avoid other constraints. 
In a general scenario, one may easily imagine that dark Higgs fields of both varieties may be present simultaneously as will be realized in a particular manifestation below.

Perhaps, the very simplest toy model/bottom-up example\cite{Rizzo:2018vlb} 
that we can construct from the observations above, which we will analyze as a test case for demonstration purposes, is then that 
the set of PM fields consisting of just two $Q_D=\pm1$, weak isosinglets states having $Q_{em}=-1$, \ie, $E_{1,2}$ (assumed to be colorless), with comparable but somewhat different masses, 
$m_1\lsim m_2$, thus rendering $\epsilon$ finite and calculable; specifically, for this color singlet scenario one obtains 
\begin{equation}
|\epsilon| \simeq 3.0~\Big(\frac{g_D}{0.3}\Big) ~\frac{ln (m_2/m_1)}{ln ~1.5} \cdot 10^{-4}\,
\end{equation}
as desired. The $E_i$ may be thought of as vector-like copies of the RH-electron (or muon or tau, but we will use the electron in this simple toy example) and 
if we take $|Q_D(h_D)|=1$ as well, then an off-diagonal $f-F$-type interaction can be generated of the form
\begin{equation}
 {\cal L}_{Mix} = ~ \lambda_1 \bar E_{1L}e_Rh_D +\lambda_2\bar E_{2L}e_Rh_D^\dagger +h.c. \,
\end{equation}
where the $\lambda_i$ are presumably $O(1)$ Yukawa couplings and, again, all potential generation indices have been suppressed. A very similar expression can 
obviously be written in the case of, \eg,  weak isosinglet quarks. Now, when the complex field, $h_D$, acquires a 
vev, \ie, $h_D \to (v_s+S+iA)/\sqrt {2}$, $v_s\neq 0$, it generates mass mixing between the PM fields and, in this example, the SM electron. It is important to note that the mixing that is induced is 
with, \eg,  the RH-electron so that these interactions are {\it chiral}.  
If the PM fields had {\it instead} been chosen to be two sets of the color-singlet, weak isodoublets, $(N,E')_i^T$, then the corresponding chiral coupling would then be with the LH-electron 
through the same $h_D$ isosinglet dark Higgs, \ie, the mixing is generated between states with the {\it same} weak isospin as $h_D$ is here assumed to be isosinglet, \ie, $h_{D_S}$. This 
simple result can be seen to be easily generalizable to cover all of the other potential scenarios for various specific choices of the PM fields; of course, in a more UV-complete picture 
something more complex is possible as we will see below. For example, if the dark Higgs involved in the coupling had instead been an isodoublet then the helicity of these couplings would 
then have been reversed. Note that when $h_D$ is the same (and, in the toy example here, only) scalar which generates the DP mass, the CP-odd field, 
$A$, then becomes the Goldstone boson, $G_V$,  associated with the 
DP's longitudinal degree of freedom and the CP-even state, $S$, is the remaining {\it physical} dark Higgs. Of course, more generally, if multiple dark Higgs fields are present, $A$, will become 
an admixture of this Goldstone and a (set of) physical CP-odd mass eigenstate(s), \etc, as one sees, \eg, in the Two-Higgs Doublet Model. Similarly, the light dark Higgs mass eigenstate will 
also become some admixture of the various CP-even fields in this more complex scalar sector. 

Following this through, it should then be noted that in more general, and likely more realistic, models the dark Higgs fields with $Q_D\neq 0$ that transform non-trivially under $SU(2)_L$, \eg, 
as isodoublets, may also appear in the spectrum as noted above. Their vevs will also contribute to the mass of the DP so are likewise constrained to be rather small in the present setup, 
$\lsim 1$ GeV, but this remains sufficient to now influence $V-Z$ mass mixing as well as the general mixing between the PM and SM fields. Clearly, the choice of the $SU(2)_L$ 
representation(s) of the PM fields will also directly influence the dark Higgs' $SU(2)_L$ transformation properties if we want to maintain gauge invariant couplings that allow all the PM fields 
to decay. One can symbolically summarize all of these possible couplings for the color singlet, $Q_{em}=-1$, PM fields via the generalized PM-SM mixing Langrangian 
\begin{equation}
 {\cal L}_{Mix}' = ~ \lambda_S \bar E_Le_Rh_{D_S} +\lambda_S'(\bar \nu,\bar e)_L^T(N,E')_R^T h_{D_S}^\dagger+\lambda_D(\bar \nu,\bar e)_L^TE_Rh_{D_D}^\dagger+\lambda_D' (\bar N,\bar E')_L^T e_R h_{D_D}+h.c. \,
\end{equation}
where, using the notation above,  $h_{D_{(D,S)}}$ here represents the weak $SU(2)_L$ isodoublet or isosinglet dark Higgs field, respectively, and the $\lambda$'s are again assumed to be $O(1)$. Note the primed and unprimed fields in this expression. 
As in the previous example, analogous interaction terms can straightforwardly be written down for other PM quantum number choices. 

In the form written above we have suppressed any flavor issues and one may ask if the PM couples predominantly only to a single generation (in some limit), mixes with all three SM generations, 
or whether or not the PM themselves come in three generations. Clearly, some more than others of these scenarios will be strongly constrained by experimental results for the flavor sector. 
All of these possibilities appear in the literature\cite{Rueter:2019wdf} but it is beyond the scope of the present work to provide an overview of all of them especially as we are here more focussed on the 
collider aspects of PM models. However, flavor issues will enter the discussion when they are clearly relevant to specific experimental signatures and production mechanisms so 
here we'll usually assume that the simplest possibilities are realized.
 
Let us now return to the simple isosinglet dark Higgs model above with isosinglet, colorless PM fermions. 
As is easily seen, interactions such as ${\cal L}_{Mix}$ will allow for the generic PM decays of the form $F\to f(S,V=V_{long}\simeq G_V)$, with rates controlled by the $O(1)$ values of the 
$\lambda_i$ parameters. The Goldstone Boson Approximation\cite{GBET} limit is applicable here as $m_{S,V}^2 << m_F^2$ since all the PM fermion masses must 
clearly be at least several hundreds of GeV or more in order to have so far evaded experimental detection. 
It is easy to see in this same Goldstone limit that $\Gamma(F\to fS)=\Gamma(F\to fV)$ provided that $m_f^2<<m_F^2$ which will always be the case except, perhaps, when $f=t$. 
One explicitly finds in this limit that 
\begin{equation}
\Gamma(F_i \to fV,fS)\simeq \frac{\lambda^2_i m_i}{64\pi}\,,
\end{equation}
under these assumptions.  As is well-known, the SM-like Higgs-induced mixing of heavy vector-like fermions with analogous SM fields with the {\it opposite} value of the 
weak isospin (\eg, singlets with doublets), such as $E'_L-e_L$ and not $E_L-e_L$ type mixing, would also yield the more familiar decays $F\to f'W, fZ$ and $fh_{SM}$ which are usually 
anticipated in performing vector-like fermion searches at the LHC.  Here, such couplings will be immediately induced once the $F-f$ mass matrix is diagonalized\cite{Rizzo:2018vlb,Kim:2019oyh}. 
In the case of an isosinglet $F$, in the approximation that $m_{f,f',W,Z,h_{SM}}^2 << m_F^2$ then the partial widths to these final states are related with the usual result being 
$\Gamma(W)=2\Gamma(Z)=2\Gamma(h_{SM})$, for $F$ being an isosinglet (the $W$ mode is absent in the isodoublet case)  and are highly suppressed by the square of the relevant 
$F_i-f$ mixing angle, here $\sim \theta_{L_i}^2 <<1$,  so that, \eg,  
\begin{equation}
\Gamma(F_i \to fh_{SM})\simeq \frac{G_F m_i^3}{16 {\sqrt 2}\pi} \theta_{L_i}^2\,,
\end{equation}
with $G_F$ being the Fermi constant, 
while the corresponding partial widths to $S,V\simeq G_V$ final states experience no such suppression and are instead proportional to the $\lambda_i^2$, which are expected to be of $O(1)$, as 
noted above. However, the $\lambda_i$ themselves at some level control the size of the $F-f$ mixing. Thus, we should expect that the decays of the PM fields, $F$'s, into dark sector fields 
will always be (very) far dominant which makes their production signature different from ordinary VLF's. This implies that it will be the decays of $S$ and $V$, which essentially always appear 
as final states in $F_i$ decay, that will determine the collider signature for PM production.  

Once a PM $\bar FF$ pair is produced from an $e^+e^-$, $\bar qq$ or $gg$ initial state collision and then decays, the resulting final state will be a simple admixture of $\bar ff+2V,2S,SV$ with the  
$\bar ff$ appearing as either a opposite-sign, same flavor lepton pair or a pair of jets (or possibly missing energy if $f=\nu$). The more complex issue is what will $V,S$ decay into: as is 
well-known, this is essentially 
determined by the relative masses of $V,S$ and the DM fields. If $m_V>2m_{DM}$, since the DM is stable, $V$ will decay invisibly to DM pairs with a rate proportional to $\sim g_D^2$; if 
this condition is not satisfied then $V$ will instead visibly decay into pairs of SM fields, \eg, $e^+e^-$, $\mu^+\mu^-$, \etc, although such decays are suppressed in rate by $\sim \epsilon^2$. 
If $m_S>2m_V$ then $S\to 2V$ happens quickly and will dominate whereas if $2m_V>m_S>m_V$ then $S$ decays as $S\to VV^*$ and can be potentially relatively long lived. This will certainly 
be the case if $m_S<m_V$ so that the completely off-shell, $S\to V^*V^*$, and the loop- or mixing-induced, \eg, $S \to e^+e^-$\cite{Batell:2009yf}, decays are relevant. Thus we must 
conclude that $S,V$ will either materialize as missing energy or highly boosted lepton-jets\cite{ljet,Rizzo:2018vlb}. For example, in the case of the missing energy mode, the existing LHC searches for 
Supersymmetry involving jets and/or leptons + MET can be recast for the case of $\bar FF$ production if $S,V$ decay invisibly which then tells us that in such a scenario, \eg,  
$m_{E(E')} \gsim 0.90(1.05)$ TeV\cite{Guedes:2021oqx,OsmanAcar:2021plv,Baspehlivan:2022qet} for first/second generation-like isosinglet $E$ (isodoublet $E'$) and 
$m_Q \gsim 1.3-1.5$ TeV\cite{Rizzo:2018vlb,Kim:2019oyh}, depending on the final state fermion flavor, values which are beyond those 
directly accessible to the ILC, CEPC or FCC-ee, as we will discuss further below. For third generation-like color-singlet PM the prospects for obtaining comparable constraints are somewhat 
poorer\cite{Kumar:2015tna,Bhattiprolu:2019vdu}. It is to be noted, however, that if the PM fields 
exist within a more complex scenario, such as those we will touch upon below, where, \eg, the dark gauge group may be enlarged, then at least the more massive PM fields may be allowed to 
suffer other decay paths than those discussed above. A simple version of such setup will be discussed in the next Section but we will see that in a significant part of the parameter space the 
essential phenomenology returns to this simple picture that we have so far described.

Although this basic scenario is perhaps adequate as an effective theory at the weak scale and below, it does not on its own provide much guidance as to how or if the PM and SM 
fields may fit together is some type of more common framework. We will return to this issue from a specific perspective below, mostly concentrating on one possible next step up the ladder in the 
bottom-up approach.

\section{An Example of a More UV-Complete PM Scenario}

Outside of the direct production of the PM fields themselves at colliders via SM processes, \ie, the usual QCD and electroweak interactions, the simple toy examples above do not lead to 
any `unusual' phenomenology other than MET and/or displaced vertices. However, if we take another step upward in the ladder to a UV-complete theory this is no longer true. Perhaps one of the 
most interesting paths is to imagine enlarging the dark gauge group beyond $U(1)_D$, \ie, to a more general $G_{dark}$ which is a single non-abelian group or is a product group with 
non-abelian and/or abelian factors.  A full UV-complete picture would then have a `unified' group, $G_U$ break down to, \eg, $G_{SM}\times G_{dark}$ at some large scale followed by 
$G_{dark}$ breaking 
to $U(1)_D$ at, say, $\sim 10$ TeV.  Of course, one can imagine all sorts of ways to grow the dark gauge group itself, \eg, a dark gauge group `orthogonal' to SM\cite{inprogress}, as an 
extension of the SM electroweak sector\cite{Rueter:2019wdf}, as a unification with an enlarged gauged flavor symmetry group\cite{Wojcik:2020wgm},  or even combined with QCD into a single 
larger group\cite{Murgui:2021eqf}. Clearly, any new physics signatures will depend to some extent upon which of these paths, if any, might be  chosen by nature.

\begin{table}[h] 
\begin{center}
\caption{Fermionic Field Content}\label{fermtab}
\vspace{-.2cm}
\begin{tabular}{ l  c  c  c  c  c  c  c }
\hline		
SU(5) & & SU(3)$_C$ & $T_{3L}$  & $Y$/2 & $T_{3I}$  & $Y_I$/2 & $Q_D$  \\ 
\hline
{\bf 10} & $Q \equiv \begin{pmatrix} u \\ d\\\end{pmatrix}_L $  & {\bf 3} & $\begin{pmatrix} 1/2 \\ -1/2 \\ \end{pmatrix}$ & 1/6 & 0 & 0 & 0 \vspace{.1cm} \\
 & $u^c_L$ & $\bf{ \bar3}$ & 0 & -2/3 & 0 & 0 & 0 \vspace{.1cm}\\
 & $e^c_L$ & $\bf{ 1 }$ & 0 & 1 & 0 & 0 & 0 \vspace{.1cm}\\
$\bf{\bar5}$& $L \equiv \begin{pmatrix} \nu \\ e\\\end{pmatrix}_L $ & $\bf{ 1 }$ & $\begin{pmatrix} 1/2 \\ -1/2 \\ \end{pmatrix}$ & -1/2 & 1/2 & -1/2 & 0 \vspace{.1cm}\\
 & $d^c_L$ & $\bf{ \bar3 }$ & 0 & 1/3 & 1/2 & -1/2 & 0 \vspace{.1cm}\\
$\bf{\bar5}$ & $H \equiv \begin{pmatrix} N \\ E\\\end{pmatrix}_L $ & $\bf{ 1 }$ & $\begin{pmatrix} 1/2 \\ -1/2 \\ \end{pmatrix}$  & -1/2 & -1/2 & -1/2 & -1 \vspace{.1cm}\\
 & $D^c_L$ & $\bf{\bar3 }$ & 0 & 1/3 & -1/2 & -1/2 & -1 \vspace{.1cm}\\
{\bf 5} & $H^c \equiv \begin{pmatrix} E \\ N\\\end{pmatrix}^c_L $ & $\bf{ 1 }$ & $\begin{pmatrix} 1/2 \\ -1/2 \\ \end{pmatrix}$ & 1/2 & 0 & 1 & 1 \vspace{.1cm}\\
 & $D_L$ & $\bf{ 3 }$ & 0 & -1/3 & 0 & 1 & 1 \vspace{.1cm}\\
\hline
\end{tabular}
\end{center}
\end{table}

\subsection{A UV-Inspired Model Framework} 

Ref.\cite{Rueter:2019wdf} provides a basic $E_6$-inspired setup in this direction with some interesting collider phenomenology. At the level relevant for our discussion here, in this example 
the dark gauge group is effectively 
enlarged to $SU(2)_I\times U(1)_{I'}\equiv 2_I1_{I'}$, which is qualitatively similar in nature to the SM, and is broken down to $U(1)_D$ at a multi-TeV scale. As we 
will see (but not make use of here) this $2_I1_{I'}$ group structure and $U(1)_D$ quantum number assignments are such that this dark product group might be embeddable into $SU(3)_I$ at an 
even higher mass scale. 
Here the gauge couplings are simply $g_I,g_{I'}$, respectably, and we identify the $U(1)_D$ gauge coupling above as $g_D=e_I=g_I \sin \theta_I$ with $x_I=\sin ^2 \theta_I$ being the analog 
of the usual SM weak mixing angle, $x_w=\sin^2 \theta_w$,  of  {\it {a priori}} unknown value which depends upon the full UV-completion of the model. In such a setup, 
$Q_D=Q_I=T_{3I}+Y_I/2$, as one might expect.  Due to the $E_6$-inspired nature of this scenario, the `exotic' PM fields 
form (at least) a complete $\bf{5}+\bf{\bar{5}}$ of the familiar $SU(5)$ with $Q_D=1,-1$, respectively, so that $Q_D(D)=1$ while $Q_D(N,E)=-1$; other fermions which are SM singlets, \eg, the 
$\nu_R,S_L$ fields which fill out the usual $\bf{27}$ of $E_6$\cite{Hewett:1988xc} as well as other exotics, will also be present and may be necessary to, \eg, cancel gauge anomalies 
and also insure that 
$\epsilon$ is finite{\footnote {For clarity we will here use first generation labels for all the fields as we have in the previous Section.}}. This 
basic matter content for this scenario is schematically shown in Table~\ref{fermtab}; note that in this setup $(d_L,D_L)^c$ and $(L_L,H_L)$ are doublets under $SU(2)_I$ with the latter 
actually being a bidoublet under $SU(2)_L\times SU(2)_I$, similar to what is encountered in the Left-Right Symmetric Model. Thus, $SU(2)_I$ is seen to directly link the $D_L^c$ 
and $H_L$, $Q_D\neq 0$ PM fields with the corresponding $d_L^c$ and $L_L$ fields of the SM both having $Q_D=0$. In such a setup, the role of $Q_D$ in the KM expressions above is now 
replaced by the combination $Q_D\to c_IY_I/2$ since it is $1_Y1_{I'}$ whose corresponding gauge fields now undergo dim-4 KM, appearing symmetrically with (so far the fermionic 
contribution) $TrYY_I=0$ thus insuring that $\epsilon$ is finite. 
\begin{table}
\begin{center}
\caption{Higgs Sector Content}\label{higgstab}
\begin{tabular}{ l c c c c c }
\hline
$\Phi$ & SU(2)$_L$ & $Y/2$ & SU(2)$_I$ & $Y_I/2$ & $\left< \Phi \right>$ \\
\hline
$H_1$ & {\bf 2} & 1/2 & {\bf 1} & 0 & $\frac{1}{\sqrt{2}}\begin{pmatrix} 0 \\ v \\ \end{pmatrix} \sim \begin{pmatrix} 0 \\ 100 \textrm{ GeV} \\ \end{pmatrix}$  \vspace{0.1cm}\\
$H_2$ & {\bf 2} & -1/2 & {\bf 2} & 1/2 &  $\frac{1}{\sqrt{2}}\begin{pmatrix} v_1 & v_2 \\ 0 & 0 \\ \end{pmatrix} \sim \begin{pmatrix} 1 \textrm{ GeV} & 100 \textrm{ GeV} \\ 0 & 0 \\ \end{pmatrix}$ \vspace{0.1cm}\\
$H_3$ & {\bf 1} & 0 & {\bf 2} & -1/2 & $\frac{1}{\sqrt{2}}\begin{pmatrix} v_3 & v_4 \\ \end{pmatrix} \sim \begin{pmatrix} 10 \textrm{ TeV} & 1 \textrm{ GeV} \\ \end{pmatrix}$ \\
\hline 
\end{tabular}
\end{center}
\end{table}   

Since this scenario is $E_6$-inspired and, as seen from Table~\ref{fermtab}, a single generation of the SM and PM fields fit into a $\bf{27}$ representation of this group, one may ask whether 
or not there is only a single set of the PM fields or there is one set for each SM generation, as would perhaps be the naive $E_6$ expectation. As noted above, there are clear flavor physics 
aspects to this choice which we will only explore from the collider side here. Note that while vector-like with respect to the SM, under $2_I1_{I'}$, the PM fields are {\it not} vector-like so that 
dark Higgs fields are necessary to generate 
their masses. In fact, we needs multiple Higgs fields with the appropriate transformations properties but which have vevs at the $\sim 10$ TeV, $\sim 100$ GeV and $\sim 1$ GeV scales. 
The minimal content of the (fine tuned) Higgs sector that accomplishes the required symmetry breaking at these mass scale is then given in Table~\ref{higgstab} and is discussed in detail in 
Ref.\cite{Rueter:2019wdf}.  Additional scalars without vevs may need to be present to maintain a finite $\epsilon$ and/or to act as a low mass scalar DM candidate. Once these three 
Higgs fields obtain the various vevs as shown in the Table, numerous new collider signatures are found to be generated.

The vev $v_3$ is the one responsible for the $2_I1_{I'}\to 1_D$ breaking, generating the dominant contribution to the masses of the $Z_I$ and $W_I^{(\dagger)}$ gauge bosons as well as 
the dominant mass terms for the $N,E$ and $D$ PM fermion fields. Here we note that while $Z_I$ has $Q_D=0$, but one finds instead that the $Q_{em}=0$, but non-hermitian, fields 
$W_I^{(\dagger)}$, carry $Q_D=\pm 1$. In such a setup the analog of the SM photon after the breaking the $2_I1_{I'}\to 1_D$, which we might call $A_I$, can be easily identified with 
the DP, \ie, $V$, discussed above. In the $v_3^2>>v^2, v_{1,2,4}^2$ and $\epsilon \to 0$ limits, as expected one finds the SM-like results that $m_{W_I}=g_Iv_3$ and 
$m_{Z_I}=m_{W_I}/c_I$, \ie, the analogous $\rho_I=1$ at tree-level due to isodoublet breaking but may suffer from significant loop effects here due to large mass splittings within the $SU(2)_I$ 
fermion representations, in principle. 
The gauge fields in this same limit couple in a manner completely analogous to the corresponding ones in the SM with the obvious substitutions, \eg, $L\to I$. 
Similarly, in the $v^2,v_2^2>>v_{1,4}^2$ and $\epsilon \to 0$ limits, the SM gauge fields obtain there masses similarly to what happens in the Two Higgs 
Doublet Model(2HDM), \ie, $m_W^2=g^2(v^2+v_2^2)/4=m_Z^2c_w^2$. Of course, all these vevs are non-zero as is $\epsilon$ so that the gauge boson mass matrices are much more complex; 
we can, however, adequately work to lowest order in $\epsilon \sim 10^{-(3-4)}$ and in the ratios of the squares of the various vevs. This leads to several interesting 
effects{\footnote {For details, see 
Ref.\cite{Rueter:2019wdf}.}}: ($i$) We observe that that the $u$ mass (and a possible a $\nu$ Dirac mass) is generated by $v$ while the $d,e$ masses are generated by 
$v_2$ similar to the 2HDM. ($ii$) The SM $Z$ and the $Z_I$ are found to (mass) mix by a small angle, \ie, 
\begin{equation}
\theta_{ZZ_I} \simeq \frac{g_I/c_I}{g/c_w}~\frac{m_Z^2}{m_{Z_I}^2}~ \frac{v_2^2}{v^2+v_2^2}\,,
\end{equation}
where the last ratio of vevs is similar to the $\cos^2 \beta$ factor in the 2HDM, and is expected to be $O(1)$. ($iii$) The SM $Z$ mixes with the DP, $V=A_I$, in this language, by a small angle 
\begin{equation}
\theta_{ZA_I} \simeq -\epsilon t_w +\frac{gg_Is_I}{2c_w} ~\frac{v_1^2}{m_Z^2}\equiv -\epsilon t_w+\sigma\,,
\end{equation}
which is induced by both KM and mass mixing. Note that the ratio of these two contributions, $r_0=\sigma/(\epsilon t_w)$, is also expected to be $O(1)$ for typical choices of parameter values: 
\begin{equation}
r_0\simeq 0.26 ~\Big(\frac{10^{-4}}{\epsilon}\Big)~\Big(\frac{v_1}{1 \rm GeV}\Big)^2 ~\Big(\frac{g_Is_I}{gs_w}\Big)\,.
\end{equation}
($iv$) The mass of the DP in the eigenstate basis to leading order in the small mixings is now given by 
\begin{equation}
m_{V=A_I}^2 =g_I^2s_I^2v_1^2+\sigma(2\epsilon t_w-\sigma)m_Z^2\,. 
\end{equation}
where, quite generally, the first term is dominant for masses $\geq 50-100$ MeV; note the absence of $v_4$ here. 
($i$) Recalling that $g_D=e_I=g_Is_I$ and $Q_D=Q_I$, the DP is now found to couple to the combination 
\begin{equation}
~~~~ g_DQ_D+e\epsilon Q_{em}-\sigma \frac{g}{c_w}(T_{3L}-x_wQ_{em})\,.
\end{equation}
While the first two vector-like terms are as above in the previous Section, 
here we see that the DP has an additional coupling to the SM fields proportional to that of the SM $Z$ boson, a well-known side effect of there 
being a dark Higgs field that obtains a vev while also carrying ordinary weak isospin, here being an $SU(2)_L$ isodoublet, \ie, $v_1\neq 0$.

As in the toy example, the set of Higgs fields in Table~\ref{higgstab} not only generate masses for the SM and PM fermions but also the mixings in the, \eg, $d-D$ and $e-E$ sectors as discussed 
in detail in Ref.\cite{Rueter:2019wdf}. These terms produce (almost) calculable values for the corresponding $\lambda$ couplings for the light dark Higgs and the DP's Goldstone boson 
which appeared in the last Section up to $O(1)$ parameter ratios, \eg, 
\begin{equation}
~ \lambda_{E,D} =\sqrt 2 ~\Big(\frac{v_4}{v_1}\Big)~\Big(\frac{2m_{E,D}}{v_3}\Big)= 0.42 ~\Big(\frac{v_4}{v_1}\Big)~\Big(\frac{m_{D,E} }{1.5 \rm TeV}\Big) ~\Big(\frac{10 \rm TeV}{v_3}\Big)~\sim O(1)\,,
\end{equation}
where $v_4/v_1$ is the ratio of the two $U(1)_D$ breaking vevs and here we see that these $\lambda$'s are indeed naturally $O(1)$ in this setup.

\section{Collider Signature Survey} 

\subsection{$e^+e^-$ Colliders}

The initial run of the ILC\cite{Baer:2013cma} at $\sqrt s=240$ GeV and/or the FCC-ee/CEPC at $\sqrt s=240-380$ GeV\cite{FCC:2018evy,CEPCStudyGroup:2018ghi} do not allow for the 
on-shell production of any of the new PM fields 
discussed in the previous section as their masses likely lie $\gsim 1$ TeV, so, at best, we must concentrate on their indirect effects which are not suppressed by powers of $\epsilon$ 
or by tiny mixing angles given the available integrated luminosities.  Similar kinematic constraints would apply for low energy $\gamma \gamma$ colliders\cite{Telnov:2020gwc} although 
they may have access to different PM sensitivity channels. As noted, since the direct production of PM states cannot occur, we concentrate on the the production of the light $V,S$ states. 
Traditionally at $e^+e^-$ colliders\cite{BaBar:2017tiz,Campajola:2021pdl}, one employs the $e^+e^-\to V\gamma$, $\epsilon^2$-suppressed process to search for the DP, but here we can 
take advantage of the higher center of mass energies to examine the $e^+e^-\to 2V(2S)$ and $e^+e^-\to VS$ processes. These reactions can now occur through $t-,u-$channel $E$ 
exchange suffering no $\epsilon$-dependent coupling suppressions with rates proportional to $\lambda_E^4$ with $\lambda_E=\lambda$ being $O(1)$ as discussed above. The $V$ that 
appears in this case is essentially the longitudinal component of $V_L\simeq G_V$ via the Goldstone Theorem. Here we will 
initially assume that both $S$ and $V$ materialize as boosted, but with otherwise visible, decay products in the $e^+e^-$ collider detector. 
 
\begin{figure}
\centerline{\includegraphics[width=4.5in,angle=0]{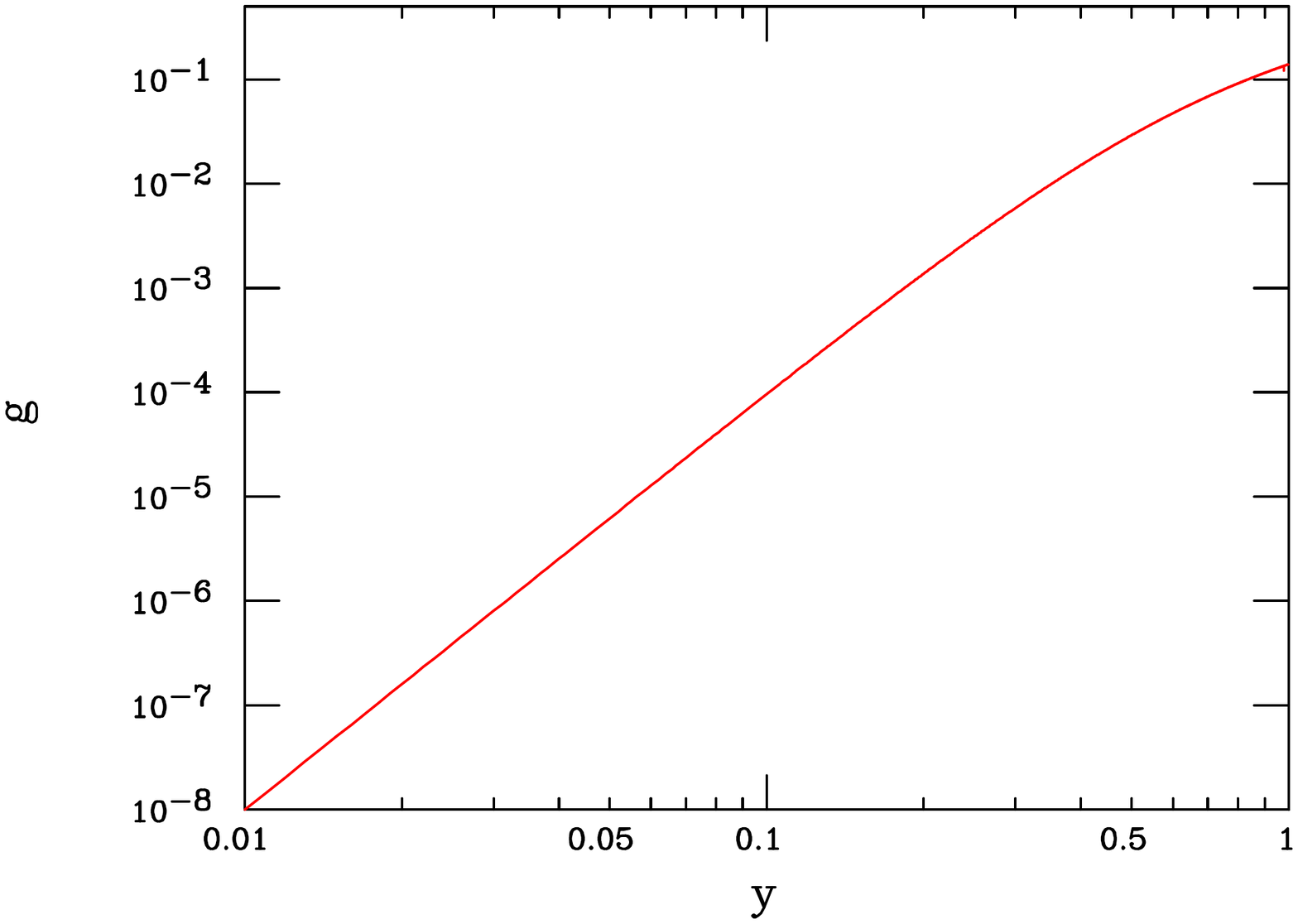}}
\vspace*{-2.3cm}
\centerline{\includegraphics[width=4.5in,angle=0]{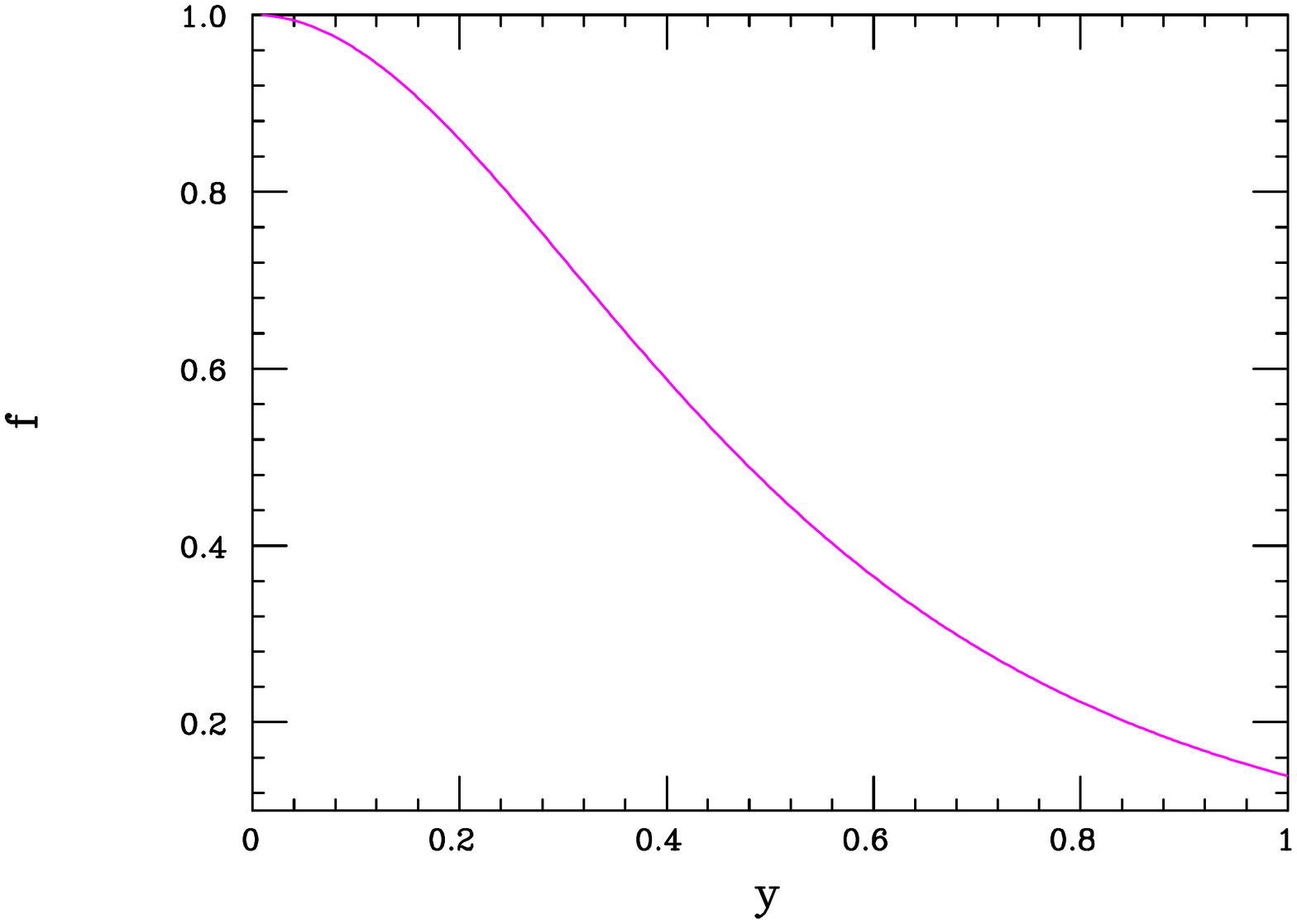}}
\vspace*{-1.30cm}
\caption{The kinematic functions $g(y)$ (top) and $f(y)$ (bottom), as defined in the text, as functions of $y={\sqrt s}/(2m_E) < 1$.}
\label{figa}
\end{figure}

The cross sections for the processes $e^+e^-\to 2V(2S)$ are identical in the $m_{S,V}\to 0$ limit that we consider, since $m_{V,S}^2<<s$,  and are given 
by (in the Goldstone Boson Approximation)
\begin{equation}
\frac{d\sigma_{2V,2S}}{dz}=\frac{\lambda^4}{256\pi s} ~\frac{z^2(1-z^2)}{(a^2-z^2)^2} \,,
\end{equation}
where $a=1+2m_E^2/s$ and $z=\cos \theta^*$, the center of mass scattering angle, while the corresponding cross section for $e^+e^- \to VS$ is given by (in the same limit)
\begin{equation}
\frac{d\sigma_{VS}}{dz}=\frac{\lambda^4}{128\pi s} ~\frac{a^2(1-z^2)}{(a^2-z^2)^2} \,.
\end{equation}
 Note that since $4m_E^2>>s$ for the PM mass range of interest to us and the ILC and FCC-ee/CEPC center of mass energies above, $a>>1$ so that both cross sections will peak at large angles 
 and  that the integrated cross section $\sigma_{2V,2S} \sim (\sqrt s/2m_E)^8$ while that for the $VS$ final state instead scales as  $\sigma_{VS} \sim (\sqrt s/2m_E)^4$ and is thus 
 significantly larger so that we will mainly be interested in this process. To emphasize this scaling behavior we can write this total cross section as 
\begin{equation}
\sigma_{VS}=\frac{\lambda^4}{24\pi s} \Big(\frac{\sqrt s}{2m_E}\Big)^4~f(y) \,,
\end{equation}
where $f(y)=y^{-4}g(y)$ with $y=\sqrt s/(2m_E)$ and, writing $a=1+(2y^2)^{-1}$ we find that
\begin{equation}
g(y) =\frac{3}{16} \Bigg[\frac{1}{2}(a+a^{-1}){\rm log}\big(\frac{a+1}{a-1}\big)-1\Bigg]\,.
\end{equation}
The functions $f$ and $g$ are shown in Fig.~\ref{figa} where as expected we see that $f$ is essentially $O(1)$ over the kinematic range of interest while $g$ displays the very strong 
power-law dependence of the total cross section on $y$. 
The $e^+e^- \to SV$ cross section can be rather significant; indeed, assuming that $\sqrt s=240(380)$ GeV, one obtains $\sigma_{VS}\simeq 7.9(5.8) [\lambda^4 g(y)] \cdot 10^4$ fb. This 
implies that even if the product $\lambda y\lsim 0.1$, a substantial event sample will be obtainable when integrated luminosities on the order of ${\cal L} \sim $ a few ab$^{-1}$ are achieved. For 
completeness we note that under the same set of assumptions, $\sigma_{2S,2V}$ is smaller by a factor of $\sim y^4 \sim 10^{-4}$ and is indeed unobservably small. 

\begin{figure}
\centerline{\includegraphics[width=5.0in,angle=0]{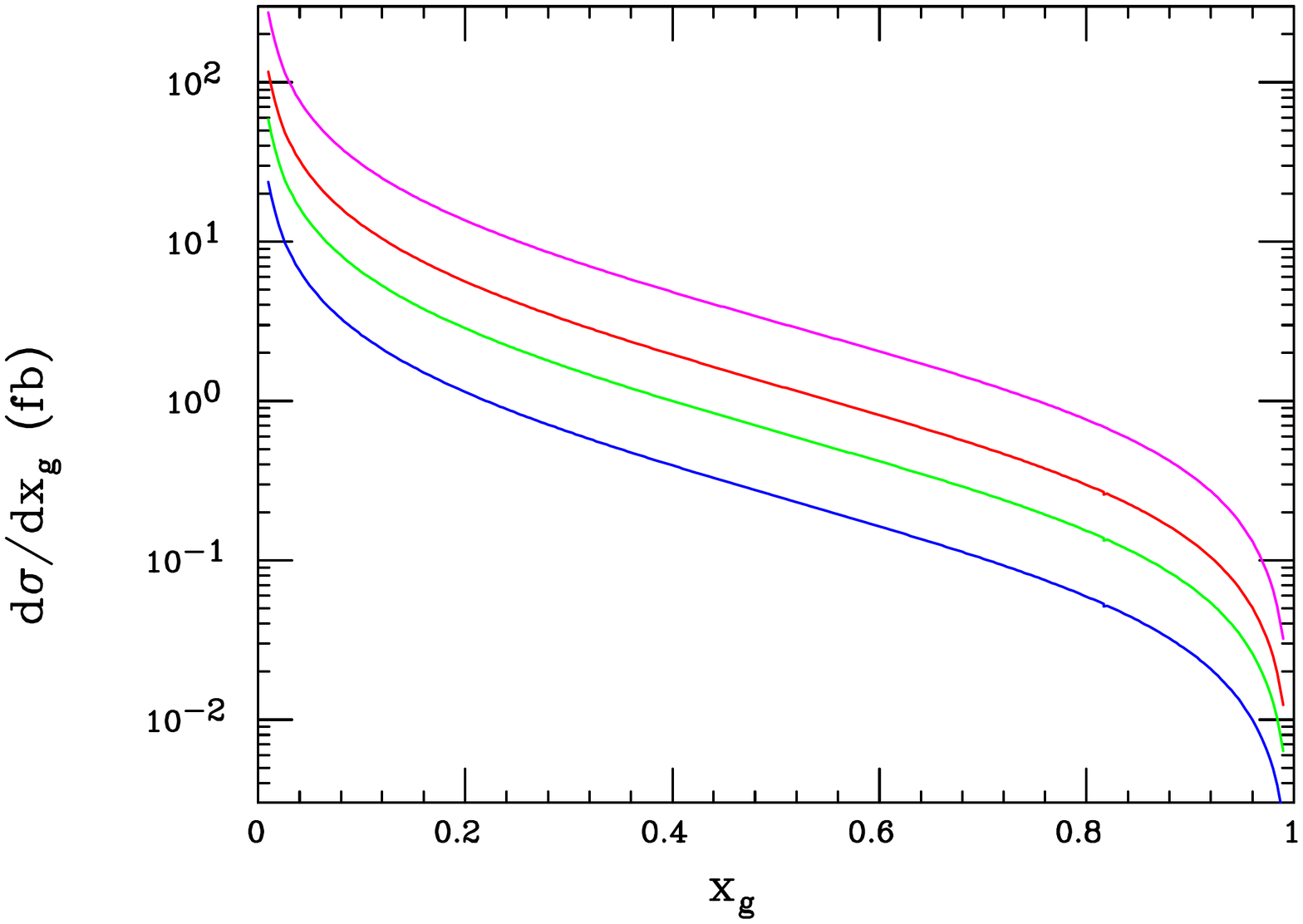}}
\vspace*{-2.3cm}
\centerline{\includegraphics[width=5.0in,angle=0]{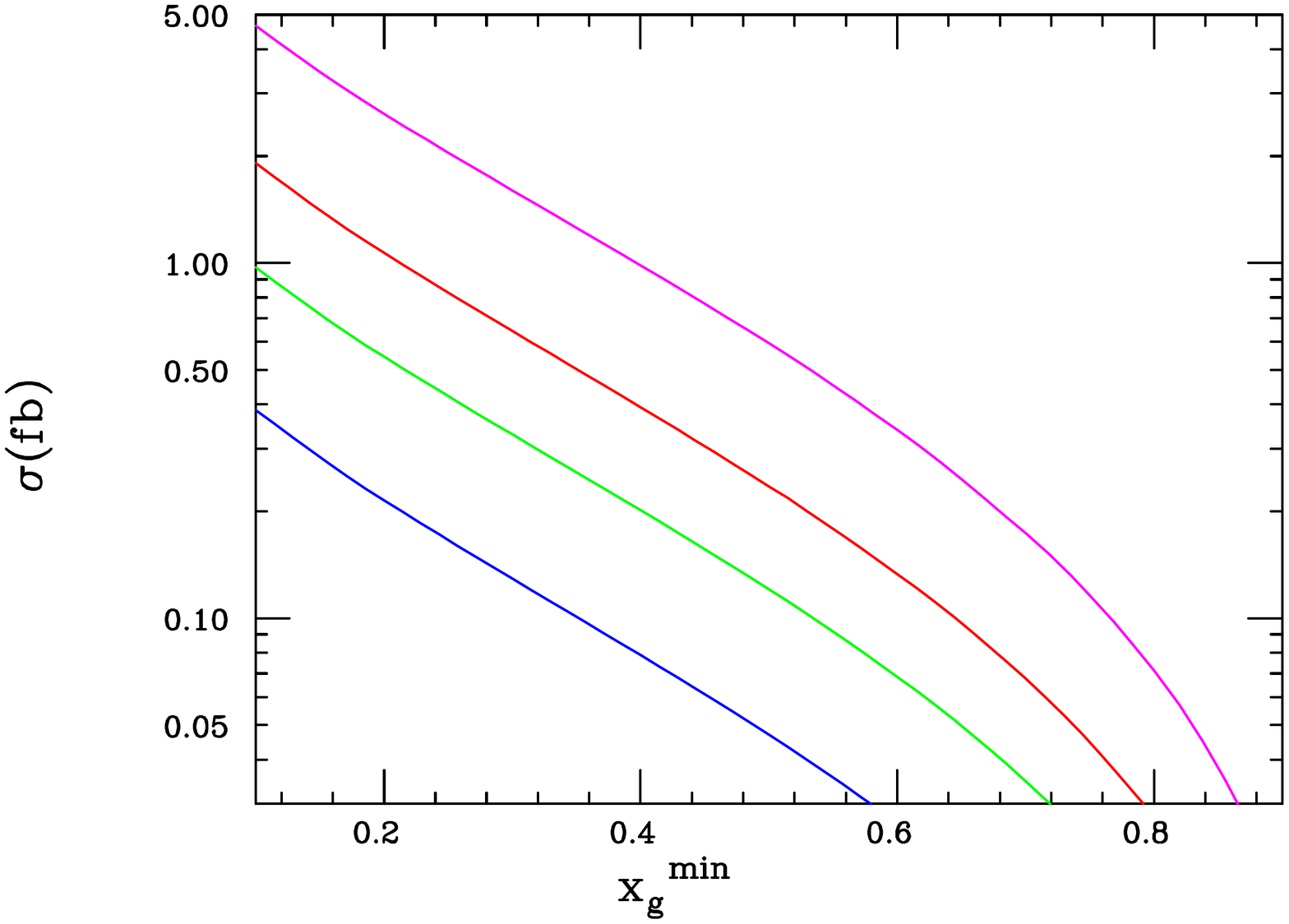}}
\vspace*{-1.30cm}
\caption{Emitted photon energy distribution (top) as a function of $x_g=2E_\gamma/{\sqrt s}$ and the corresponding integrated cross section above a photon energy cut, $x_g^{min}$,  
(bottom) for the process $e^+e^-\to SV$ as described in the text assuming that $\lambda_E=1$. From top to bottom the curves assume that $\sqrt s=380$ GeV, $m_E=1$ TeV; 
$\sqrt s=240$ GeV, $m_E=1$ TeV; $\sqrt s=380$ GeV, $m_E=1.5$ TeV and $\sqrt s= 240$ GeV, $m_E=1.5$ TeV, respectively.}
\label{figb}
\end{figure}

So far we have assumed that $S,V$ materialize as visible objects in the detector but if they are long-lived or $V$ decays to DM then there is no way to really observe such events. The 
usual approach, which we follow here, is then to require an ISR photon as a tag as part of the $SV$ production process which will clearly reduce the event rate before cuts by a factor of order 
$\sim (\alpha/\pi)log(s/m_e^2)$. The final state will now appear as arising from $e^+e^- \to \gamma+$missing energy which has a significant well-known background from the 
SM process $e^+e^- \to \gamma \bar \nu \nu$ arising from $s$-channel $Z$ exchange as well as $t-$channel $W$ exchange when $\nu=\nu_e$. The resulting cross section 
assuming that, \eg,  $\lambda_E=1$ and for fixed values of $\sqrt s$ and $m_E$ is then directly obtainable; the results of this calculation are shown in Fig.~\ref{figb} for $d\sigma_{SV}/dx_g$, 
where $x_g=2E_\gamma/\sqrt s$. Also shown here in the integrated cross section above a given minimum cut on the photon energy, \ie, $x_g^{min}$, for the same values of the other parameters.
Here one sees that the shape of the photon energy distribution is as expected from a typical ISR process and that the overall rate is quite small, $\lsim 5$ fb, even for judicious choices of the 
parameters $m_E,\lambda_E$. As is well-known from past studies\cite{Dreiner:2007vm}, even with well chosen cuts and with both beams polarized it will be difficult to lower the SM background 
to with a factor of $\sim 10-100$ times the expected signal rate for $SV$ production in this scenario. Clearly, if $S,V$ do decay invisibly or are very long-lived the usual search approaches 
for signals such as this will very likely fail.

\subsection{$\gamma \gamma$ Colliders}

Another way one might imagine to possibly probe the PM sector indirectly 
is via the process $\gamma \gamma \to VV$ in analogy to the well-studied SM $gg,\gamma \gamma \to ZZ$ reactions\cite{Glover:1988rg,Jikia:1993di} which we will briefly 
consider. Again, we see that this reaction is $\epsilon$-independent with an amplitude proportional to the product $N_c(eg_DQ_{em}Q_D)^2$ for each of the PM fields circulating in the box graphs. 
However, like the 
$e^+e^-\to SV$ tree level process, this $\gamma \gamma$ reaction is, as might be expected, highly suppressed by the small values of $\sqrt s/m_{PM}$ that we encountered previously 
as we will see below. 

Following the analyses as presented in Refs.\cite{Preucil:2017wen,Quevillon:2018mfl,DeFabritiis:2021qib,Ghasemkhani:2021kzf}, and suppressing all
 Lorentz indices, we note that the $\gamma \gamma \to 2V$ process in this limit can be described by the effective Lagrangian of the form 
\begin{equation}
{\cal L}_{eff} =\frac{\alpha \alpha_D}{90m_0^4} ~ \Bigg[\gamma_1(F^2)(V^2)+\gamma_1'(FV)(FV)+\gamma_2(F\tilde F)(V\tilde V)+\gamma_2'(F\tilde V)(f\tilde V)\Bigg]\,,
\end{equation}
where $m_0$ is a reference scale $\sim 1$ TeV typical of the PM masses, $F(V)$ is the photon(DP) field strength tensor and $\tilde F(\tilde V)$ its corresponding dual. 
For fermionic and scalar PM appearing in the box graph, one finds that\cite{Preucil:2017wen,Quevillon:2018mfl,DeFabritiis:2021qib,Ghasemkhani:2021kzf} $\gamma_i'=2\gamma_i$, 
where the $\gamma_i$ themselves are $O(1)$ sums over the various PM in the loop: 
\begin{equation}
\gamma_i=\sum_k ~\eta_{(F,S),i_k}(Q_D^2Q_{em}^2N_c)_k~\frac{m_0^4}{m_k^4}\,,
\end{equation}
where the sum is over all fields in the box graph with masses $m_k$, \etc, and where $\eta_{(F,S),1}=(1,7/16)$ and $\eta_{(F,S),2}=(7/4,1/16)$ are the relevant coefficients 
for (fermionic, scalar) PM fields. This leads to the $\gamma \gamma \to VV$ differential cross section in the center of mass frame, assuming that $m_{V,S}^2<<s$,  of 
\begin{equation}
\frac{d\sigma}{dz}=\Big(\frac{s^3}{8\pi}\Big)~\Bigg[\frac{\alpha \alpha_D}{90m_0^4}\Bigg]^2~(3+z^2)^2~ \Big[(\gamma_1-\gamma_2)^2+2(\gamma_1^2+\gamma_2^2)\Big]\,,
\end{equation}
where $z=\cos \theta^*$ as above. Defining, similar to the above,  the dimensionless ratio $y_0=\sqrt s/(2m_0)$, this yields an extremely small total cross section of 
\begin{equation}
\sigma \simeq \frac{0.29~{\rm fb}}{s ~(\rm{in~ TeV^2})}~\alpha_D^2~y_0^8 ~X\,,
\end{equation}
where X is the just combination of the $\gamma_i$'s appearing in the final bracket in the previous expression and is expected to be $O(1)$ or perhaps slightly larger. 
Forgetting the effects of the photon energy 
distributions relevant for the $\gamma \gamma$ collider for now, by taking the suggestive values $\sqrt s=0.2$ TeV, $\alpha_D=X=1$ and $y_0=0.1$, we find that $\sigma=7.3\cdot 10^{-5}$ 
ab, which is essentially invisible at any foreseeable collider luminosity. Outside of some radical developments it's clear that we can forget about this particular possibility for now. However, 
other related $\gamma \gamma$ processes\cite{tgrgg} may lead to somewhat enhanced production cross sections in comparison to the one we have just obtained.

\subsection{LHC and HL-LHC}

The most obvious new channels at the LHC (and future hadron colliders) are the production of the PM fields themselves along with any of the new gauge bosons associated with an extended dark 
gauge group. In almost all cases, the cross sections for the various processes we consider below can be found in Ref.\cite{Rueter:2019wdf} and which we heavily quote below.

\begin{figure}
\centerline{\includegraphics[width=5.0in,angle=0]{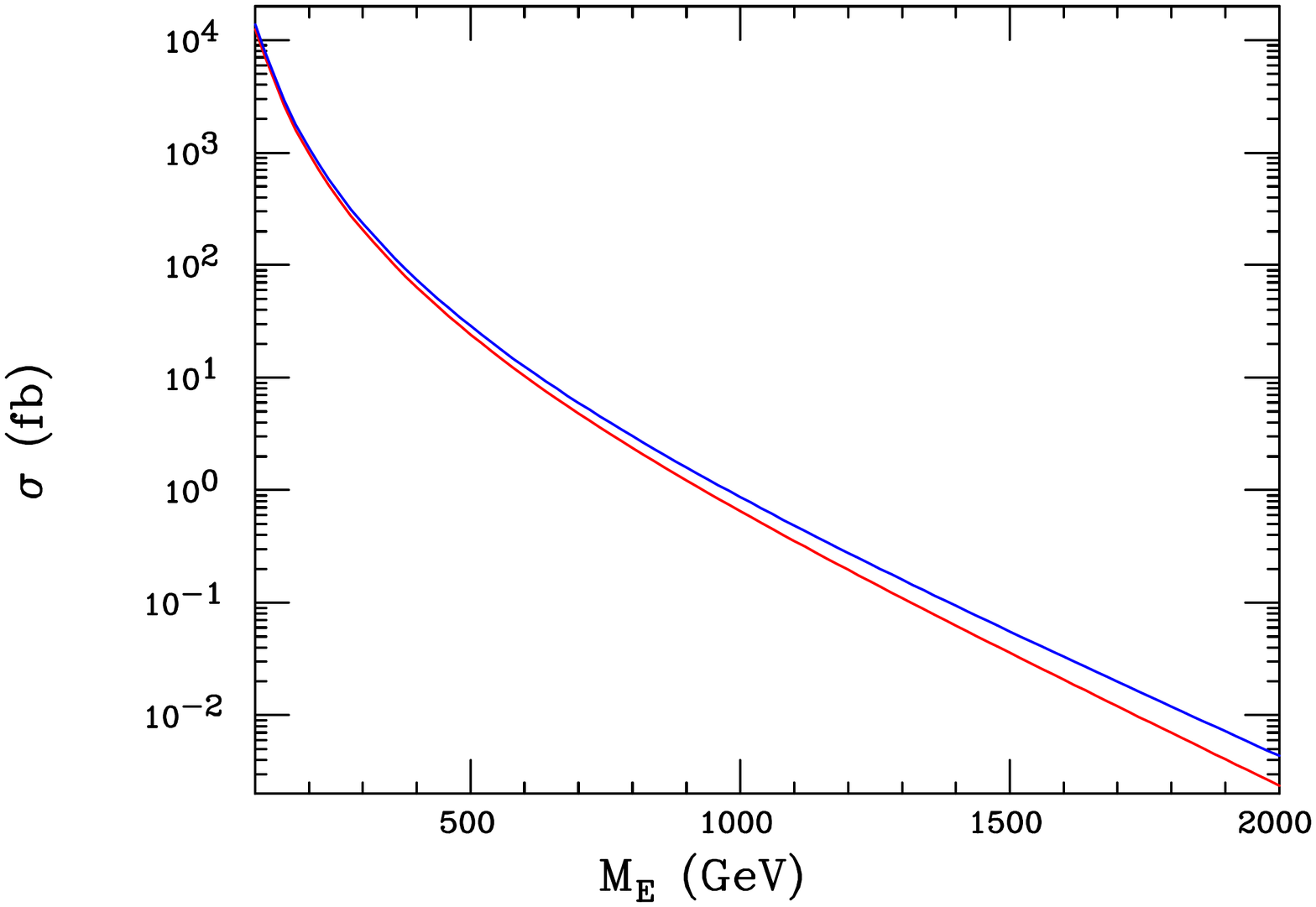}}
\vspace*{-2.3cm}
\centerline{\includegraphics[width=5.0in,angle=0]{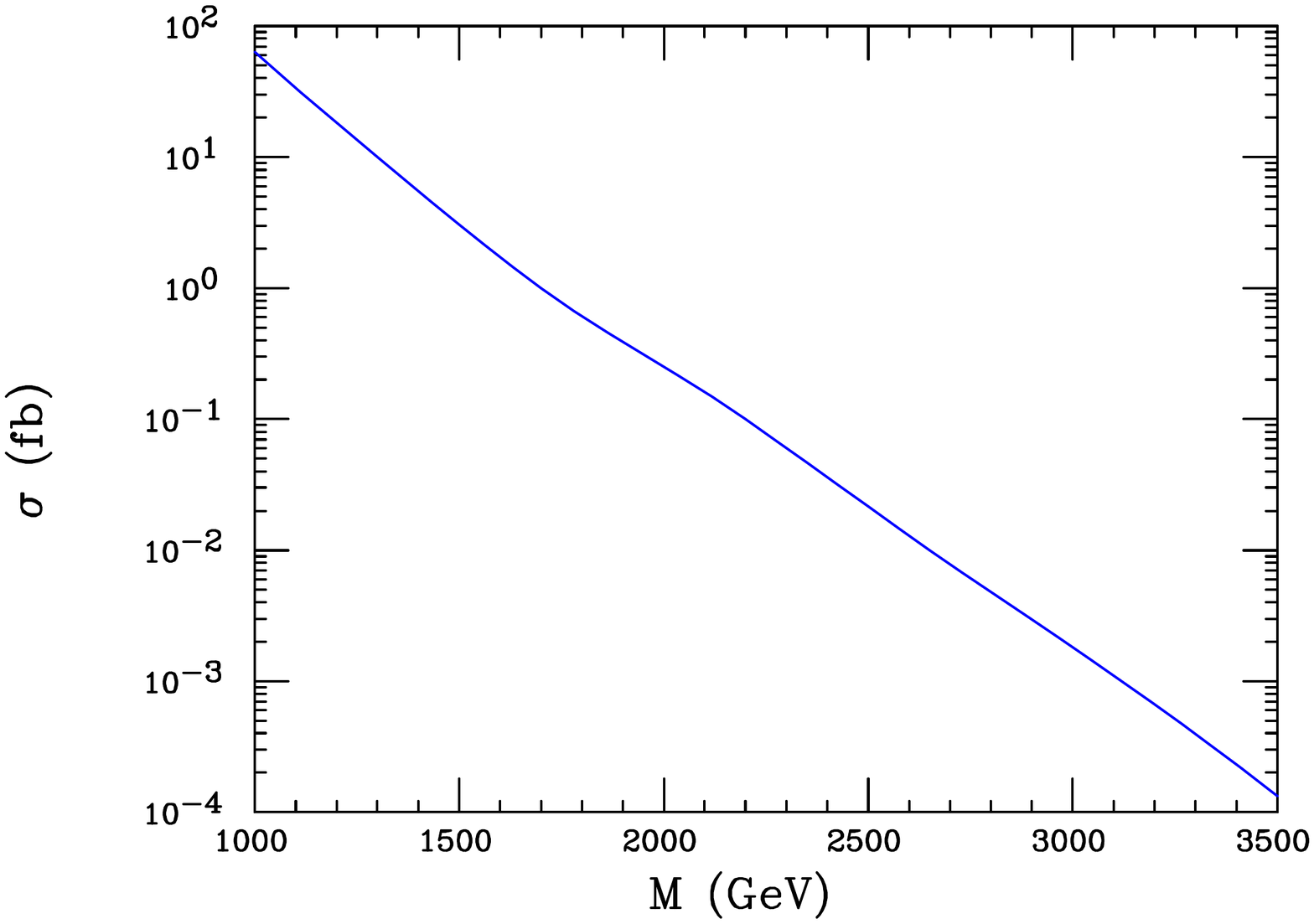}}
\vspace*{-1.30cm}
\caption{(Top) $q\bar q \to W^{\pm*}_{SM}\to E\bar N+N\bar E$ production cross section at the $\sqrt s=13$ (red) and 14 (blue) LHC as a function of $m_N=m_E$. (Bottom) Inclusive QCD 
NNLO $\bar DD$ production cross section at the $\sqrt s=14$ TeV LHC as a function of $m_D$ using HATHOR following Ref.~\cite{hpair}.}
\label{figENDD}
\end{figure}

Clearly, we need to begin this discussion by considering $\bar DD$ production via QCD (under the usual assumptions) 
and $\bar EE$ production via SM $\gamma,Z$ exchange; these cross sections are well-known as they are 
identical to those for the conventionally examined vector-like quarks and leptons. For example, Fig. 2 (left) in Ref.\cite{OsmanAcar:2021plv} shows the total $\bar EE$ production cross section for both the 
isosinglet and, the somewhat 
larger, isodoublet PM cases. In the isodoublet case, the process $\bar q q\to W^\pm \to \bar NE+\bar EN$ can also occur with $m_N=m_E$, up to small mixing and radiative corrections, and with a 
somewhat larger cross section at the LHC as is shown in the top panel of Fig.~\ref{figENDD} from Ref.\cite{Rueter:2019wdf}. The lower panel of this Figure also shows the 
total $\bar DD$ cross section at the 
$\sqrt s=14$ TeV LHC from this same paper.  As noted above, the signatures associated with the production of the PM states at colliders is dependent upon whether or not the DP 
decays invisibly, either due to a very long lifetime or via a kinematically allowed decay into DM. In the case of invisible decays, SUSY searches for sleptons and squarks can be recast to obtain 
the corresponding limits on PM fermions. In the case of a pair of isosinglet PM leptons decaying into $e$ or $\mu$ pairs plus MET, this analysis has been performed 
in Ref.\cite{Guedes:2021oqx} for the 
$\sqrt s=13$ TeV LHC with ${\cal L}=139$ fb$^{-1}$ leading to a lower bound of 895 GeV with a limit which is expected to be $\simeq 180$ GeV greater in the isodoublet case due to the 
larger production cross section\cite{OsmanAcar:2021plv} resulting from the alternate couplings to the SM $Z$.  
Somewhat weaker results would be expected in the case of the $\tau$ final state due to a smaller $\tau$ ID efficiency\cite{Kumar:2015tna,Bhattiprolu:2019vdu}. A similar analysis for the 
$W^\pm$-initiated $e^\pm$+MET final state, as would occur for $\Bar N E+ \bar EN$ production,  has not yet been 
performed. Corresponding analyses for color-triplet PM point to limits in the range of $\simeq 1.3-1.5$ TeV or so depending somewhat upon the the quark flavor in the final 
state\cite{Rizzo:2018vlb,Kim:2019oyh}. In the 
cases where the {\it very} highly boosted ($\gamma > 10^3$) DP decays visibly in the detector to lepton-jets\cite{Rizzo:2018vlb} the analyses have not yet been performed but are 
expected to yield comparable limits but which are more strongly dependent on the model parameters and other details.

It is interesting to note that $\bar DD$ production at hadron colliders can, in fact, be altered in scenarios similar to the ones we have consider in the previous Section in a way not 
experienced by ordinary vector-like quarks. For example, in the $E_6$-inspired case, although 
the $gg, \bar uu, \bar cc\to \bar DD$ processes are unaffected, $\bar dd, \bar ss, \bar bb \to \bar DD$ may be modified by $t-$channel exchanges of light dark Higgs fields and longitudinal DP's  
due to the couplings seen above in Eqs.(5) and (6), especially so if these $\lambda$ couplings are sufficiently large. Since it is easily imagined that $\lambda_i \sim g_s$, the QCD coupling, these 
exchanges may make substantial alterations to both the angular distributions for $\bar DD$ production, pushing it more forward, as well as to the overall total cross section, influencing search 
sensitivities, especially when the coupling is to the first generation quarks. These issues are briefly examined in Ref.\cite{Rueter:2019wdf} but a detailed study of this possibility is certainly 
warranted.

\begin{figure}[htbp]
\centerline{\includegraphics[width=5.0in,angle=0]{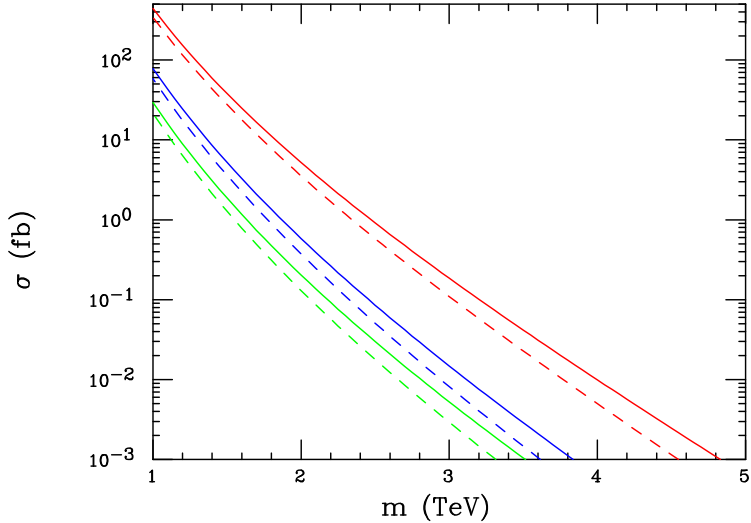}}
\vspace*{-1.50cm}
\caption{The $gq\to D(G_V,S)$ production cross section at the $\sqrt s=14$(solid) and 13(dashed) TeV LHC in the Goldstone Boson Approximation as a function of the mass of the 
$D$ assuming that $\lambda_D=1$. 
From top to bottom, the curves are for $q=d,s,b$, respectively.}
\label{singleD}
\end{figure}

Finally, before moving on to the signatures of the extended gauge sector of the $2_I1_{I'}$ scenario, one might consider the possibility of single production of the PM $D$ state via its mixing 
with the $q=d,s$ or$b$ quarks. In the usual picture of single VL quark production via a SM $W$ charged-current interaction, this process will have a cross section which sensitively 
dependent upon the relevant mixing matrix 
element. Unlike the direct Yukawa couplings of the form $\bar qD(G_V,S)$ discussed above, where the action of the Goldstone theorem rescues us from a possible extremely mixing 
suppressed interaction as we will shortly see, this salvatory effect is absent for this mechanism of single $D$ production so that this cross section is {\it indeed} quite highly suppressed. 
However,  the $gd(s,b)\to D(S,G_V)$ associated production process cross section can indeed be quite significant, dependent only upon the values of $\lambda_D$ and $m_D$ as well as the 
choice of generation with which quark $D$ is partnered within the $SU(2)_I$ doublet. We note that the enhancement seen here is similar to that observed for the decay $D\to qG_V$ in 
comparison to, \eg, $D\to Zq$ that was discussed above. The $gq\to DG_V$ sub-process differential cross section for this reaction employing the same notation as above is given in the 
limit that $m_{V,S}^2<< \hat s$ by
\begin{equation}
\frac{d\sigma}{dz}=\frac{\alpha_s \lambda_D^2}{16\hat s}~\Big[\frac{\hat s}{-t}+\frac{-t}{\hat s}-2 -2m_D^2\big(1+\frac{m_D^2}{t}\big) \big(\frac{1}{t}+\frac{1}{\hat s}\big)\Big] \,,
\end{equation}
where $\hat s$ is the the sub-process center of mass energy squared as usual and where $t=\frac{-\hat s}{2}(\tilde E-\tilde pz)$ with $\tilde E,\tilde p=1\pm\frac{m_D^2}{\hat s}$, respectively.  
Fig.~\ref{singleD} shows the total cross section for this process at the $\sqrt s=13$ and 14 TeV LHC as a function of $m_D$ assuming that $\lambda_D=1$ for ease of rescaling. Here we see 
that, indeed, 
at smaller values of $m_D$ (again, for $\lambda_D=1$), this cross section is quite comparable to the familiar $\bar D D$ cross section (when $q=d$), this being enhanced by the larger 
gluon parton density while simultaneously being suppressed by the relative heavy $D$ phase space. For larger values of $m_D$, this associated production process may yield a larger rate 
when $q=d$. Of course, for $q=s,b$, the cross section for all values of $m_D$ is seen to be somewhat smaller as would be expected. Assuming that $S/V$ produces MET as discussed above, 
the experimental constraints from a lack of any monojet signature at the LHC can be used to place constraints in the $\lambda_D-m_D$ parameter plane for the different choices of $q$; 
this analysis has yet to be performed. 

We now turn to the production of the new gauge bosons present in this model. 
The production of and signatures for the $Q_{em}=Q_D=0$, $E_6$-inspired gauge boson, $Z_I$, are in many ways similar to a host of familiar $Z'$ 
scenarios\cite{Hewett:1988xc,Leike:1998wr,Rizzo:2006nw,Langacker:2008yv} but with some potentially 
very interesting differences depending upon the PM mass spectrum, its flavor nature, as well as the value of the parameter $x_I$ which appears in its couplings and controls the $Z_I-W_I$ 
mass relationship, \ie, if $x_I>3/4$ the decay $Z_I \to W_IW_I^\dagger$ can occur on-shell due to the usual non-abelian coupling. Recall that $Z_I$ couples to the familiar-looking combination 
$\frac{g_I}{c_I}(T_{3I}-x_IQ_D)$ so that the $T_{3I}$ assignments of the SM fermions and PM fields will play an important role in its phenomenology. Note that in all cases, the $Z_I$ 
does {\it not} couple to the $u,c,t$ quarks. It is at this point where the question of whether only a single SM generation 
(and indeed {\it which} one) or all three generations have PM partners and thus carry $SU(2)_I$ quantum numbers perhaps becomes the most critical. For example, if $Z_I$ only couples to the 
second generation of the SM, then it can only be produced via $\bar ss$ annihilation and decay to the unique charged lepton $\mu^+\mu^-$ final state.  This is also directly related to the question 
of how, \eg, $D$ mixes with $d,s,$ or $b$ or whether there is a $D$ for each generation. To cover all these possibilities, we consider 
two extreme cases: one where only a single specific generation is augmented by PM fields (which has important flavor physics implications) or instead, where all three of the SM 
generations are augmented and so that the usual family universality is enforced.

\begin{figure}[htbp]
\centerline{\includegraphics[width=5.0in,angle=0]{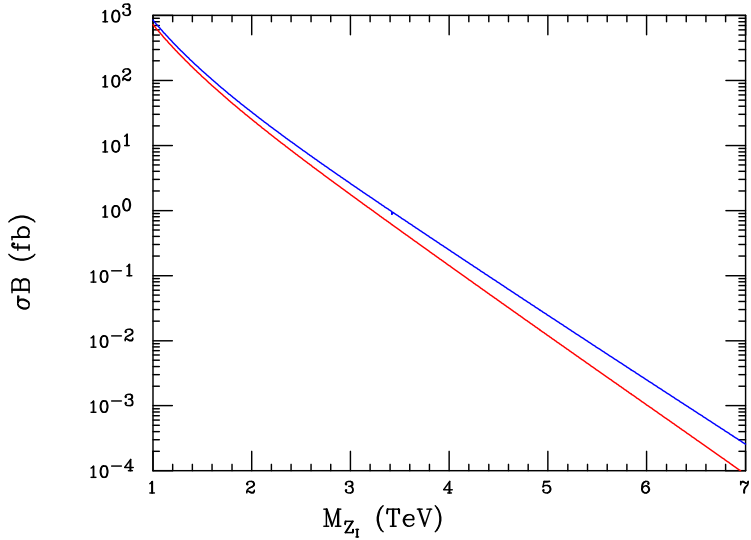}}
\vspace*{-1.50cm}
\caption{The values of $\sigma B_\ell$ for $Z_I$ production as a function of mass at the $\sqrt s=13$ (red) and 14 (blue) TeV LHC. Here, only decays to SM fields are assumed to be kinematically 
allowed and $r=1$ is assumed. It is assumed here that the $Z_I$ couples in a generation-independent manner. }
\label{ZI}
\end{figure}

Perhaps the simplest possible scenario is where the $Z_I$ can decay only to SM particles by kinematic constraints, \ie, $m_{Z_I}<2m_{PM},2m_{W_I}$. Since the SM fields by definition have 
$Q_D=Q_I=0$, the Drell-Yan signal rate for $\bar q q\to Z_I \to \ell^+\ell^-$, in the narrow-width approximation (NWA), $\sigma B_\ell$, with $B_\ell$ being the leptonic branching fraction for 
$Z_I\to \ell^+\ell^-$, is independent of $x_I$ but does depend upon the overall coupling strength $g_I/c_I$ which is unknown {\it a priori}. Writing $g_I/c_I=r~g/c_w$, $\sigma B_\ell$ is then 
totally determined up to an overall factor of $r^2$ as a function of $m_{Z_I}$ since only SM final states are involved. Here we will assume that $r=1$ so that our results can easily 
be appropriately rescaled by this 
overall factor. The first case we will consider is that where all 3 SM generations carry $2_I1_{I'}$ quantum numbers, \ie, generation-independence/family universality exists as is typically the 
case in $Z'$ models with the results shown in Figs.~\ref{ZI} and ~\ref{Dylan1} from Ref.\cite{Rueter:2019wdf}.  As can be seen here, the present ATLAS null searches\cite{Aad:2019fac} 
employing 139 fb$^{-1}$ of $\sqrt  s=$13 TeV integrated 
luminosity presently excludes $Z_I$ masses below $\simeq 5.2$ TeV under these set of assumptions. A similar null search performed at the $\sqrt s=$14 TeV HL-LHC with 3 ab$^{-1}$ of 
luminosity\cite{ATLASNote} would increase the exclusion limit on a $Z_I$ to masses $\simeq 5.9$ TeV\cite{Rueter:2019wdf} under the same set of assumptions.

\begin{figure}[htbp]
\vspace*{0.5cm}
\centerline{\includegraphics[width=4.5in,angle=0]{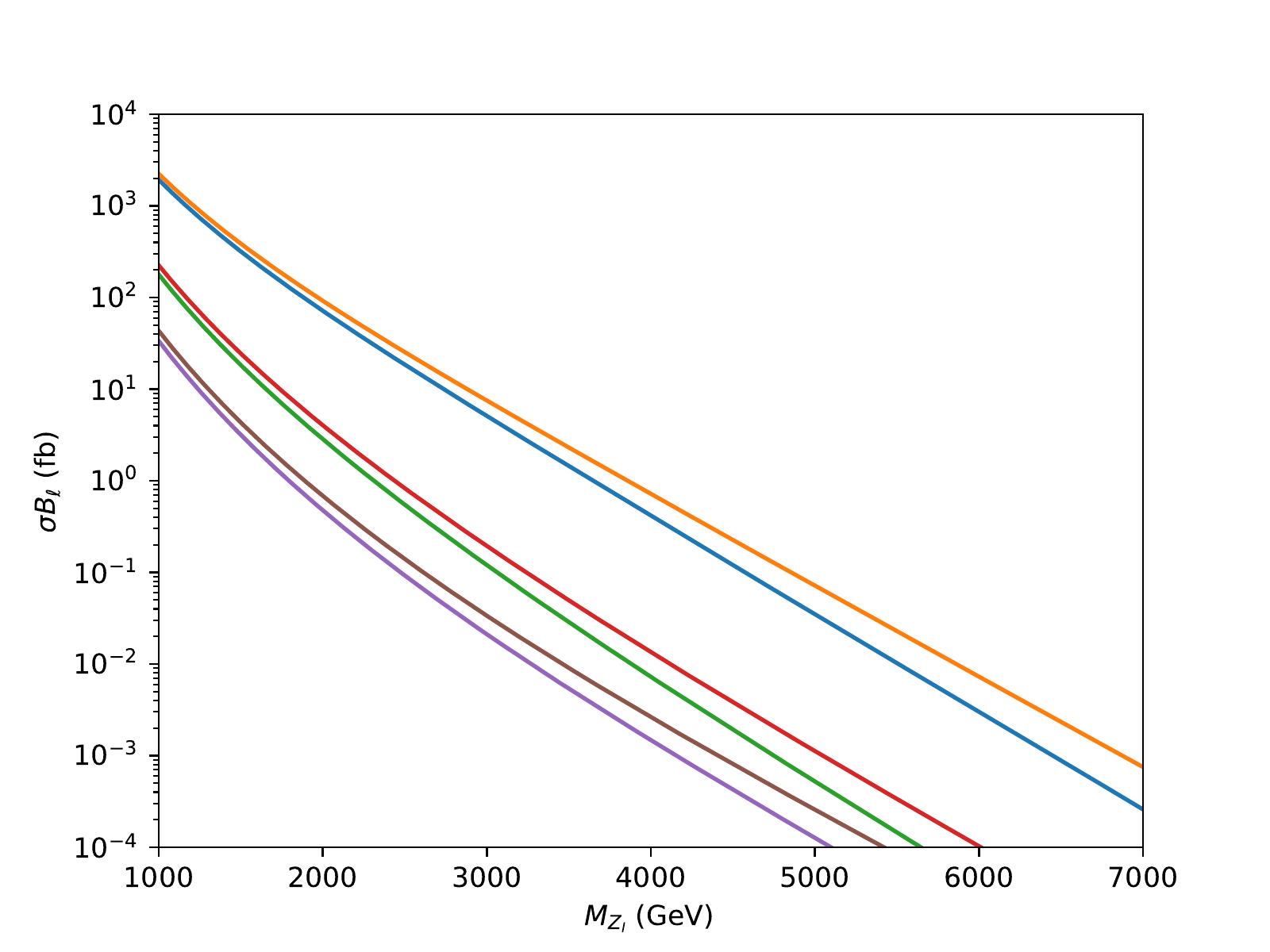}}
\vspace*{0.30cm}
\caption{$\sigma B_\ell$ for $Z_I$ production at the $\sqrt s=13$ (lower) and 14 TeV (upper) LHC as a function of mass with $r=1$. For each pair of curves it is assumed that the $Z_I$ 
only couples to a single SM generation and that decays to PM and $W_I$ are kinematically forbidden. From top to bottom, the curves correspond to a $Z_I$ coupling to the first, second, 
or third generation of the SM, respectively, in the absence of mixing.}
\label{Dylan2}
\end{figure}

\begin{figure}[htbp]
\vspace*{0.5cm}
\centerline{\includegraphics[width=3.7in,angle=0]{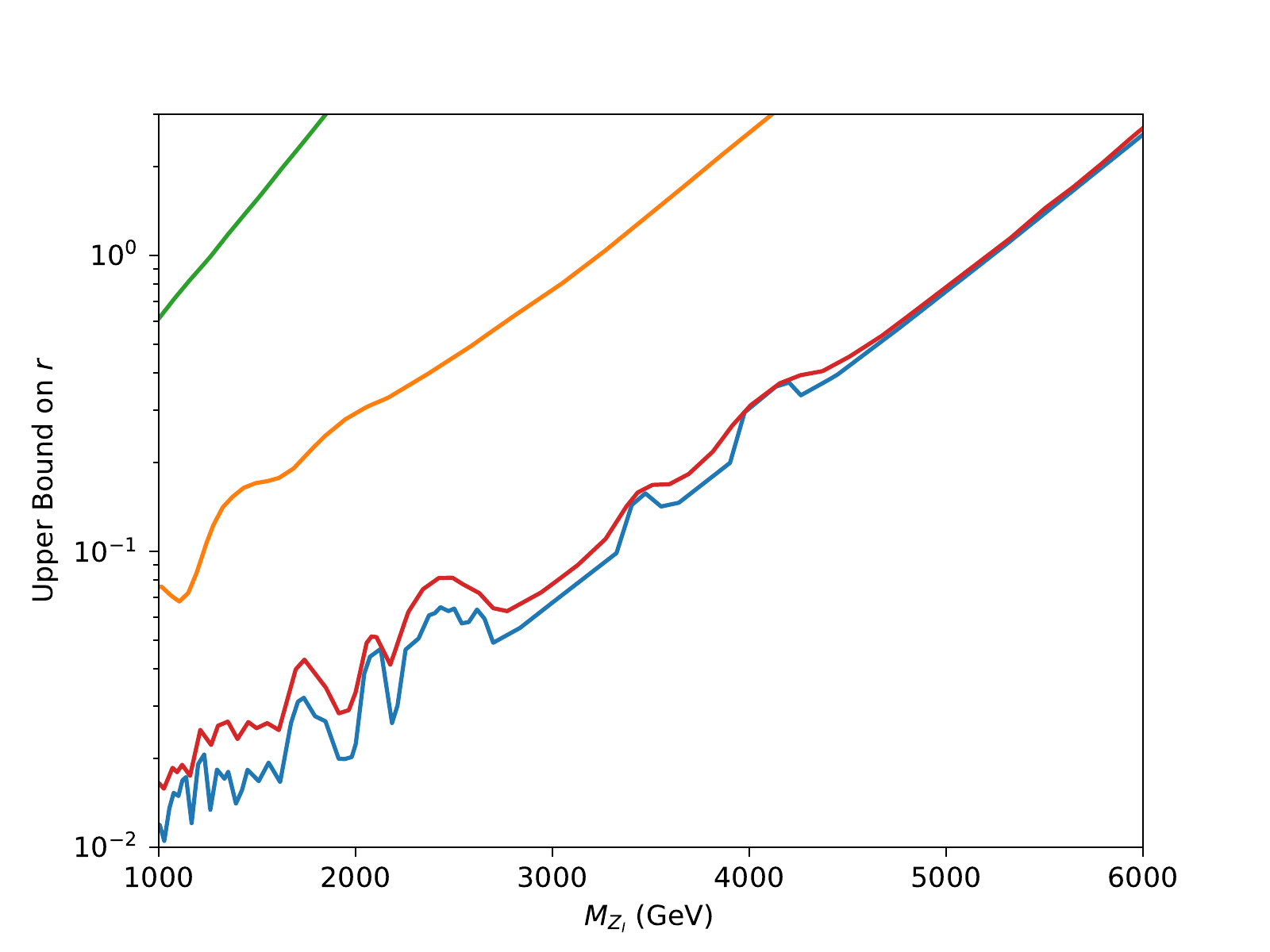}
\hspace*{-0.7cm}
\includegraphics[width=3.7in,angle=0]{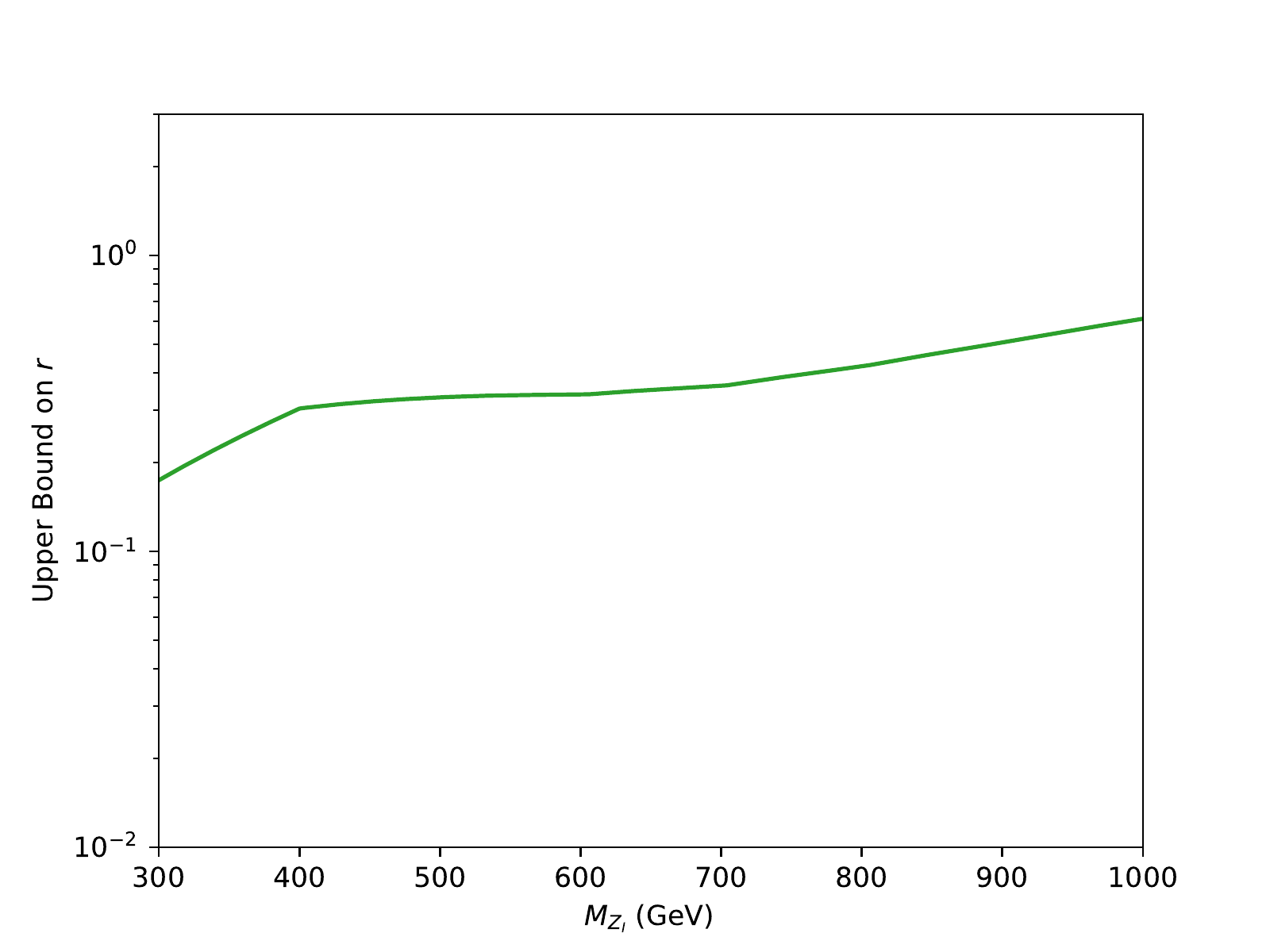}}
\vspace*{1.0cm}
\centerline{\includegraphics[width=3.7in,angle=0]{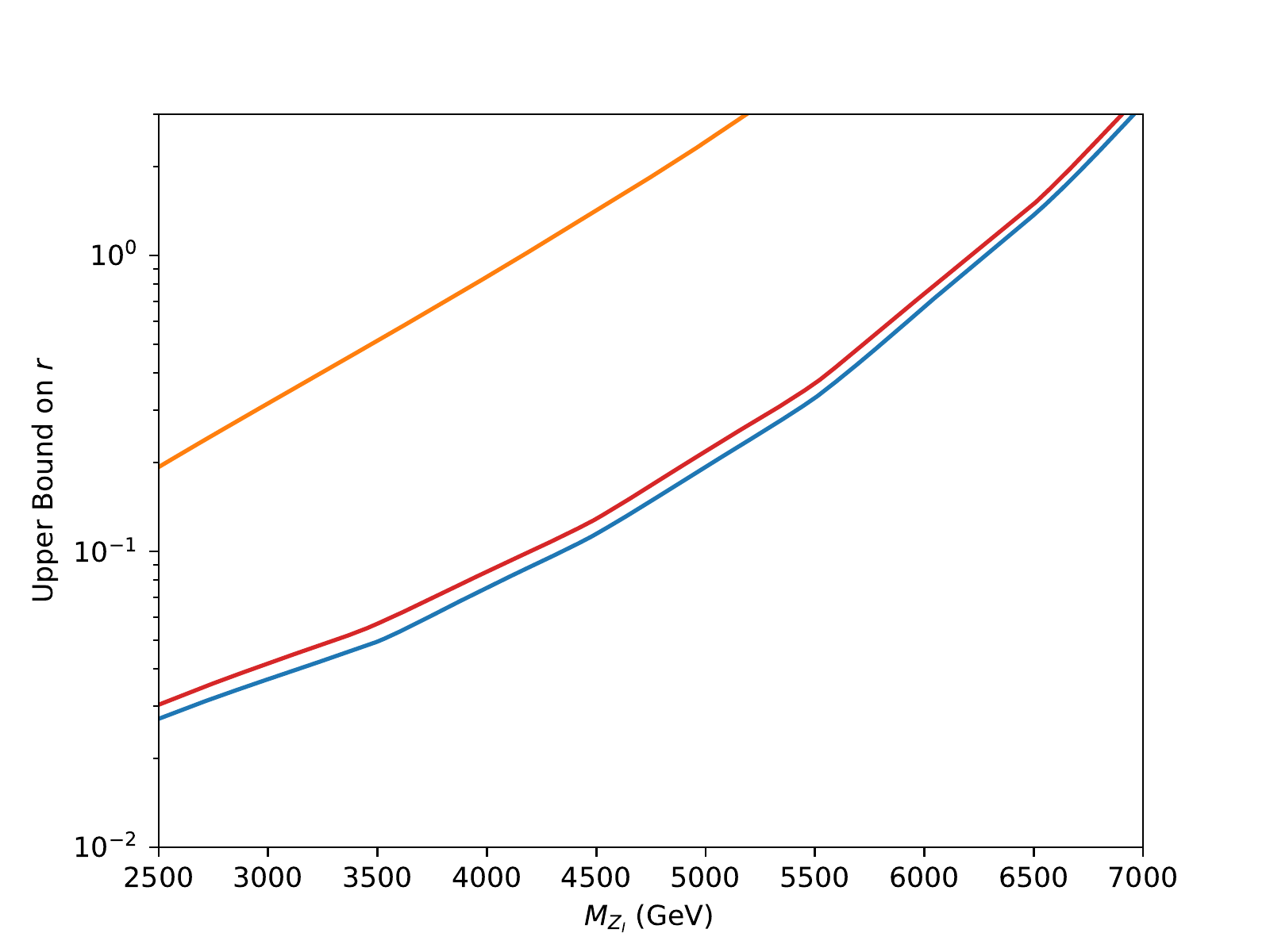}
\hspace*{-0.7cm}
\includegraphics[width=3.7in,angle=0]{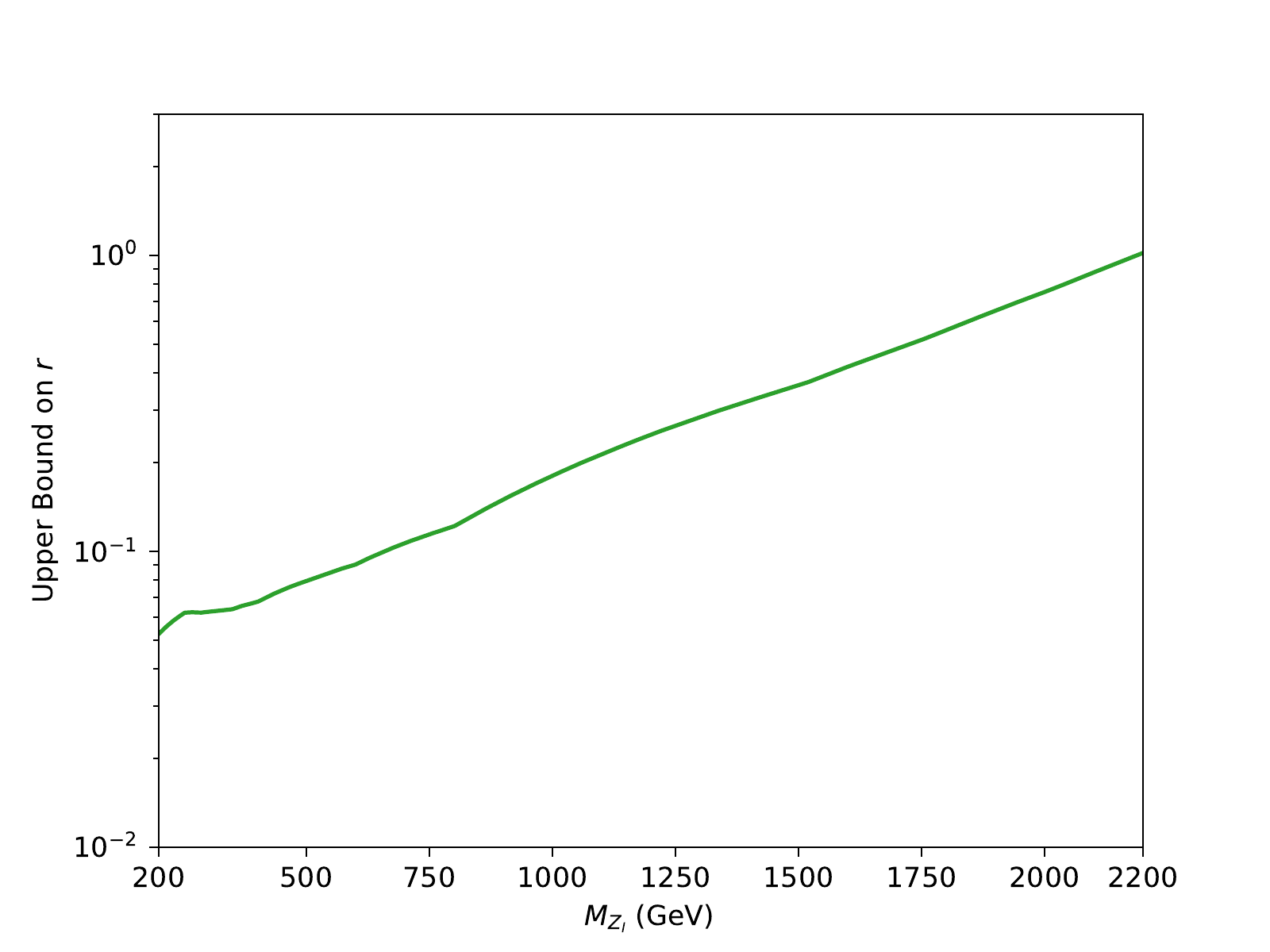}}
\vspace*{0.50cm}
\caption{(Top Left) Limits from the $\sqrt{s} = 13$ TeV LHC on the values of the parameter $r$ as described in the text, as a function of the mass of the $Z_I$, employing the results from ATLAS searching for dilepton decays \cite{Aad:2019fac,ATLAStau36}. From left to right the curves correspond to only third generation couplings (green), only second generation couplings (gold), 
universal couplings (red), and only first generation couplings (blue). (Top Right) Extrapolation of the results in the previous panel for the case of third generation couplings to lower $Z_I$ 
masses. (Bottom Left) Same as the top left panel, but now employing an ATLAS analysis assuming a null result at the HL-LHC with $\sqrt s=14$ TeV and L=3 ab$^{-1}$ \cite{ATLASNote}. 
(Bottom Right) The corresponding limit in the $\tau^+\tau^-$ case employing the ATLAS heavy Higgs study\cite{ATLASNote2} with acceptance corrections included for a spin-1 state.}
\label{Dylan1}
\end{figure}

The next logical case is where only a single SM generation carries non-trivial values of $T_{3I}$ and so couples to the $Z_I$. While the value of $B_\ell$ is the same in all three case (being a 
factor of 3 larger than in the universal case just discussed), the production cross section itself is not as the $\bar dd$, $\bar ss$ and $\bar bb$ parton luminosities are all very different. 
Fig.~\ref{Dylan2} from \cite{Rueter:2019wdf} shows the values of $\sigma B_\ell$ in these three individual cases, here ignoring any effects from inter-generational mixing for simplicity. 

Fig.~\ref{Dylan1} from Ref.\cite{Rueter:2019wdf}  then shows 
how these $Z_I$ production cross sections in the various dilepton channels translate into current LHC search limits and the expectations for 
the HL-LHC assuming either universal couplings or coupling to only one of the SM generations. In order to obtain these results in the case of the $\tau^+\tau^-$ third generation couplings, 
the results from an ATLAS $b\bar b\to H\to \tau^+\tau^-$ study\cite{ATLAStau36,ATLASNote2} were recast, making corrections for the detector acceptance differences between spin-0 and 
spin-1 resonances. 

\begin{figure}[htbp]
\centerline{\includegraphics[width=5.0in,angle=0]{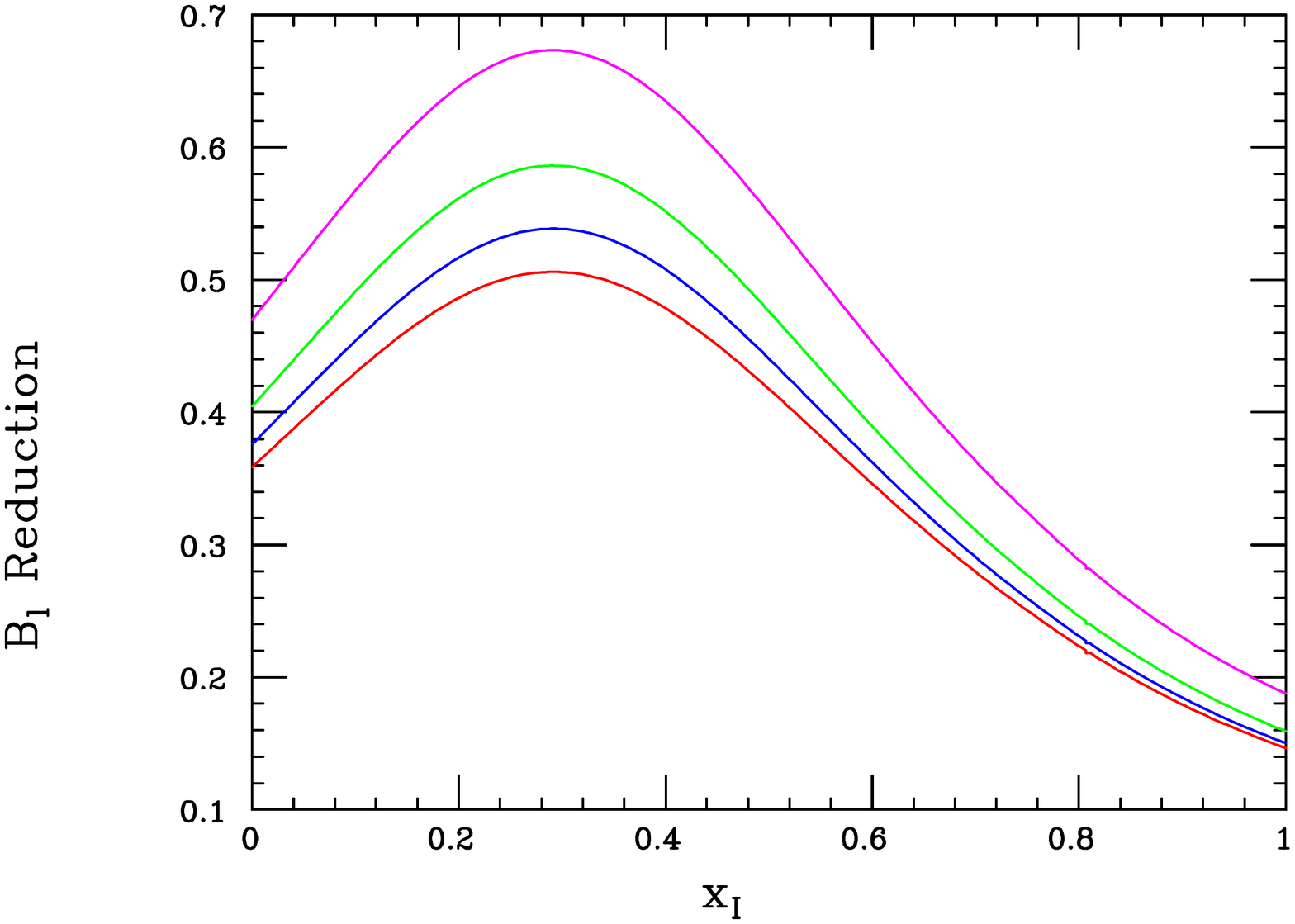}}
\vspace*{-1.50cm}
\caption{Multiplicative $B_\ell$ suppression factor for the $Z_I$ as a function of $x_I$ due to the additional non-SM decays into 3 generations of degenerate exotic/PM fermions 
plus $W_IW_I^\dagger$ (for $x_I>0.75$) as discussed in the text. From bottom to top the curves assume a common value of $m_{PM}/m_{Z_I}=0, 0.2, 0.3$ and 0.4, respectively.}
\label{exo}
\end{figure}

As far as the Drell-Yan signal channel is concerned the final layer of complexity arises when decays into the PM and other exotic fermions as well as $W_I$ pairs (when $x_I>3/4$) 
are kinematically accessible. In the NWA, this has no effect on the the $Z_I$ production cross section itself but leads to a suppression of $B_\ell$ as other decay channels are now open. 
To be specific, let us consider the generation-independent cases and ask how this suppression of $B_\ell$ can depend upon the value of $x_I$; recall that $m_{W_I}^2=(1-x_I)~m_{Z_I}^2$ is fixed 
but that the PM and other exotic particle masses relative to $m_{Z_I}$ are unknown.  For simplicity, we will here assume that all of the PM fields have a common mass value. Figure~\ref{exo} 
from Ref.\cite{Rueter:2019wdf}  shows the resulting suppression of $B_\ell$ as a function of $x_I$ under these assumptions. It is seen that the impact of these additional decay modes will not 
have a huge effect on the $Z_I$ signal rate expected. 

Unlike the $W'$ of, \eg,  the familiar Left-Right Symmetric model\cite{LRM}, the {\it neutral} - but non-hermitian - $W_I^{(\dagger)}$ has $Q_{em}=0$ 
but now carries $Q_D=\pm 1$ so cannot be singly produced 
in the absence of the fermion mixing that is induced by $U(1)_D$ breaking and so is very highly suppressed by factors of order $v_{1,4}^2/v_3^2 <<1$. This implies that we must instead 
examine the pair- and/or associated-production channels, \eg, $W_IW_I^\dagger$, $W_I^\dagger+D+h.c.$, \etc, that conserve $Q_D$ to lowest order. While $W_IW_I^\dagger$ production 
proceeds via $Z_I$ and $D$ exchange, associated production occurs via the process $g(d,s,b)\to W_I^\dagger+D+h.c.$. Both of the rates for these processes will  clearly depend upon whether 
only a single generation of PM exists or if there are PM partners for each generation. Interestingly, when $x_I>3/4$ happens, resonant $W_I$-pair production can occur via the $Z_I$ making 
for a significant rate enhancement; in either case these are both interesting production mechanisms to consider. Due to the Goldstone Theorem, a third process is also possible, \eg, 
$\bar dd \to W_I+V/S +h.c.$, and similarly for $d\to s,b$, but with suppressed rates due to the differing parton luminosities.  We will examine each of these mechanisms in turn.  Note that 
the associated production process is similar in nature to the $gd(s,b) \to DV_L$ process examined above.

We note that there will always be an {\it indirect} bound on the $W_I$ mass within this setup arising from the corresponding constraint on the $Z_I$ mass from the previous discussion due to 
the mass relationship  
$m_{W_I}^2=(1-x_I)m_{Z_I}^2$. Below we will concern ourselves with {\it direct} production constraints but this relationship and its possible impact on searches should always be kept in mind.

\begin{figure}[htbp]
\centerline{\includegraphics[width=4.0in,angle=0]{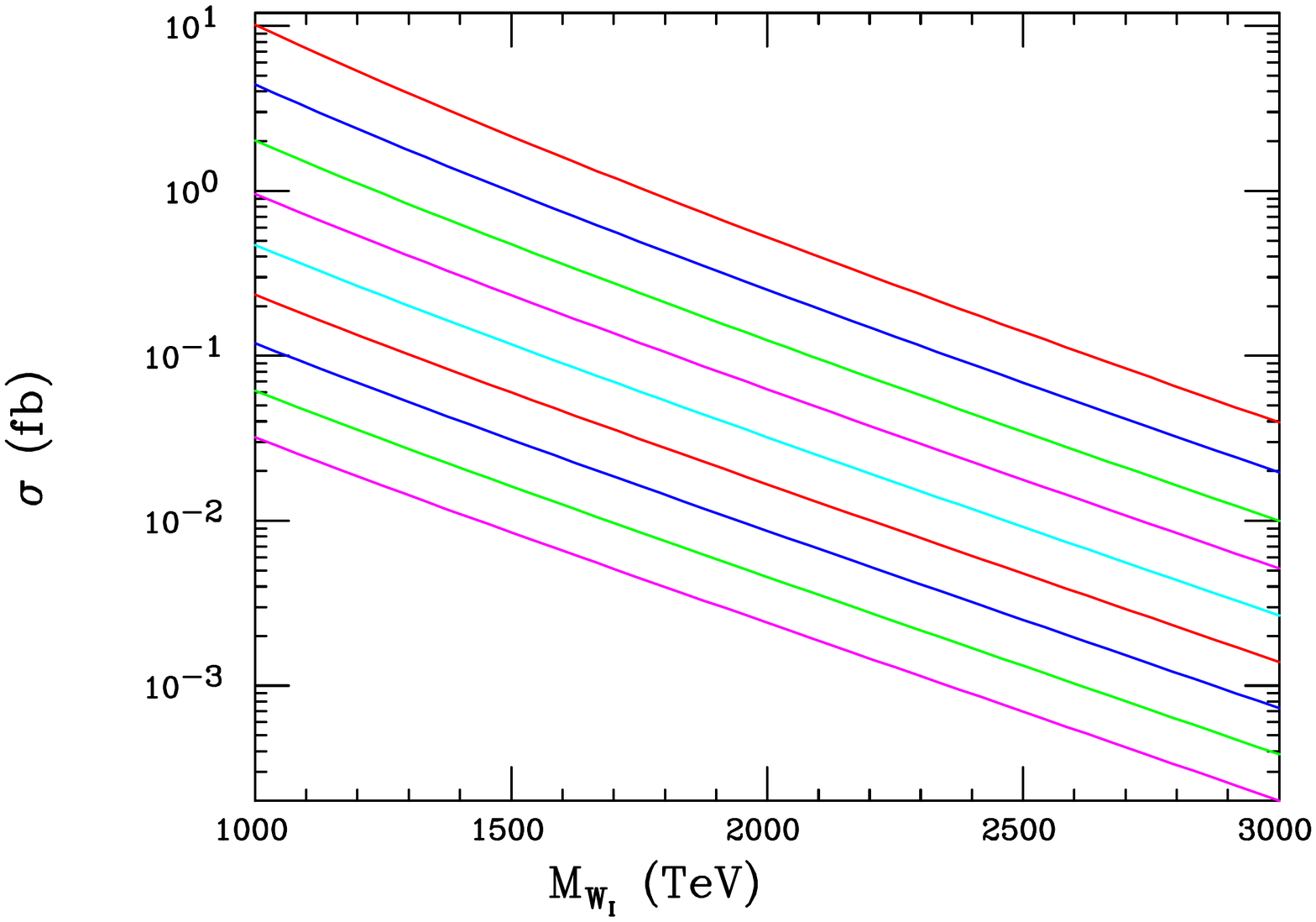}
\hspace*{-1.8cm}
\includegraphics[width=4.0in,angle=0]{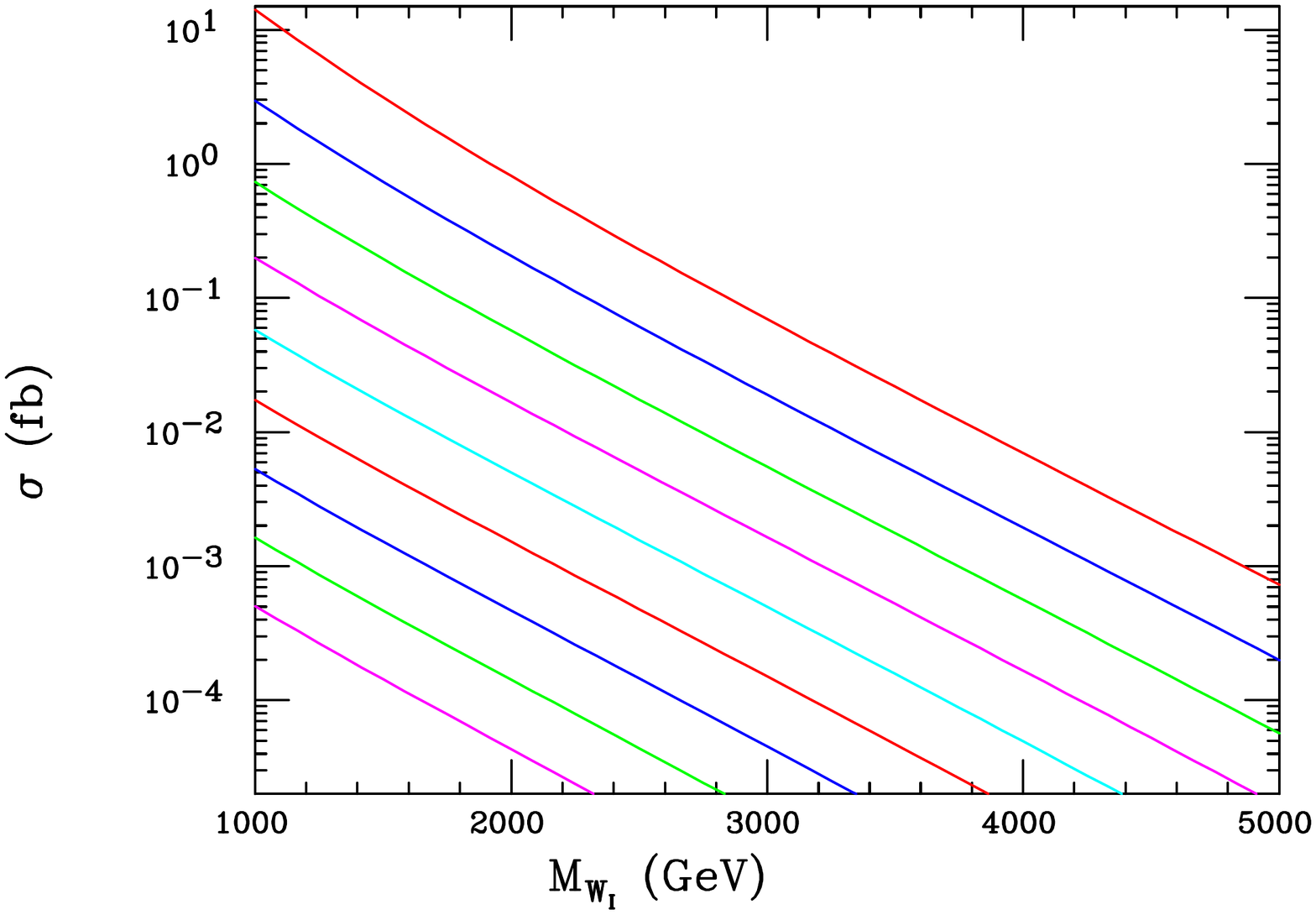}}
\vspace*{-1.5cm}
\centerline{\includegraphics[width=4.0in,angle=0]{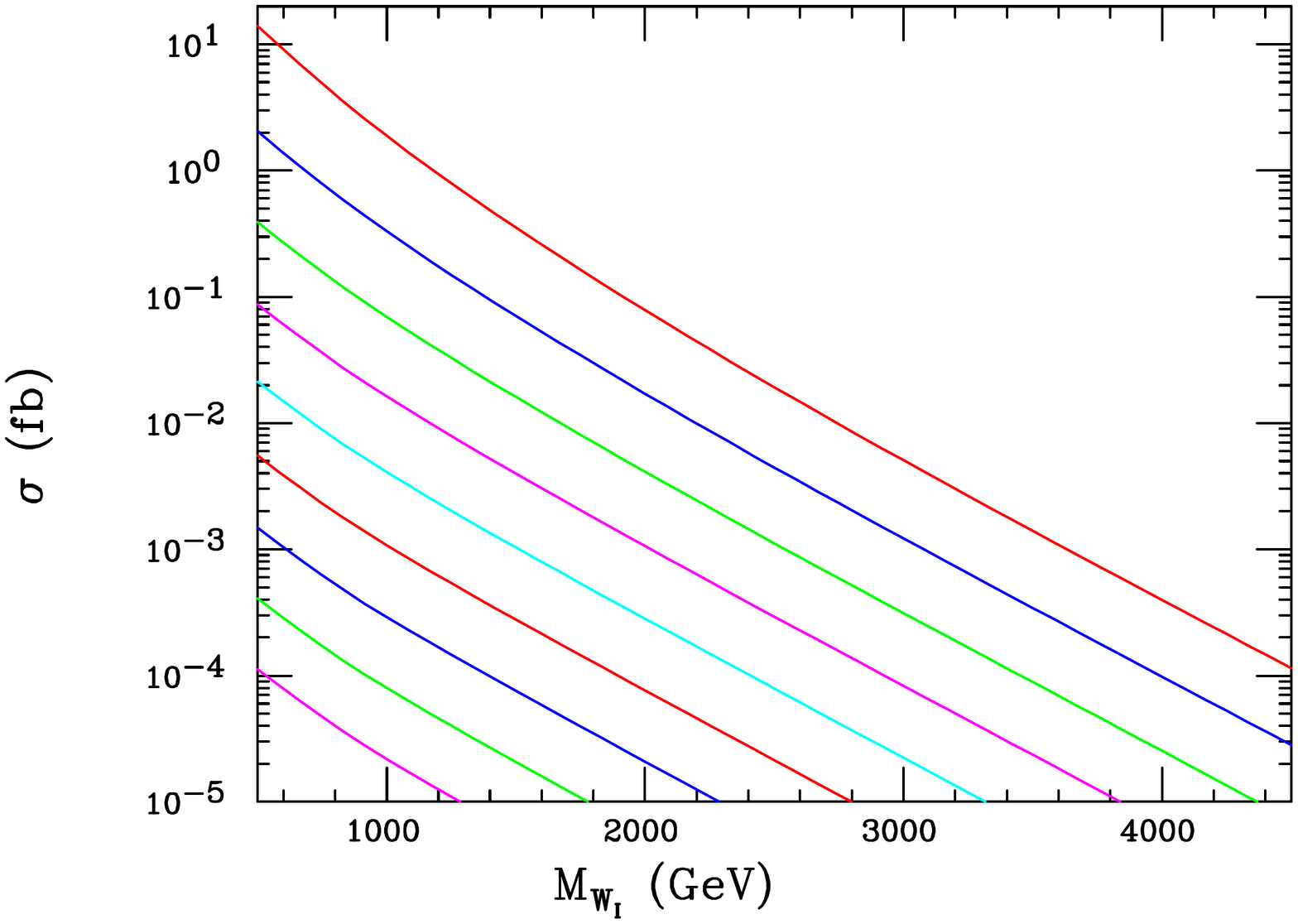}
\hspace*{-1.8cm}
\includegraphics[width=4.0in,angle=0]{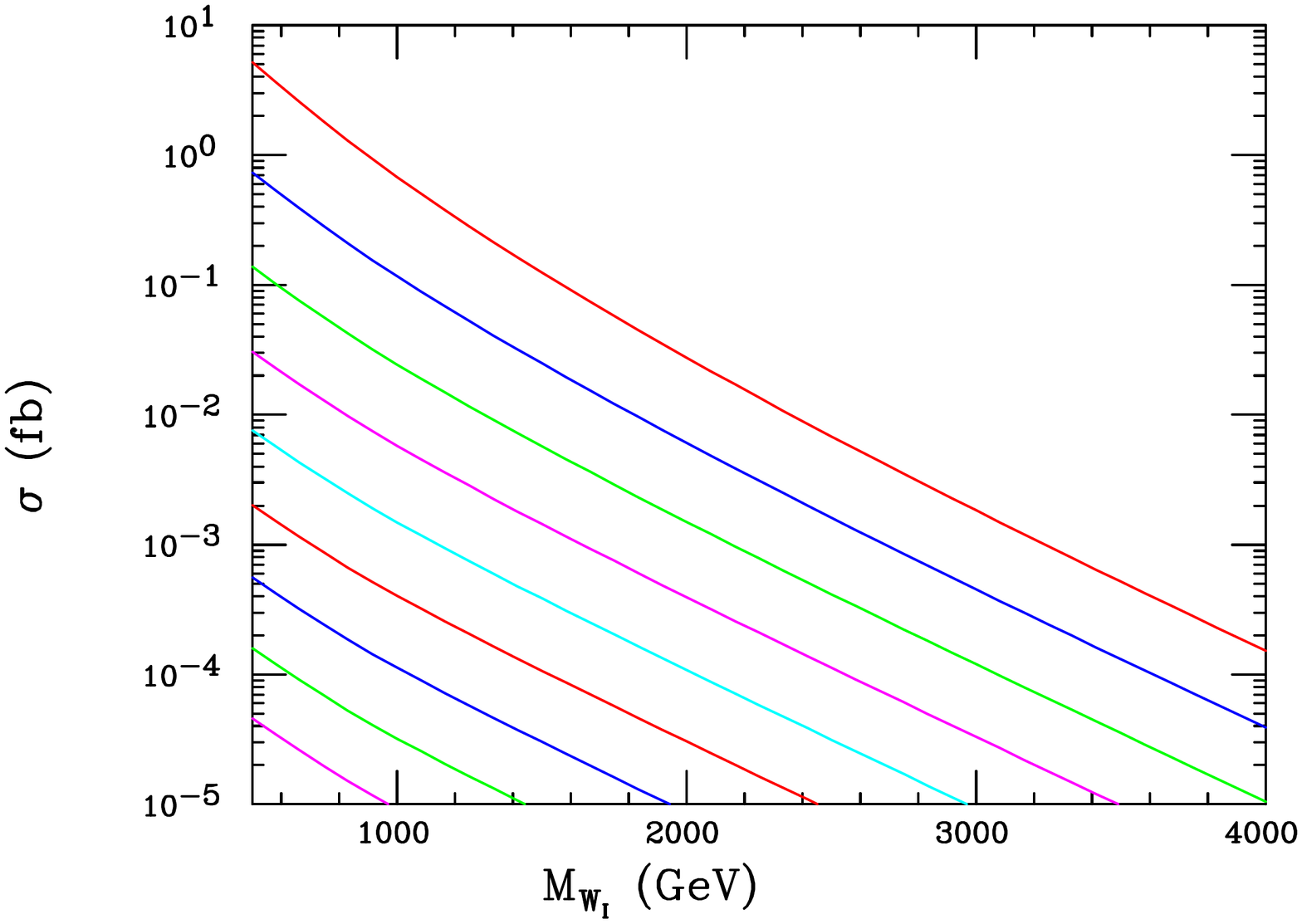}}
\vspace*{-1.20cm}
\caption{(Top Left) $gd \to DW_I+h.c.$ associated production cross section, taking $(g_I/g)^2=1$, as a function of the $W_I$ mass assuming, from top to bottom, that 
$m_D=1,1.25,..,3$ TeV, respectively, at the 13 TeV LHC with $d$ being $D$'s SM partner. (Top Right) Same as the previous panel but now for 14 TeV LHC and for 
$m_D=1,1.5,..5$ TeV. (Bottom Left) and (Bottom Right) Same as the previous panel but now assuming $s(b)$ is the SM partner coupling to $D$, respectively. }
\label{mhmwi}
\end{figure}

The cross section for the associated production process, assuming a given flavor of initial state quark, depends rather sensitively on the combined heavy $D$ and $W_I$ masses as well 
as the overall coupling, $g_I^2$, which we can scale to the usual SM $SU(2)_L$ gauge coupling, $g$, leaving an overall unknown scale factor similar to what we did above for the case of $Z_I$. 
The resulting cross sections as a function of the $W_I$ mass for various choices of $m_D$ are shown in Fig.~\ref{mhmwi}. The two upper panels correspond to the cases where $q=d$ 
for $\sqrt s=13$ and 14 TeV, respectively, while the lower panels are for 14 TeV with $q=s$ or $b$, respectively; in all cases one sees that a large part of the model parameter space 
is potentially kinematically accessible for all choices of $q$ at $\sqrt s=14$ TeV.  
Once $W_ID$ is produced, then $D\to qA_I(=V),qS$ (with $q=d,s$ or $b$) as discussed above and, if 
$m_{W_I}>m_{D,E}$, then the $W_I$ will rapidly decay as $W_I\to Dq,Ee$ with $D,E$ then decaying as previously described. It is to be noted that if $W_I$ is less massive than 
$D$ (or $E,N$, \etc) it could possibly decay into a 3-body final state, \eg, $W_I\to \bar e E^*\to e^+e^- +V/S$ with a consequently longish lifetime. For $W_I D$ production and hadronic 
$W_I$ decay, we observe that the final state is somewhat similar to that for $\bar DD$ production as discussed above 
(which provides a significant background) but with a mandatory extra jet which, if not flavor-tagged, could be easily by mimicked by QCD ISR. Within that region of parameter space where 
$V,S$ decay inside the detector so that the $D$'s can be reconstructed, the corresponding reconstruction of the $W_I$ mass peak using the extra 
$q$ jet would then lead to a substantial reduction of the significant QCD background. This production process requires a further, more detailed study.  Of course, in the case where 
$W_I\to \bar eE$, the final state will appear as a lepton plus jet combined with MET.

The $\bar qq \to W_IW_I^\dagger$ process for $q=d,s,b$ takes place via $s-$channel $Z_I$ exchange as well as $t-$channel $D$ exchange which destructively interfere to maintain unitarity and  
is similar to the case of $W^+W^-$ production in the SM. The overall cross section will, again, be highly sensitive to the choice of $q$ due to the parton luminosities. Recall that the $W_I$ 
and $Z_I$ masses are 
correlated via the familiar-looking $m_{W_I}=m_{Z_I}c_I$ relationship{\footnote {Also recall that $c_I^2=1-x_I$ and that the on-shell decay $Z_I\to W_IW_I^\dagger$ becomes kinematically allowed 
when $x_I>3/4$.}. Due to this mass relationship, this production cross section depends upon the parameters $m_{W_I},m_D, x_I$ and, as an overall factor, the ratio $(g_I/g)^4$. As in the 
case of associated production, since a heavy $W_I$-pair is produced the discovery reach for $W_I$ in this channel will be significantly reduced in comparison to the case of single, 
resonant $Z_I$ production. For fixed masses (and the ratio $g_I/g$), as $x_I$ increases from a small value we expect the cross section to grow as more and more as the cross section 
increasingly probes the effects of the $Z_I$ resonance until on-shell resonant production becomes possible. In that case, the value of the $Z_I$ reduced width, $\Gamma_{Z_I}/m_{Z_I}$, will 
also become relevant as the $W_IW_I^\dagger$ cross section will be related to height of the resonance peak, hence, the other $Z_I$ decay modes. This reduced width, for $g_I/g \sim 1$, 
is expected to be roughly $\simeq 0.01-0.03$ based on the discussion above and we will assume this range to be the case in the analysis below. Obviously,  the larger the reduced width 
becomes, the smaller will be the effect of the resonance enhancement on the $W_I$ pair production cross section.

\begin{figure}[htbp]
\centerline{\includegraphics[width=4.0in,angle=0]{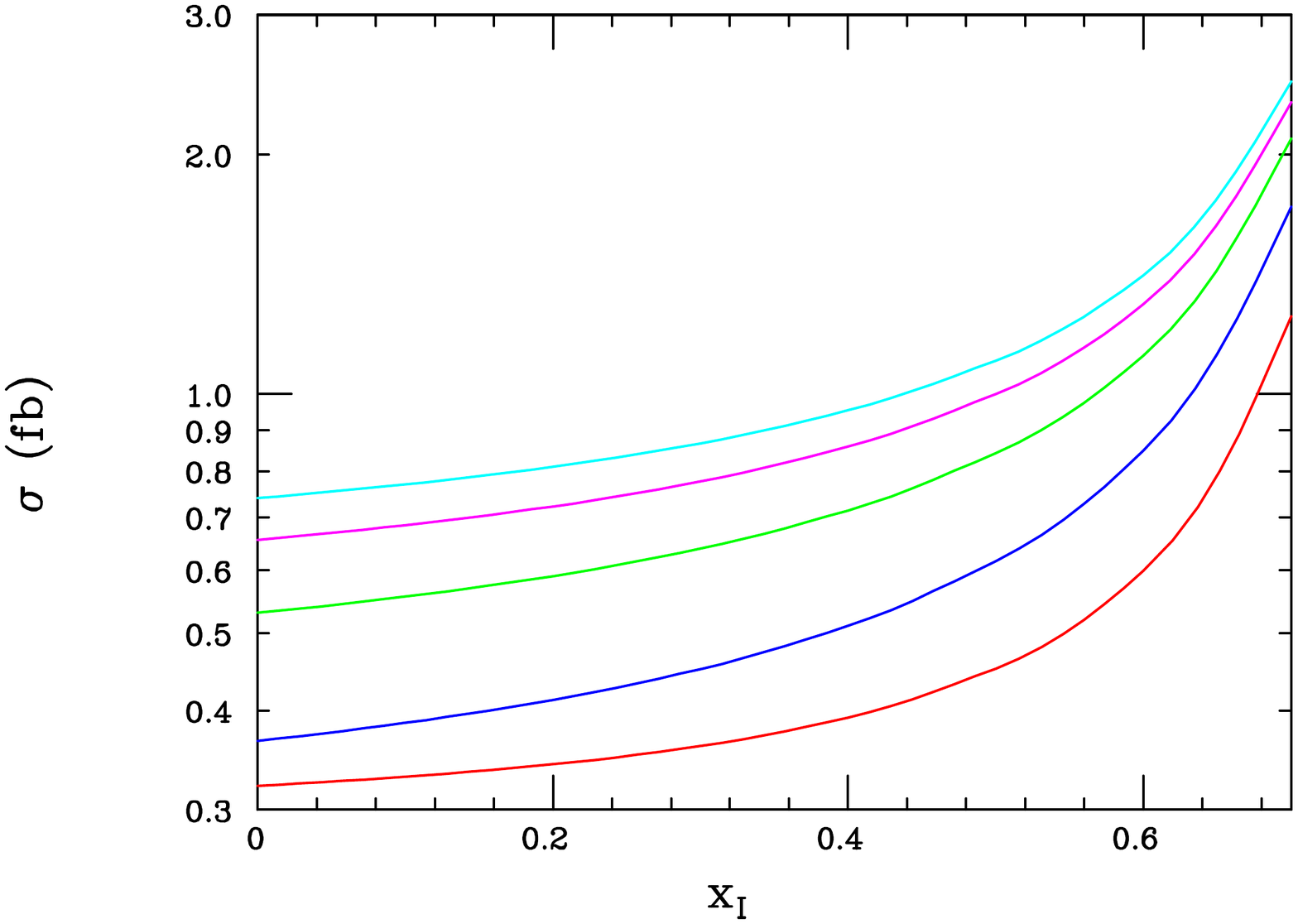}
\hspace*{-1.8cm}
\includegraphics[width=4.0in,angle=0]{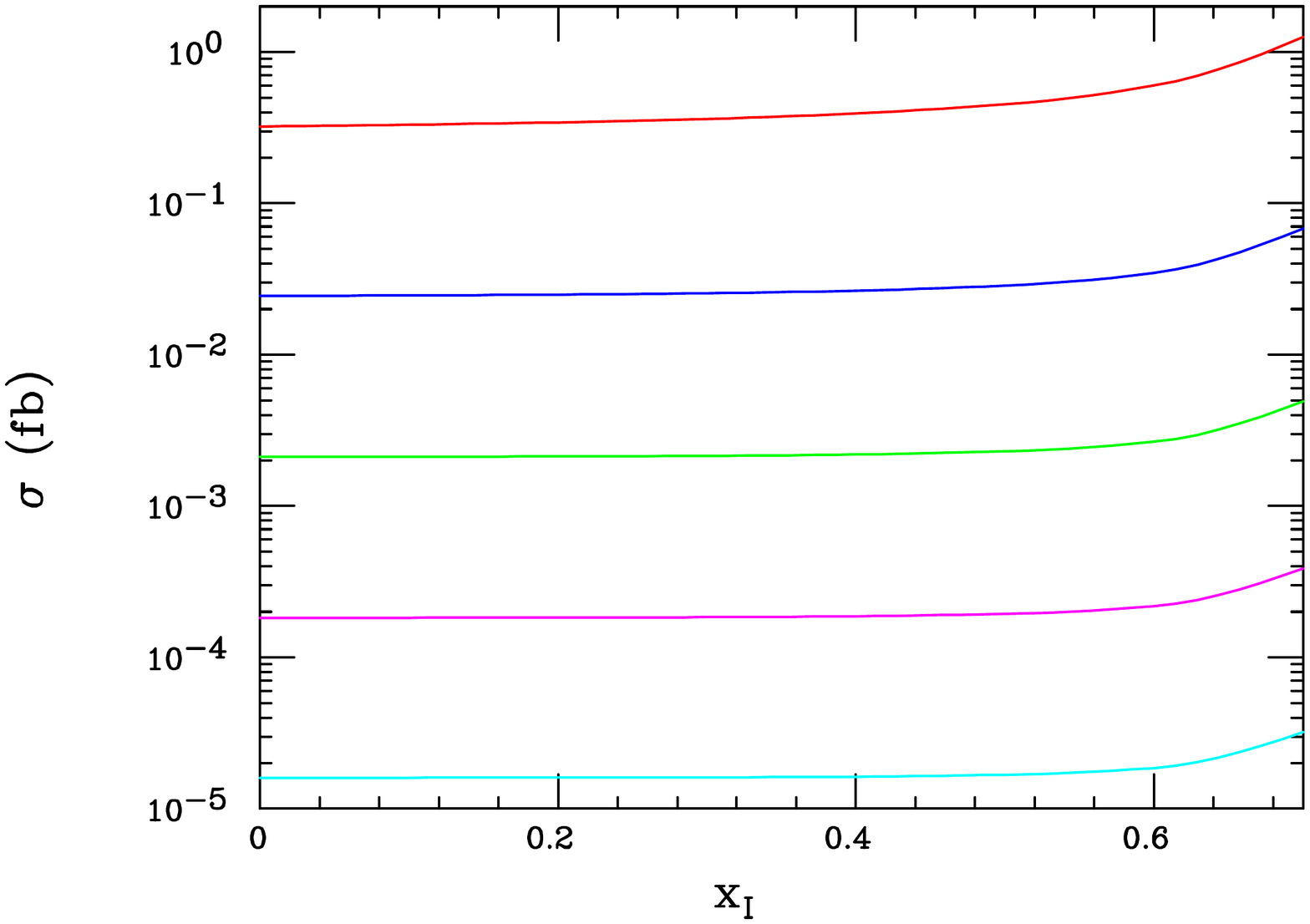}}
\vspace*{-1.5cm}
\centerline{\includegraphics[width=4.0in,angle=0]{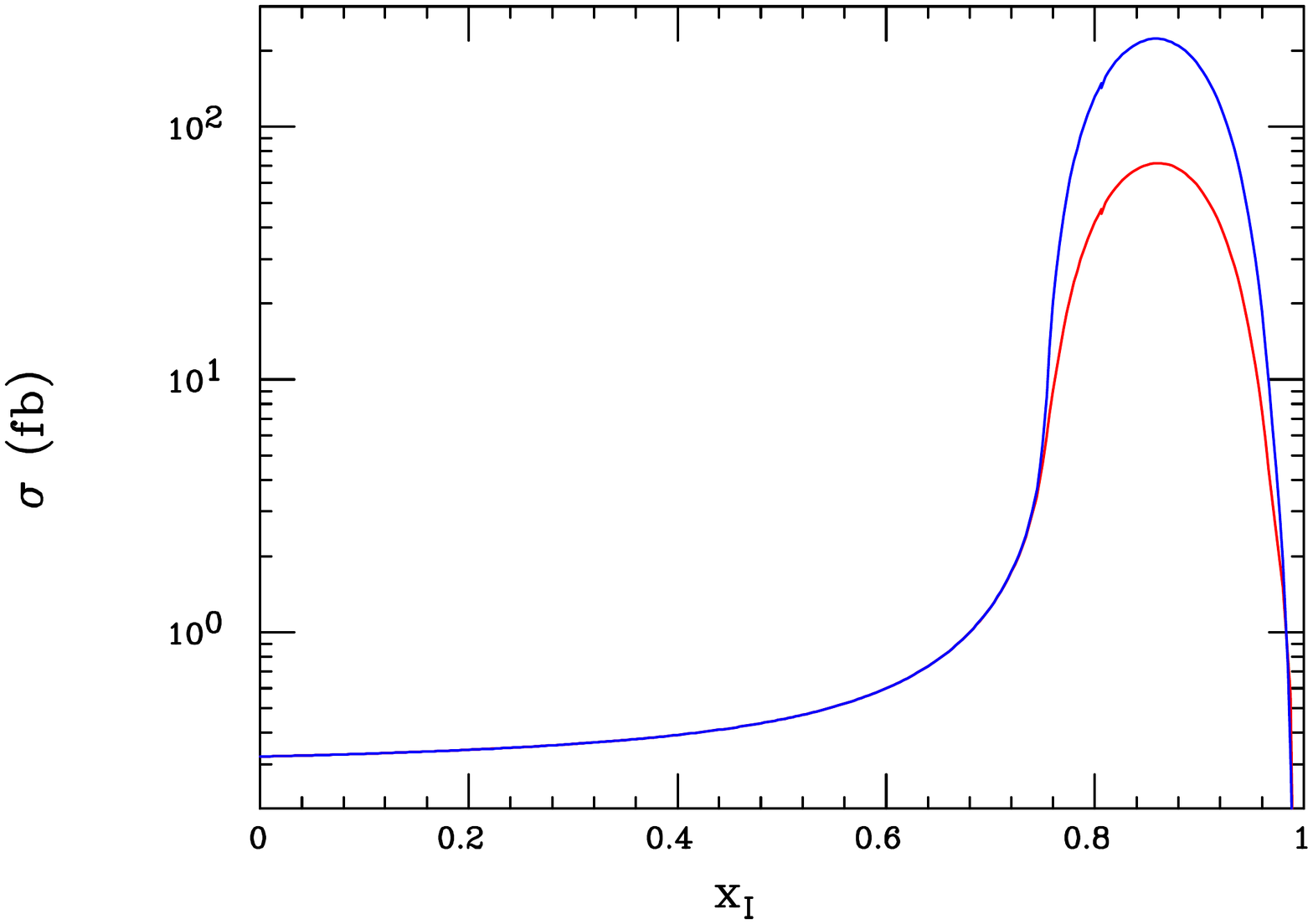}}
\vspace*{-1.20cm}
\caption{(Top Left) $q\bar q \to W_IW_I^\dagger$ total cross section at the $\sqrt s=14$ TeV LHC as a function of $x_I=s_I^2$ assuming $M_{W_I}=1$ TeV and, from bottom to top  
$m_D=1,2,..,5$ TeV, respectively,  with $d$ being $D$'s SM partner. Here an overall scaling by $(g_I/g)^4$ is still required as in the case of associated production. (Top 
Right) Same as the previous panel but now with $m_D=1$ TeV and, from top to bottom, $m_{W_I}=1,1.5,..,3$ TeV, respectively. (Bottom) Same as the previous panel but 
now assuming that $m_{W_I}=m_D=1$ TeV and including the $Z_I$ resonance region at large $x_I$ assuming that $\Gamma_{Z_I}/m_{Z_I}=0.01(0.03)$ as the top 
blue(bottom red) curve.} 
\label{wipair}
\end{figure}

Some examples from these considerations can be found by examining the cross sections as shown in both Figs.~\ref{wipair} and ~\ref{wipair2} from Ref.\cite{Rueter:2019wdf}.  
A strong $x_I$-dependence is 
clearly observed at larger values, as we anticipated, due to the action of the $Z_I$ resonance. Interestingly, \eg, we note that with $d$ and $D$ sharing an $SU(2)_I$ isodoublet, 
assuming for purposes of demonstration, \eg, $m_D=m_{W_I}=1$ TeV, the associated production process leads to the larger of the two cross sections at $\sqrt s=14$ TeV by over 
an order magnitude when 
$x_I \leq 0.7$ as might be expected from a mixed QCD-electroweak process. Of course, once $Z_I$ resonance is substantially probed the pair-production cross section is seen 
to easily dominate. We also note that as the value of $m_D$ increases, we effectively turn off the $t$-channel exchange and thus the resulting interference between the two contributions 
essentially goes away so that the cross section {\it rises} in the example seen in the top panel of Fig.~\ref{wipair}. As in case the ratio $m_t/m_W$ in the SM, we know, tree-level unitarity will 
impose a constraint on the ratio of these masses of roughly $m_D/m_{W_I}\lsim 10$, depending upon the value of $g_I$.  The sensitivity of the cross section to the reduced $Z_I$ width is shown 
in the lower panel of Fig.~\ref{wipair} while the sensitivity to $m_{W_I}$ for fixed $x_I=0.25$ for different choices of $q=d,s,b$ is shown in Fig.~\ref{wipair2}.

\begin{figure}[htbp]
\centerline{\includegraphics[width=4.0in,angle=0]{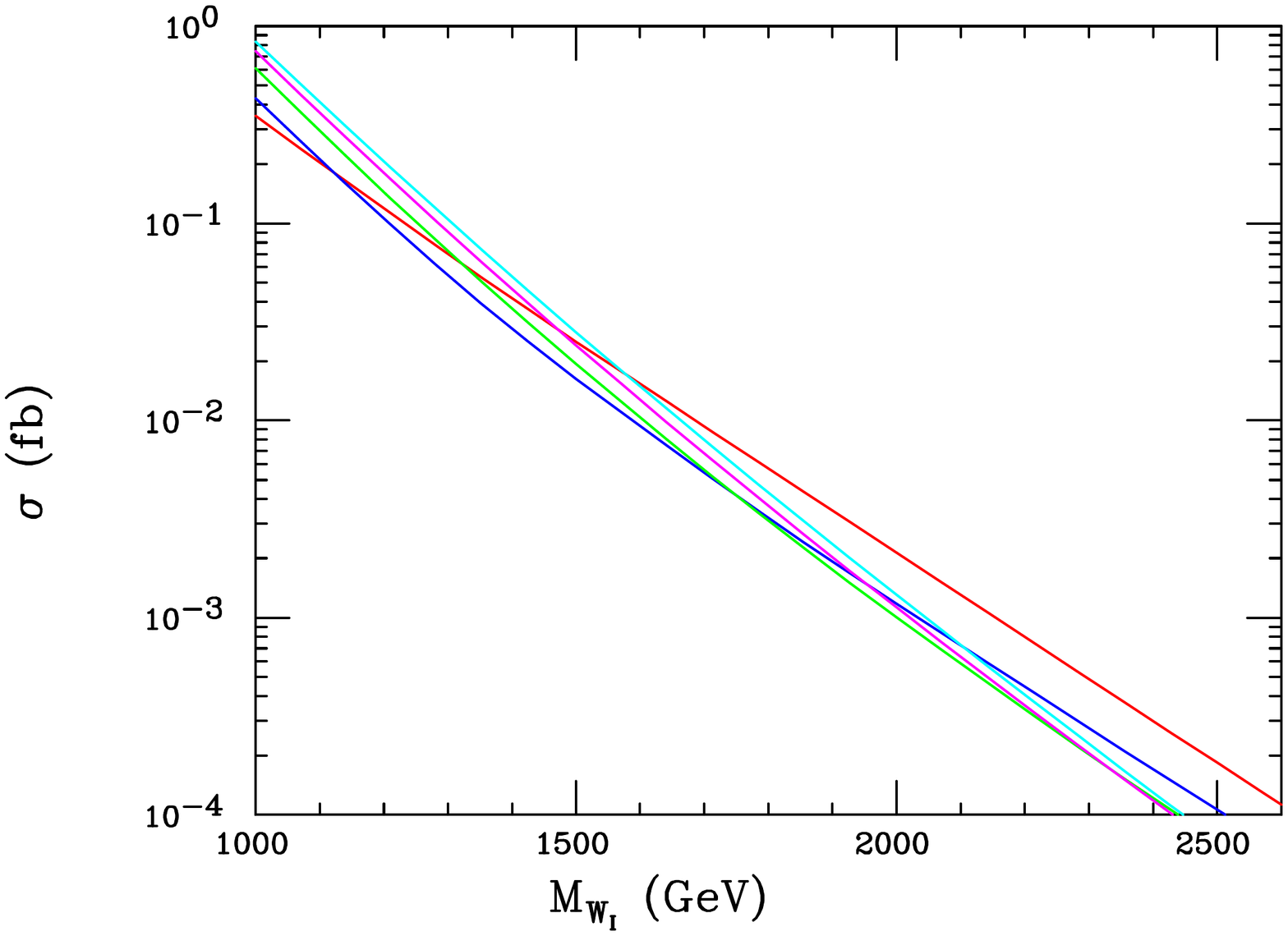}
\hspace*{-1.8cm}
\includegraphics[width=4.0in,angle=0]{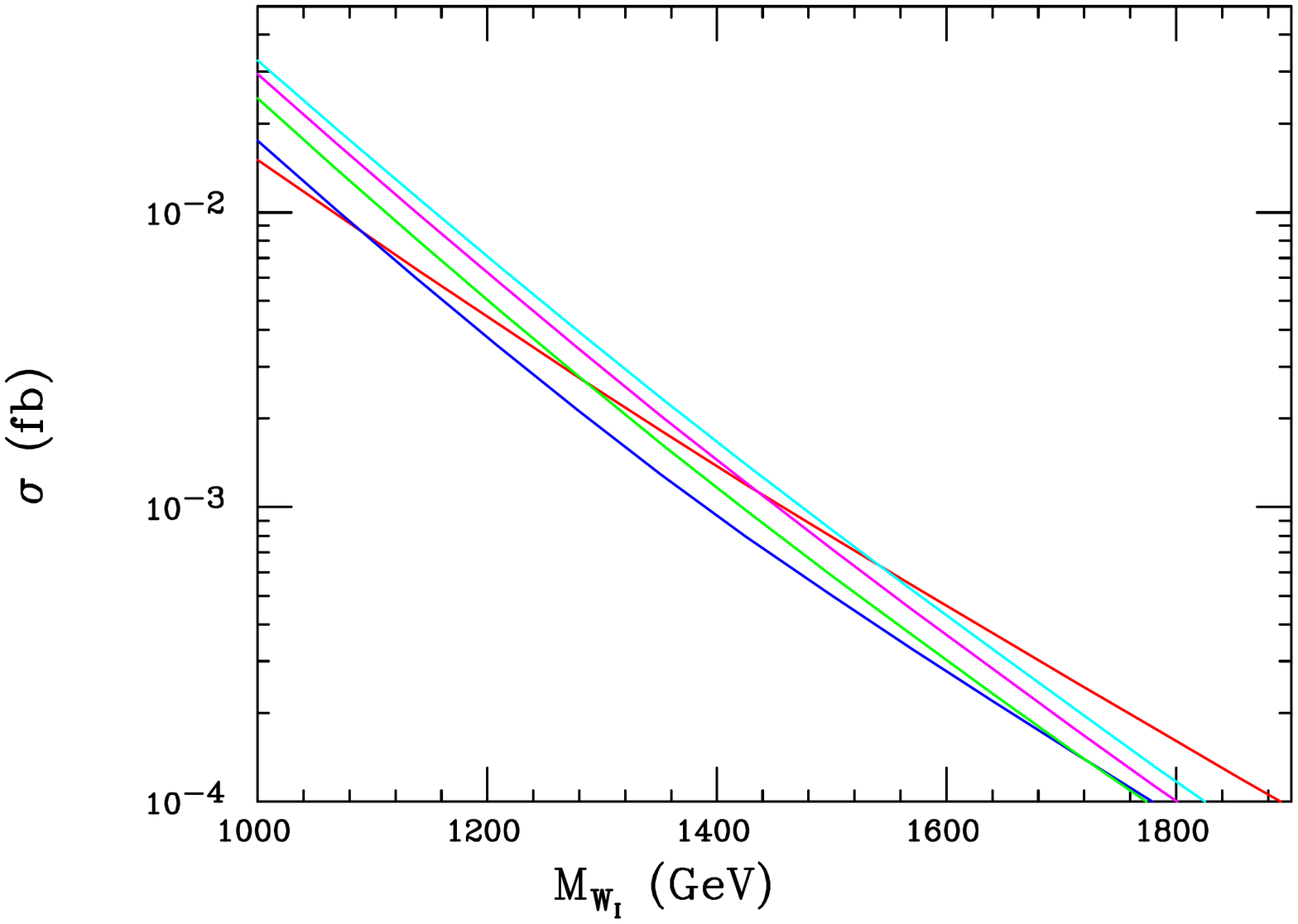}}
\vspace*{-1.5cm}
\centerline{\includegraphics[width=4.0in,angle=0]{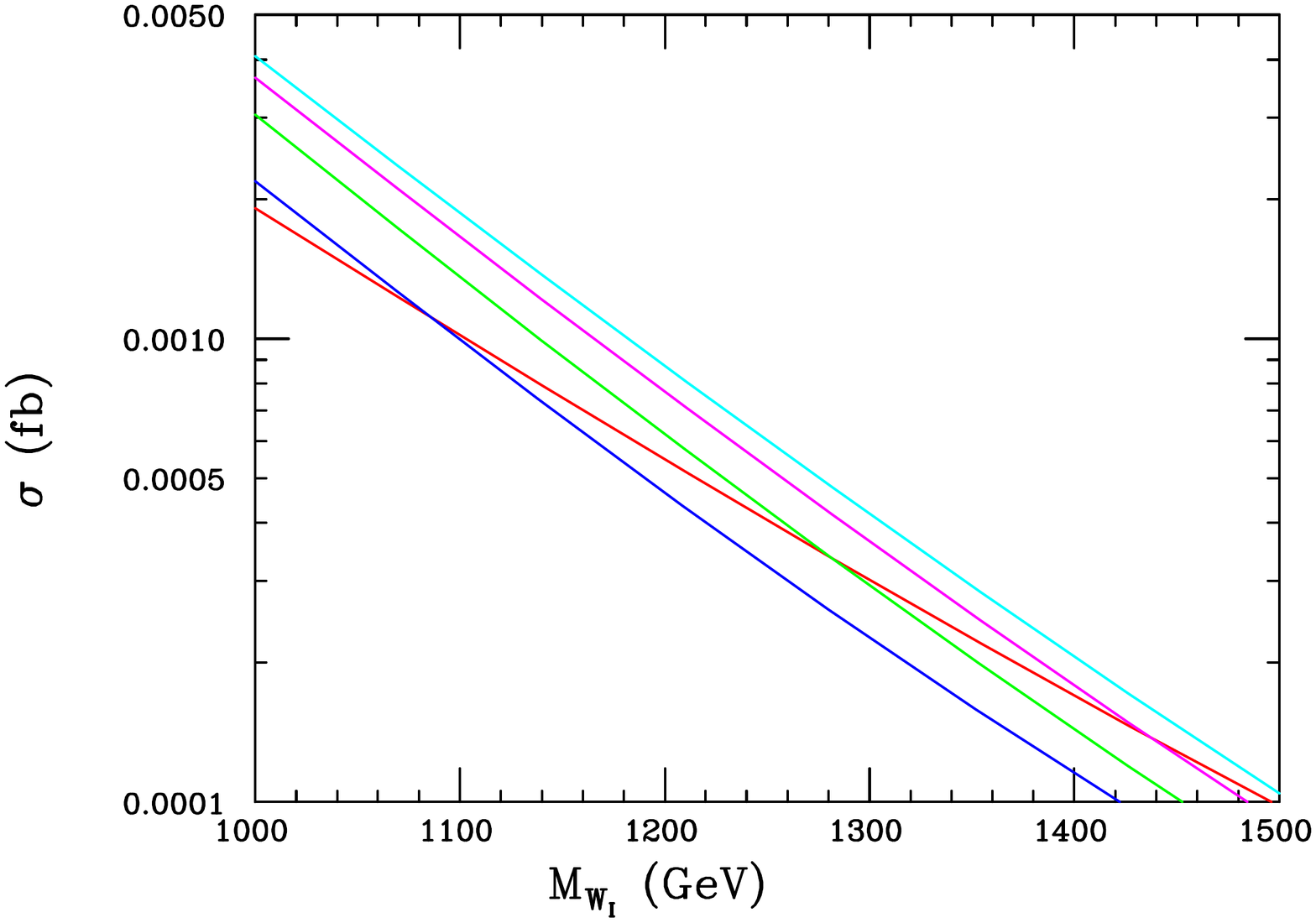}}
\vspace*{-1.20cm}
\caption{(Top Left ) Same as the in the previous Figure but now assuming that $x_I=0.25$ and displayed as a function of $m_{W_I}$ with, from bottom to top at the left axis, 
$m_D=1,2,..,5$ TeV. (Top Right, Bottom) Same as the previous panel but now assuming $s(b)$ is the SM partner to $D$ in the $SU(2)_I$ doublet, respectively.}
\label{wipair2}
\end{figure}
The signatures here, as was in the case of $W_ID$ associated production, will depend on the kinematically allowed decay modes of the $W_I$. As observed above, these are rather 
straightforward so  
long as, generically, $m_{W_I}> m_{PM}$ so that a rapid tree-level decay is allowed. As noted, when this is {\it not} the case, the $W_I$ may be potentially relatively long-lived as only off-shell 
decays via PM will be allowed, \eg, $W_I\to eE^*\to e^+e^-V$. If $W_I$ can decay to both $E$ and $D$ final states, interesting mixed signatures of lepton+jets+MET type can arise from $W_I$ 
pair production.

\begin{figure}[htbp]
\centerline{\includegraphics[width=4.0in,angle=0]{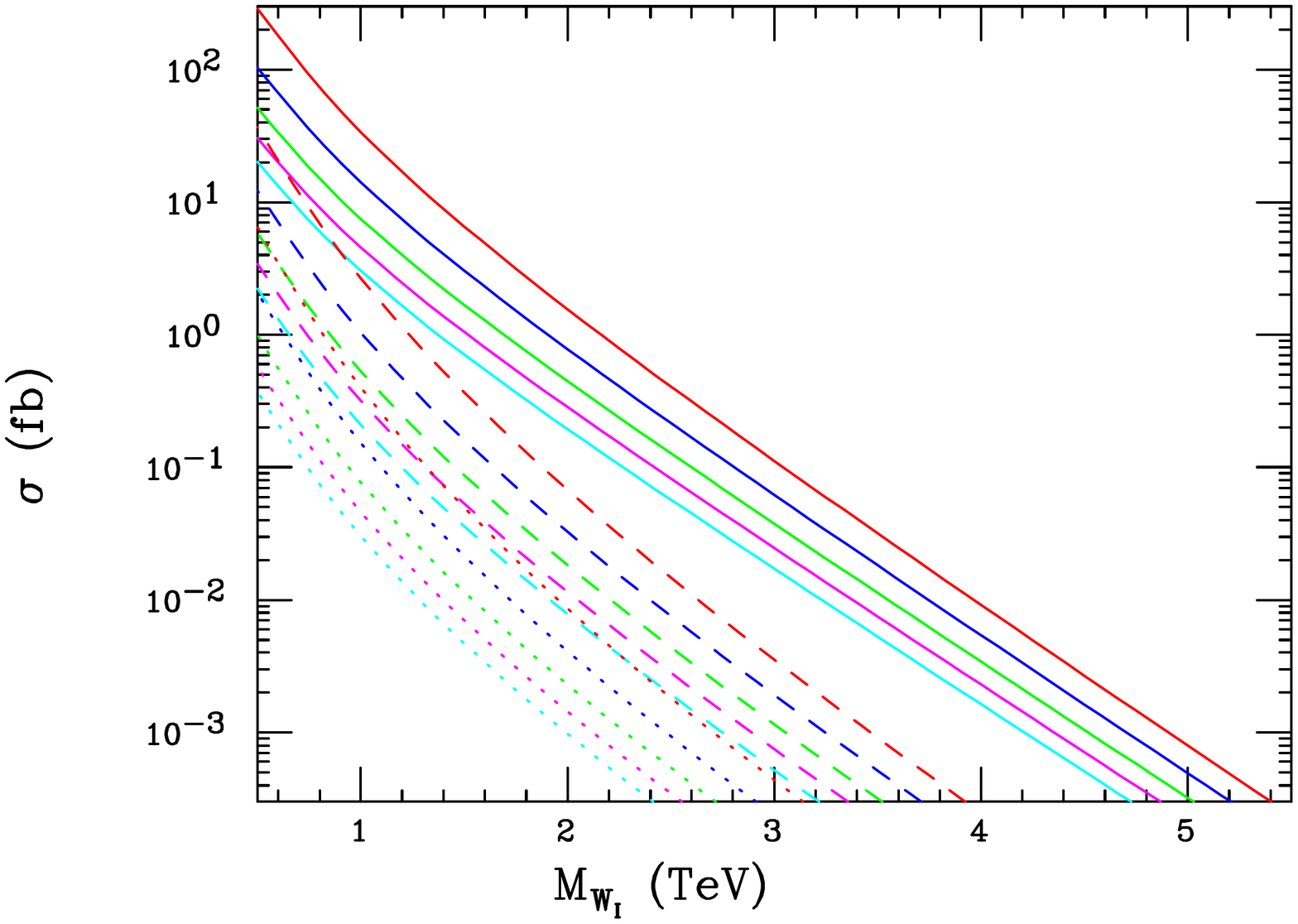}}
\vspace*{-1.5cm}
\centerline{\includegraphics[width=4.0in,angle=0]{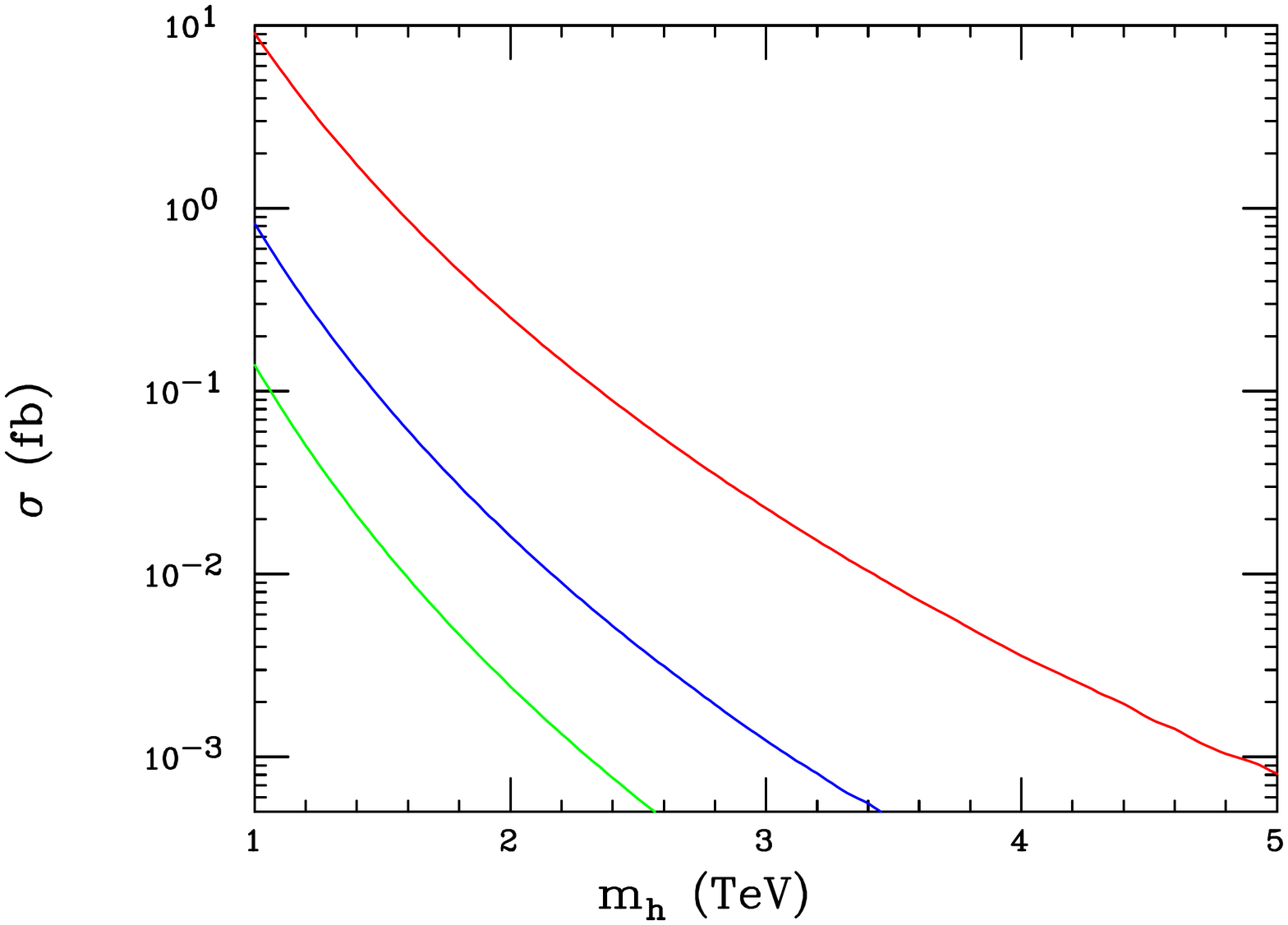}
\hspace*{-1.8cm}
\includegraphics[width=4.0in,angle=0]{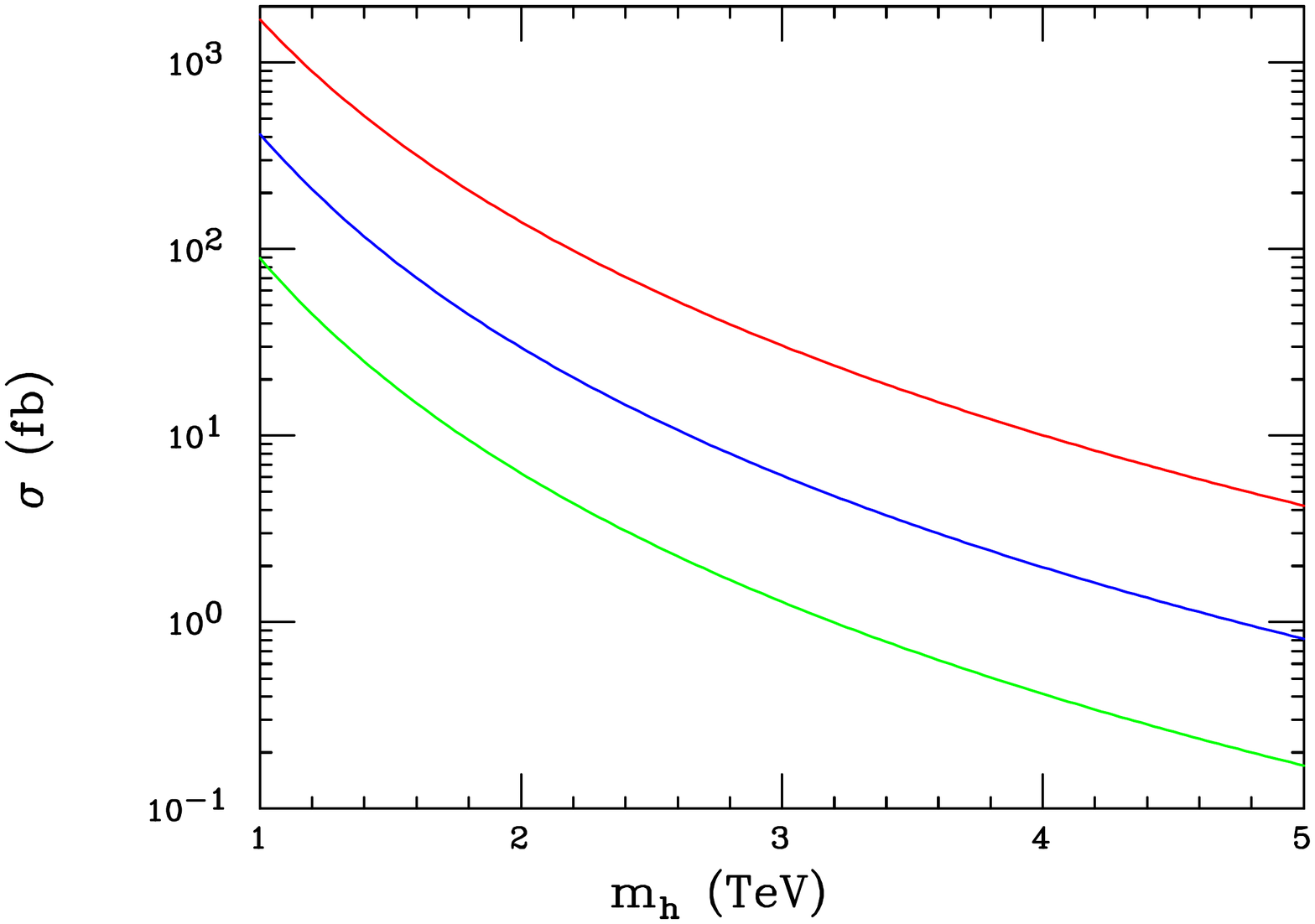}}
\vspace*{-1.20cm}
\caption{(Top) The $W_I^{(\dagger)}V/S$ associated production cross section as a function of $m_{W_I}$ and in units of $(g_I\lambda_D/g)^2$ at the $\sqrt s=14$ TeV LHC. Here the 
curves are for the choice of $d\bar d$ (solid),  $s\bar s$ (dashed) or $b\bar b$ (dotted) initial states assuming, from top to bottom in each set, that $m_h=m_D=1,..,5$ TeV, respectively. 
(Bottom Left) $V$ or $S$ pair production cross section in units of $\lambda^4$ as a function of $m_h=m_D$ assuming, from top to bottom, $q=d,s,b$, respectively.  
(Bottom Right) $VS$ associated production cross section in units of $\lambda^4$ as a function of $m_h=m_D$ assuming, from top to bottom, $q=d,s,b$, respectively. }
\label{wis}
\end{figure}

The $2_I1_{I'}$ gauge structure also allows for the  $\bar qq\to W_IV/S+h.c.$ production mechanism for $q=d,s,b$ via $t-$channel $D$ exchange with a rate scaling as 
$(g_I\lambda)^2$\cite{Rueter:2019wdf}. 
For $\lambda$'s which are $O(1)$ as is expected, this single $W_I$ production process can be advantageous since only a single heavy particle appears in the final 
state and so has an edge kinematically over both $W_IW_I^\dagger$ and the $W_ID$ mechanisms. Cross sections for this reaction are shown in the top panel of Fig.~\ref{wis} as functions of 
$m_{W_I}$ for the three different choices of $q$ and various values of $m_D$. Here we do indeed see larger production rates for at least part of the parameter space yet associated production 
remains competitive due the larger gluon parton density and the partial QCD associated  production mechanism.

\begin{figure}[htbp]
\centerline{\includegraphics[width=3.5in,angle=0]{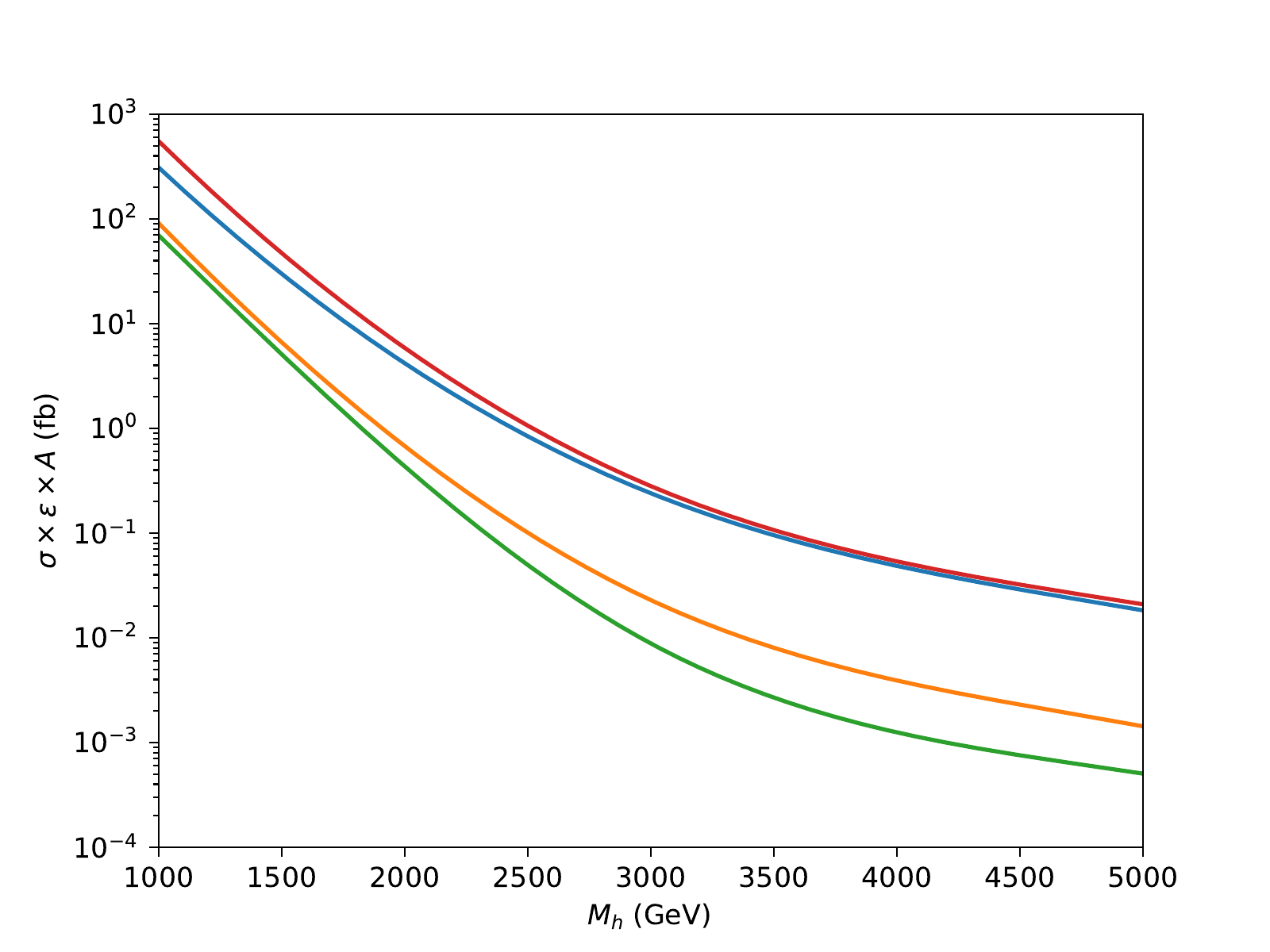}
\hspace*{-0.85cm}
\includegraphics[width=3.5in,angle=0]{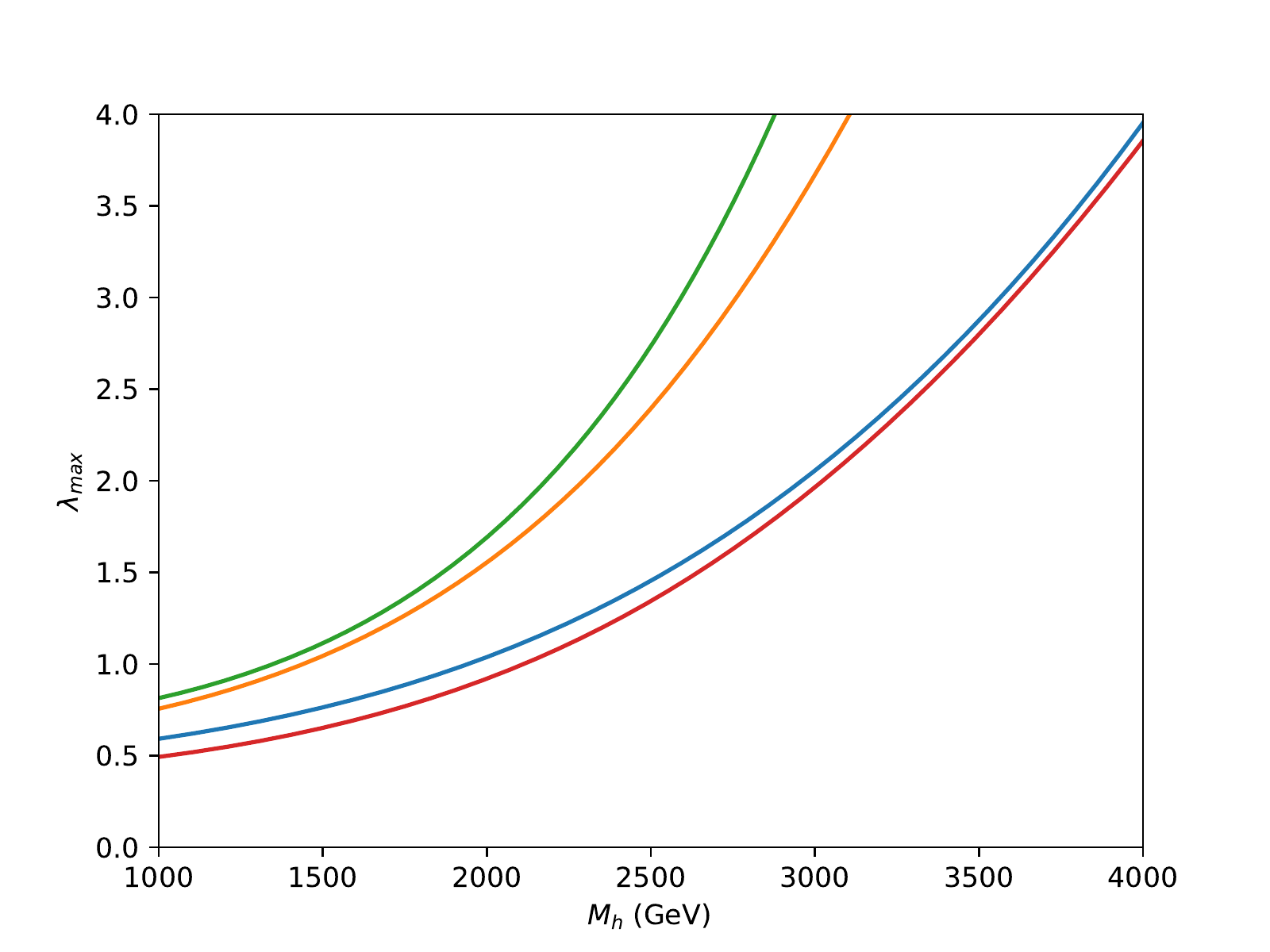}}
\caption{(Left) The signal cross section $\sigma \times \epsilon \times A$ in the ATLAS IM1 search region\cite{aaboud2018search} for $pp \rightarrow 2V, 2S,  VS + 1-4j$ as a function of 
$m_{h}=m_D$ up to an overall factor of $\lambda_D^4$ at the $\sqrt s=13$ TeV LHC. IM1 corresponds to $E_T^{miss} > 250$ GeV in addition to the cuts on jets and leptons 
described in detail in Ref.\cite{Rueter:2019wdf}. 
The red (blue, gold, green) line corresponds to $D$ coupling equally/universally to all generations (only $d$, only $s$, only $b$). As $m_D$ increases, 
the case of universal couplings becomes increasingly dominated by the first generation. (Right) Upper bounds on $\lambda_D$ from the monojet search of Ref.\cite{aaboud2018search}. As 
$m_D$ increases, the case of universal couplings is dominated by the first generation result and the upper bounds on $\lambda_D$ converge. }
\label{mono}
\end{figure}

Finally, in analogy to the previously examined $e^+e^-\to 2V,2S,VS$ set of processes discussed above, the corresponding $\bar qq$-initiated processes for $q=d,s,b$ can be probed at the 
LHC but now taking place now via $D$ exchange in the $t-$ and $u-$channels. The cross sections in this case will, of course, now scale as $\lambda_D^4$ as is shown in the lower two panels of 
Fig.~\ref{wis}. Here, apart from an overall factor of $\lambda_D^4$, the two relevant cross sections depend only upon the value of $m_D$ and the choice of $q=d,s,b$. As we observed in the 
case of the $e^+e^-$ initial state, the equal $2V$ and $2S$ cross sections are suppressed relative to that for $VS$ due to the destructive interference of the $t-$ and $u-$channel amplitudes. 
However, in the likely case that $S,V$ are either long-lived or decay to DM particles, this final state alone will be invisible, so, in analogy with the $e^+e^-$ case, we use QCD ISR to act as a 
trigger and then search for jets plus MET final states. However, we can go further and allow for $gq$ and/or $gg$ initial states that will yield additional hard jets as part of the overall production 
process.  This analysis was performed in detail in Ref.\cite{Rueter:2019wdf} (to which we refer the interested Reader) and then compared in detail with the monojet results obtained by ATLAS in 
Ref.\cite{aaboud2018search} for a number of distinct signal regions resulting in the constraints as shown in Fig.~\ref{mono}. The ATLAS IM1 signal region essentially always provided the tightest 
set of constraints yielding the upper limit on $\lambda$ as a function of $m_D$ (denoted by $m_h$ here) as shown in the right panel of this Figure.

\section{Scalar PM}

So far we have concentrated on the case where the PM fields are dominantly fermionic in the previous Sections although they are just as likely to be, \eg, color-singlet scalars as was 
mentioned above. In a bottom-up approach, as in the earlier fermionic toy models first discussed above, we seek an addition to the SM scalar spectrum that allows for PM instability, 
generates a finite and 
calculable value for $\epsilon$ as well as having a means to break the $U(1)_D$ symmetry. These constraints are rather non-trivial. If we add only a pair of, \eg, $Q_{em}=1,Q_D=\pm 1$ scalars, 
$\phi_i^+$, similar to the $E_i$ above, we can mix them with the SM Higgs 
doublet so that they can decay and $\epsilon$ will be finite; however, $U(1)_D$ will remain unbroken. If these were instead 
neutral scalars so that they could obtain $U(1)_D$-breaking vevs, then $\epsilon=0$. Clearly some combination of charged and neutral scalars carrying $Q_D\neq 0$ need to be added to 
satisfy our goals. Given this, as shown in Ref.\cite{Rueter:2020qhf} which we closely follow here, it is easy to convince oneself that the simplest possibility simultaneously satisfying 
all the model building constraints is to add 
two additional $SU(2)_L$ isodoublet Higgs representations, $\eta_{1,2}$, to the SM that have $Q_D=\pm 1$, respectively, but whose neutral members also obtains the small vevs, 
$v_i \leq 1$  GeV (with $t=v_1/v_2$ so that we can take $\geq 1$ without loss of generality), that are necessary to break $U(1)_D$. In such a case these new doublets act not only as PM but as 
the source of the Higgs fields in the dark sector. In the 
case of more complex scalar sectors with multiple new fields, however, it is possible to separate these two roles so that the scalars with $U(1)_D$ breaking vevs and those that circulate in the 
$U(1)_Y-U(1)_D$ vacuum polarization-like graphs are distinct. Here we will focus on this simpler scenario which is quite highly constrained for numerous reasons: the $U(1)_D$ breaking vevs 
must be small, $\lsim 1$ GeV, the Higgs potential must be bounded from below, the various unitarity, electroweak and Higgs(125) coupling constraints must all be satisfied, as must also be 
the bounds from the non-observation of new, heavier scalar states at the LHC. Additional constraints are obtained from measurements/bounds on the invisible decay widths of the SM $Z$, 
which is proportional to $t^2-1$ (thus forcing the value of $t$ to be not too far above unity), as well as those 
from the 125 GeV SM-like Higgs. Although this model has many free parameters because of the complexities of the scalar potential, it is quite likely that due to these numerous constraints 
this scenario will either be realized or excluded once an additional rather small amount of integrated luminosity at $\sqrt s=14$ TeV is accumulated by the LHC.

As discussed in detail in Ref.\cite{Rueter:2020qhf}, after 
spontaneous symmetry breaking, in the limit of a CP-conserving scalar potential with real vevs, the physical scalar spectrum will consist of the (almost) 
SM Higgs, $h_{SM}$ with $m_{h_{SM}}\simeq 125$ GeV, two pair of charged Higgs, $H_{1,2}^\pm$~{\footnote {Note that $H_i^\pm$ arises from the doublet $\eta_i$ so that they are {\it not} mass 
ordered}}, one 
CP-odd state, $A$ (the remaining ones playing the roles of the SM and $U(1)_D$ Goldstone bosons), as well as two additional CP-even states, one of which is heavy and close in mass to the 
CP-odd state ($A$), $H$, and one of 
which is quite light with a mass $\lsim 1$ GeV that we can identify with the dark Higgs, $h_d=S$, in the discussion above.  The size of the mixing between the SM Higgs and the other
CP-even states in $\eta_{1,2}$ is quite small as $v_{1,2}^2/v_{SM}^2 \sim 10^{-4} \sim \epsilon$.  Due to the various small parameter ratios it is easy to diagonalize the scalar mass matrices and 
rewrite the 14 parameter Higgs potential\cite{Rueter:2020qhf} 
in terms of eigenstate masses (which are required to be positive definite), $v_{SM}, t, v^2=v_1^2+v_2$ and $\epsilon$ plus some additional parameters.  
As in the $E_6$-inspired model above, the DP in this scenario also picks up a coupling proportional to those of  the SM $Z$ since the dark Higgs fields are both in $SU(2)_L$ isodoublets. 
Since the charged Higgs act as the only PM in the vacuum polarization loop, we can immediately see that $\epsilon$ is indeed finite, calculable (and can be of either sign) and is given by
\begin{equation} \label{eq:eps}
\epsilon= \frac{g_De}{48 \pi^2}\textrm{ln} \left(\frac{m_2^2}{m_1^2} \right),
\end{equation}
with $m_i$ being the mass of $H_i^\pm$ and is of the expected magnitude, $|\epsilon| \lsim 3\times10^{-4}$; recall the values of $m_{1,2}$ are {\it not} mass ordered here. It is important 
to note that in the absence of $U(1)_D$ breaking, gauge invariance prevents any of these new Higgs states from coupling to the SM fermions and so these interactions 
are suppressed by factors of order $v_i/v_{SM}$ and can generally be safely neglected in comparison to usual gauge interactions in practice. 

Perhaps the best way to examine this rather large parameter space is via a Monte Carlo scan\cite{Rueter:2020qhf}, randomly generating points with it, 
requiring that all of the many constraints discussed above (apart from those from the 
LHC specifically to be discussed below) are satisfied simultaneously and then study those points which survive. As part of this scan, it was also required that $m_A>120$ GeV and $m_{1,2}>200$ 
GeV as the spectrum tends to be rather light and such light states would likely be rather easily excluded by LHC searches. A sample of $\simeq $7k points survive this scan which we make 
use of below. An idea of the mass spectra that are possible in this scenario 
after the constraints are applied can be gleaned from Fig.~\ref{mam1m2} where we see that {\it both} charged Higgs states must lie below $\simeq 420$ GeV, both only partially 
concentrated in the lower 
mass range,  while we also observe that $m_A\simeq m_H \lsim 200$ GeV with, again, smaller masses preferred. Combinations of the new heavy scalars can be produced at the LHC via their 
SM gauge interactions with the $W^\pm,~\gamma$ and $Z$ via $s$-channel gauge boson exchanges as we will make use of below. 

\begin{figure}
\vspace*{0.5cm}
\centerline{\includegraphics[width=3.2in,angle=0]{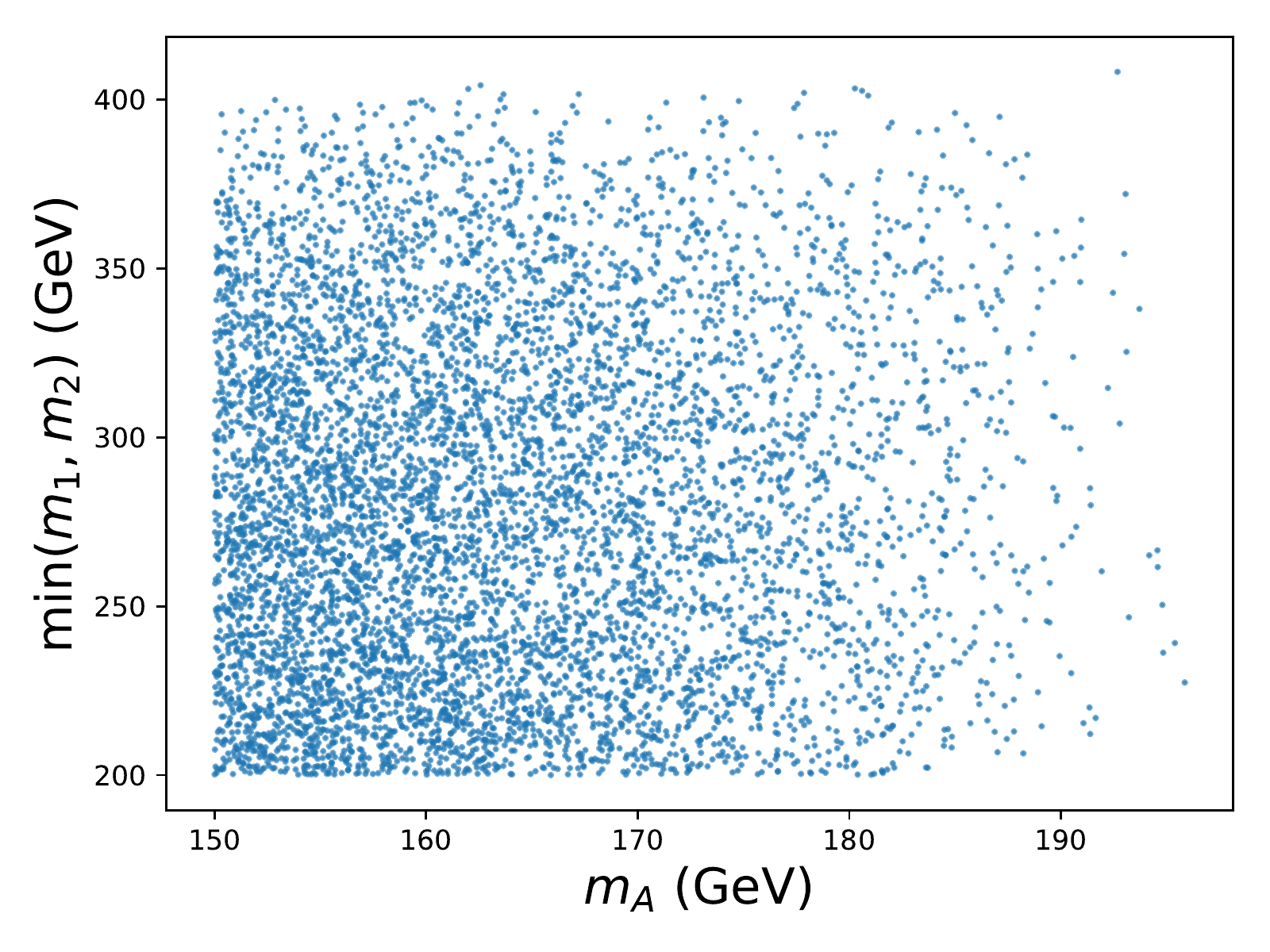}
\hspace*{-0.4cm}
\includegraphics[width=3.2in,angle=0]{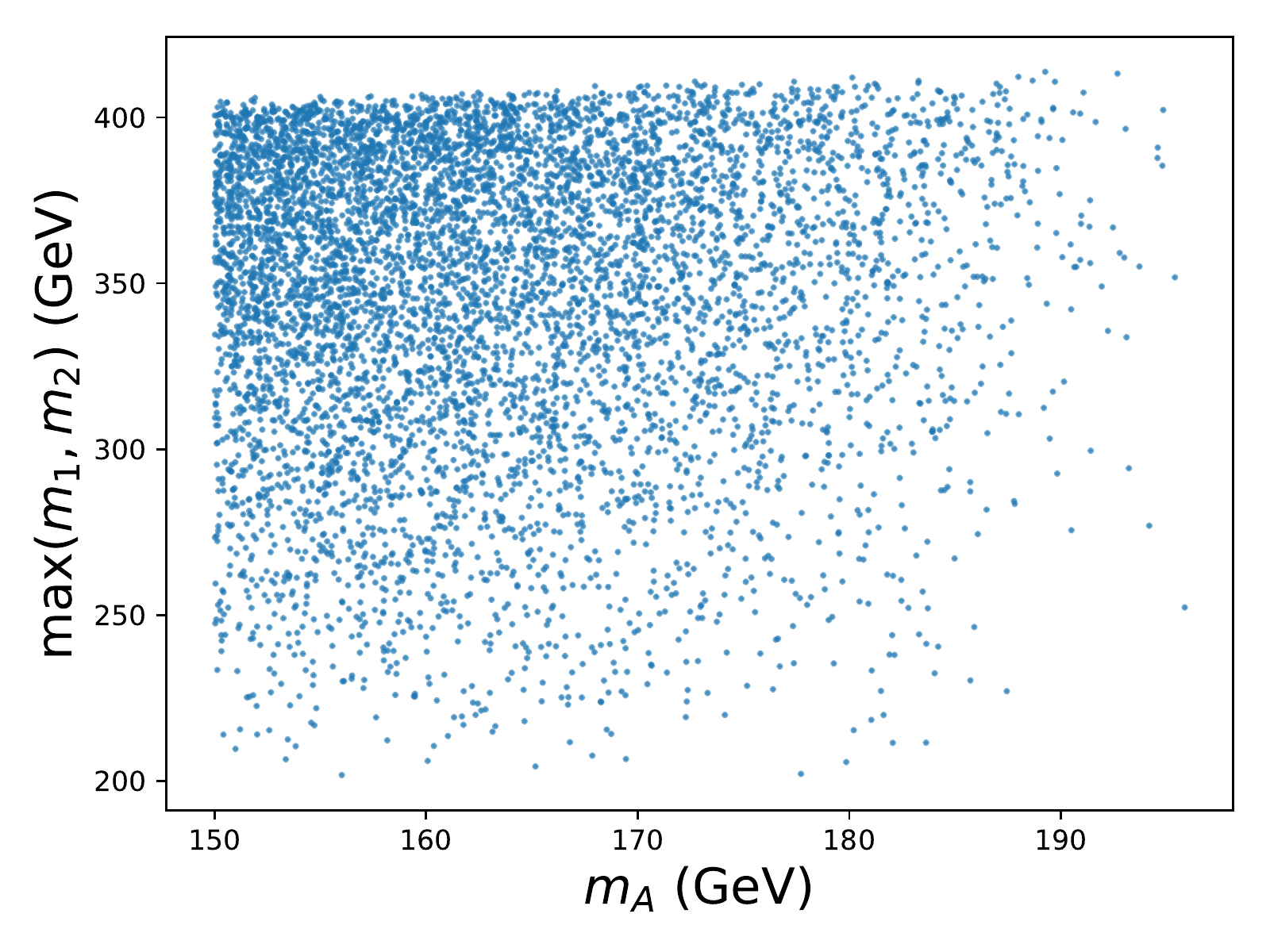}}
\caption{$m_A\simeq m_H$ plotted versus min($m_1,m_2$) (left) and max($m_1,m_2$) (right). The preference for smaller $m_A$ can be seen by the relative over density of points near 
150 GeV, and we see that there is no comparable preference for low $m_{1,2}$, as the points are relatively uniform above the constraint $m_{1,2}>200$ GeV. }
\label{mam1m2}
\end{figure}

Once produced, the decays of these new heavy scalars provide for discovery signatures at the LHC. Note that since they do {\it not} couple to the SM fermions at leading order, many of the 
typical searches performed at the LHC for heavy scalars, \eg, in the Two Higgs Doublet Model, do not apply to this scenario. Both charged Higgs will decay as $H_i^\pm \to W^\pm h_d/V$ with 
equal rates and to $W^\pm H/A$, also with equal rates, if kinematically allowed. The heavy CP-even state, $H$, will decay to either $ZV$ or to $h_{SM}h_d$ while we find that 
$\Gamma (A\to h_{SM}V)\simeq \Gamma(H\to h_{SM}h_d)$ and 
$\Gamma (A\to Zh_d)\simeq \Gamma(H\to ZV)$, essentially due to the Goldstone Theorem. Clearly, the ratio of partial widths, 
$R = \Gamma(H\rightarrow ZV)/\Gamma(H\rightarrow h_{SM} h_d)$, then determines which decay mode is dominant and thus what final states should be searched for at colliders. 
Fig. \ref{mhR} shows this ratio $R$ as a function of $m_{H}$ for the $\simeq 7$k model points, and we find that for $\simeq 72\%$ of the cases in the parameter scan $H/A \rightarrow Z+V/h_d$ is 
the dominant decay mode. We also see from the Figure that for $m_H \gtrsim 175$ GeV or so, almost all the parameter space points lead to $R>1$, so that the decay into $H \to Z+V/h_d$ 
will dominate. We will refer to the points in 
parameter space with $R>1$ as ``$Z$-dominant", and those with $R<1$ as ``$h_{SM}$-dominant" in the following discussion. Further, in all the cases considered below 
it will be assumed that both $V$ and $h_d$ will decay invisibly, either to DM or will have a sufficiently long enough lifetimes so as to escape the detector and so lead to MET signatures.

\begin{figure}
\centerline{\includegraphics[width=4.5in,angle=0]{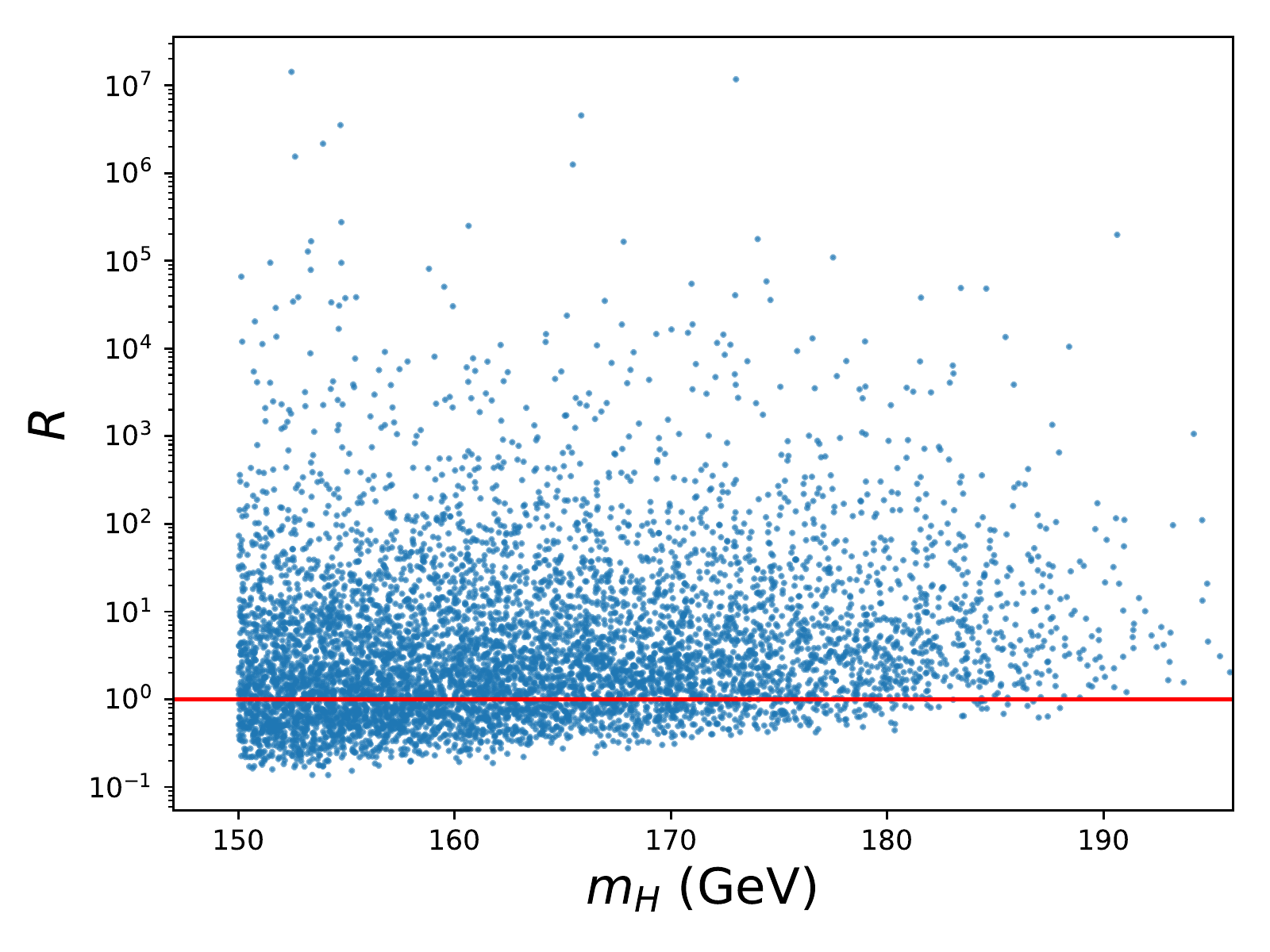}}
\caption{The ratio of partial decay widths, $R$, as defined in the text, as a function of $m_H$, showing that most of the parameter space is $Z$-dominant rather than 
$h_{SM}$-dominant, especially for large values of $m_H$. The red line corresponding to $R$=1 to guide the eye.}
\label{mhR}
\end{figure} 

In order to further our understanding of the numerous possibilities within this parameter space, four benchmark points(BPs) were selected\cite{Rueter:2020qhf} 
with either lighter of heavier Higgs states and with either $Z$-dominant or 
$h_{SM}$-dominant decays; these are presented in Table~\ref{tab:BP}. These BPs were then subjected to several recasted LHC searches (for SUSY or DM) which we only briefly discuss here; 
for a full discussion of analysis details and further information about the benchmark points, see Ref.\cite{Rueter:2020qhf}. As one example to give a flavor of these analyses, 
an obvious signal channel to consider is that provided by the $Z+$MET final state which can arise from both the $HV$ and $Ah_d$ associated production. These both occur via $Z^*$ 
$s$-channel exchange when both $H/A$ subsequently decay to an on-shell $Z$ which itself then decays to jets or charged leptons. This type of search is clearly most sensitive to model 
points similar to BP1 and BP2 but other final states, \eg, $h_{SM}+$MET, are more sensitive to the other pair of benchmarks. It is to be noted that since the final state contains a $Z$ plus 
two particles each producing MET into opposite hemisphere the {\it overall} MET is, on average, somewhat reduced making it somewhat easier for points to survive searches with high 
MET thresholds unless lots of luminosity is available to probe tails of distributions. ATLAS\cite{ATLAS2018hadronic,ATLAS2018leptonic} 
has performed MET searches in the $Z\to $ jet/lepton channels while CMS\cite{CMS2020leptonic} concentrates on the $Z\to $ leptons channel 
but also using higher integrated luminosity. Once these searches are recast, the various signal regions are examined to determine which one shows the greatest model sensitivity, \ie, the 
smallest background in comparison to the anticipated signal rate; in almost all cases examined this results in obtaining a small value for $S/B$. 

\begin{table}[b] 
\begin{center}
\begin{tabular}{ |c |c |c| c| c|}
\hline
Benchmark Point & $m_H$ & $m_1$ & $m_2$ & $Z$ or $h_{SM}$ dominant\\
\hline
BP1 & 180.8 GeV & 371.0 GeV & 333.2 GeV & $Z$ \\
BP2 & 154.7 GeV & 203.9 GeV & 249.0 GeV & $Z$ \\
BP3 & 187.8 GeV & 305.6 GeV & 346.2 GeV & $h_{SM}$ \\
BP4 & 155.7 GeV & 210.5 GeV & 275.3 GeV & $h_{SM}$ \\
\hline
\end{tabular}
\end{center}
\caption{Four benchmark points and their mass parameter values used to analyze the efficiency of LHC searches for the model. These roughly span the range of $m_H$, $m_1$, and 
$m_2$ produced by the full parameter scan, with two $Z$-dominant points and two $h_{SM}$-dominant points.} 
\label{tab:BP}
\end{table}
\begin{figure}
\centerline{\includegraphics[width=4.5in,angle=0]{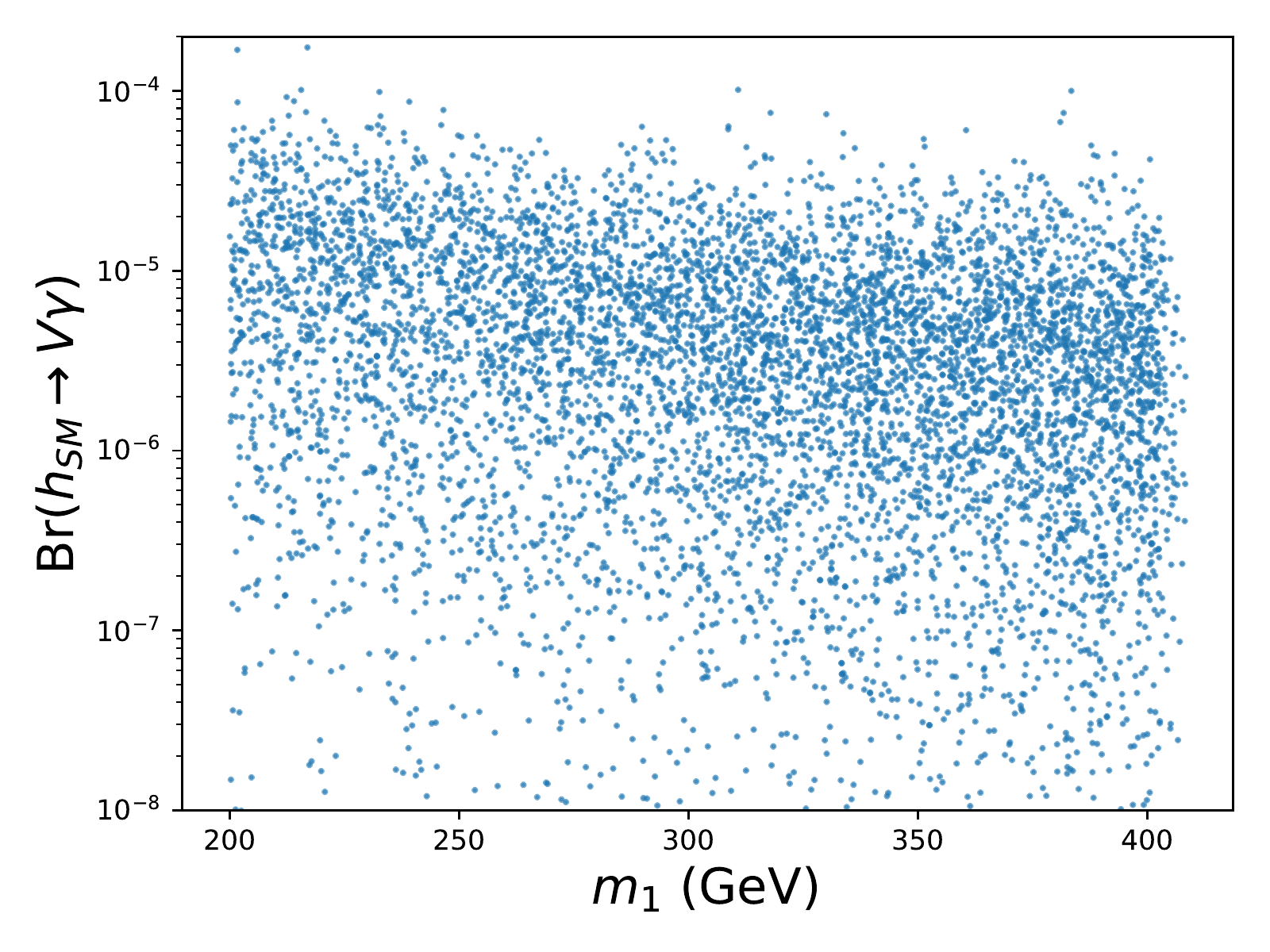}}
\caption{ Br$(h_{SM} \rightarrow V \gamma)$, up to an overall factor of $\frac{g_D^2}{e^2}$, vs $m_1$. Larger $m_1$ generally decreases the branching fraction of $h_{SM} \rightarrow V \gamma$, as expected.}
\label{brhvgam}
\end{figure}

The results of this and several other ATLAS and CMS searches applied to the 4 benchmark points is summarized in Table~\ref{tab:searches}. Overall, as noted, the constraints are seen to 
lie far from the benchmark point predictions with the obvious exception being in the case of the full-luminosity CMS $Z+$MET search with the $Z$ decaying leptonically. In this case, both BP1 
and BP2 yield $S/B$ values which are close to unity implying this is the most sensitive search of the $Z$-dominant models, which, since they dominate the population overall, implies an ability 
to probe much of the full parameter space. Clearly, a further far more detailed study of this particular search sensitivity is warranted especially as significant integrated luminosity at 14 TeV is 
obtained.

\begin{table}
\begin{center}
\begin{tabular}{ |c |c |c| c| c|}
\hline
Model & $Z(q \bar q)$+MET \cite{ATLAS2018hadronic} & $Z(l^+ l^-)$+MET \cite{CMS2020leptonic} & $h_{SM}(b \bar b)$+MET \cite{ATLAS2017hbb} & $h_{SM}(\gamma \gamma)$+MET \cite{ATLAS2017hgamgam} \\
\hline
BP1 & 13 & 1.3 & -- & -- \\
BP2 & 12 & 1.0 & -- & -- \\
BP3 & -- & -- & 9 & 3.8 \\
BP4 & -- & -- & 6 & 2 \\
\hline
 & $W^+W^-$+MET \cite{ATLAS2020wmet} & $WZ$+MET \cite{ATLAS2018wzmet} & $W h_{SM}$+MET  & $h_{SM}(\gamma \gamma)$+MET \cite{ATLASwhmetgamgam}\\
\hline
BP1 & 70 & 8 & -- & -- \\
BP2 & 22 & 6 & -- & -- \\
BP3 & 48 & -- & 28 ($b \bar b$)\cite{ATLAS2020whmetbb} & 2 \\
BP4 & 19 & -- & 14 ($\gamma \gamma$) \cite{ATLASwhmetgamgam} & 1.2 \\
\hline
\end{tabular}
\end{center}
\caption{The ratio $\sigma_{\textrm{vis,lim}}/\sigma_{\textrm{vis,BPx}}$ for the analysis bin providing the strongest constraint arising from the searches for the various final states produced 
for  scalar PM models at the LHC. Note that the limits for $Z \rightarrow l^+ l^-$+MET are estimates from searches for $Zh_{SM}\rightarrow l^+ l^- + \textrm{inv.}$ rather than 
model-independent limits. For final states with multiple searches, here is displayed the result of the search with smallest value of $\sigma_{\textrm{vis,lim}}/\sigma_{\textrm{vis,BPx}}$. 
The $h_{SM}\rightarrow \gamma \gamma$+MET search in the lower half of the table reflects the Category 12 signal region of Ref. \cite{ATLASwhmetgamgam} applied to $H/A+V/h_d$ 
associated production events.}
\label{tab:searches}
\end{table}

In addition to these SUSY-like searches, rare decays of the SM-like Higgs boson, \eg, $h_{SM}\to V\gamma$ (\ie, $\gamma$ + MET) and $h_{SM}\to Zh_dV$ (\ie, $Z$+MET), can be used 
to probe this 
model parameter space. In the first case, a SM background appears at the level of $B\simeq 3\cdot 10^{-4}$ due to the process $h_{SM}\to Z\gamma \to \gamma \bar \nu\nu$ while 
for the second process one expects $B\simeq 4.3\cdot 10^{-3}$ in the SM from $h_{SM}\to ZZ^*\to Z\bar \nu\nu$. In the familiar KM approach, the $h_{SM}\to V\gamma$ process proceeds through 
loops of SM particles via the same graphs as does $h_{SM}\to 2\gamma$ but now a finite KM allows us to make the replacement $\gamma \to V$, hence, this amplitude is always both 
$\epsilon$ as well as loop suppressed.
Here, the $h_{SM}\to V\gamma$ process proceeds via the charged Higgs triangle and loop graphs so that only $m_{1,2}$ and $g_D/e$ are unknown factors in the calculation. For this 
model set the resulting values of the branching fraction are shown in Fig.~\ref{brhvgam} where we see the predictions all lie quite safely below the expected SM background 
value when $g_D/e\simeq 1$; this is partly due to the cancellation between the two charged Higgs contributions.  
The measured upper limit on this branching fraction from the LHC for this process is presently 1.8\% from Ref.\cite{ATLAS:2021pdg}.  In the usual discussion, the process $h_{SM}\to ZV$,  which 
produces a similar final state as does $ZVh_d$ as far as the detector is concerned, can occur at tree level via KM plus mass mixing and so is doubly suppressed by 
$\sim \epsilon m_V^2/m_Z^2$. Alternatively, this process can proceed via the same SM loops as does $h_{SM}\to Z\gamma$ but with $\gamma \to V$ due to a finite $\epsilon$ as above 
so that it is again doubly suppressed.  Here, the $h_{SM}\to Zh_dV$ process occurs at {\it tree-level} via 
virtual $Z^*,A^*$ or $H^*$ exchanges, which can interfere destructively, and where the values of $t$, $m_H=m_A$, $g_D/e$ as well as the trilinear $Hh_{SM}h_d$ coupling, denoted here by 
$\tilde \lambda$, are now the unknowns. The model predictions for the branching fraction can be reasonably sizable in this case, commonly $B\sim 0.1-1 \%$, particularly if $H$ is light as is 
shown in Fig.~\ref{BrBSMscan}, and yield values somewhat comparable to the SM background prediction. For larger values of the rate, it may be possible to partially separate the SM 
background from the predicted model signal by employing cuts on the reconstructed $Z$ boson energy distribution in the $h_{SM}$ rest frame\cite{Rueter:2020qhf}. 

\begin{figure}
\vspace*{0.5cm}
\centerline{\includegraphics[width=5in,angle=0]{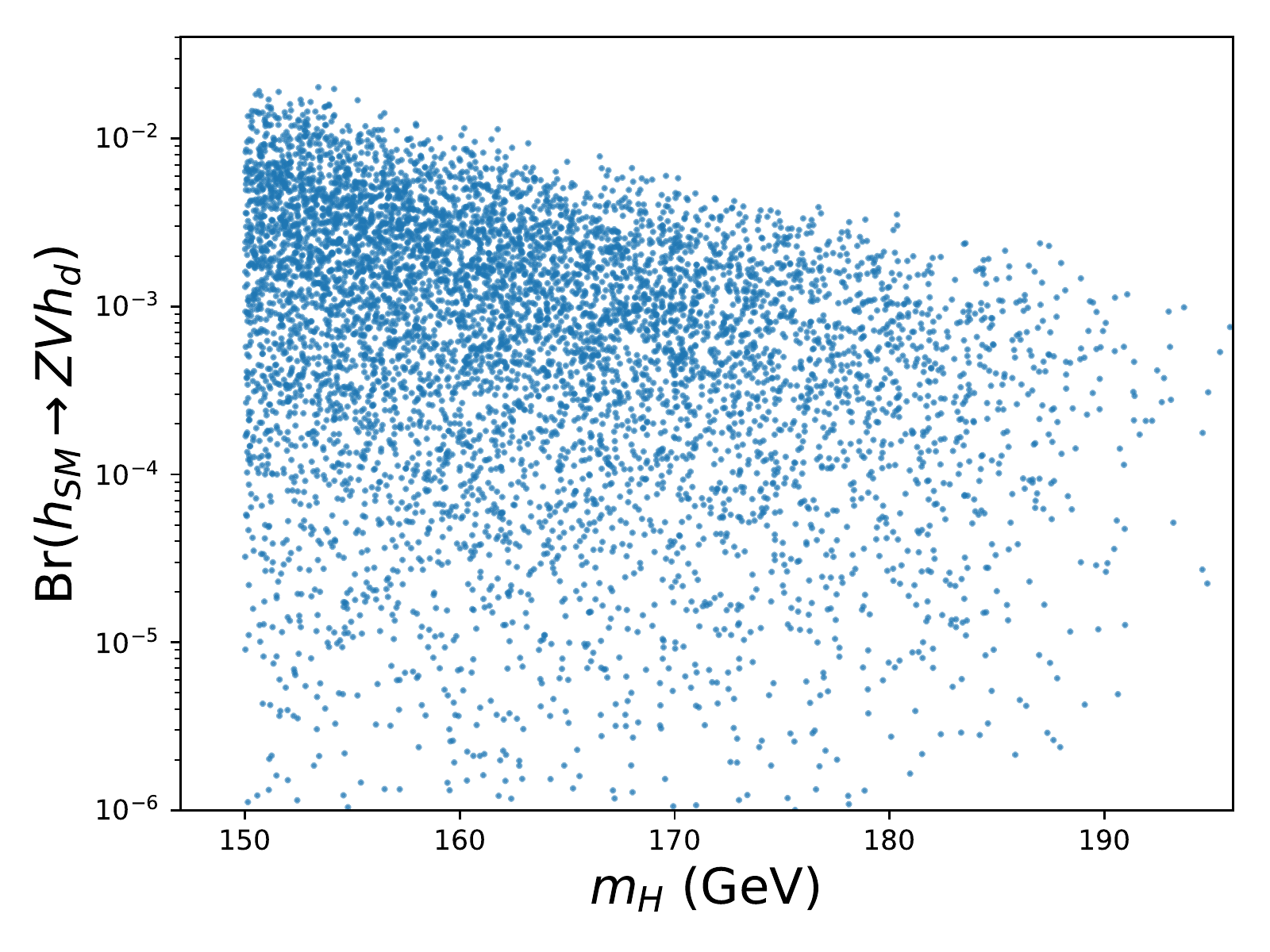}}
\caption{ Branching fraction for $h_{SM}\rightarrow Z V h_d$ vs. $m_H$, assuming the SM value of the Higgs width $\Gamma(h_{SM})=4.07$ MeV. While the majority of models have Br$(h_{SM}\rightarrow Z V h_d)\simeq0.01-1\%$, it is possible for the branching fraction to be quite suppressed due to small values of $\tilde \lambda$ and/or large destructive interference between the three decay amplitudes.}
\label{BrBSMscan}
\end{figure}

It is clear from this brief overview of this rather simple scalar PM scenario with a relatively light mass spectrum that much of the parameter space should be accessible quite soon to LHC analyses 
and so warrants further study. It would also be interesting to examine a more complex scenario where the roles of the dark Higgs multiplets and PM are played by different scalar representations.

\section{Discussion and Conclusions}

Simple renormalizable kinetic mixing models posit that the $U(1)_Y$ hypercharge gauge boson of the SM mixes with the $U(1)_D$ dark photon gauge boson through vacuum 
polarization-like graphs thus allowing for dark matter to interact with ordinary matter with suppressed couplings. Portal matter particles, which carry both SM and dark sector quantum numbers, 
play a very pivotal enabling role in these kinetic mixing/dark photon scenarios being necessary to make them function as they are responsible for generating the necessary one-loop graphs. If 
such particles exist, because they necessarily couple to the SM gauge interactions, the only way that they could have avoided detection so far is if they are relatively massive, $\gsim 1$ TeV, 
and/or decay in an elusive manner. Such particles are likely beyond the direct pair production mass reach of the $\sqrt s=240-380$ GeV ILC, FCC-ee or CEPC $e^+e^-$ colliders but can 
be probed in such machines through indirect tree-level processes. These considerations thus imply that higher energy colliders such as the HL-LHC are the only places where the direct 
production of such states can be examined and probed in detail. There are some good reasons to expect that at least some of the PM fields, as well as the heavy gauge bosons of an extension 
of the dark gauge group, may exist in the region of the TeV mass scale, especially when the value of dark gauge coupling, $\alpha_D$, is large and/or the PM are scalars as their masses 
are then directly linked to the SM vev, at least in the simplest models. In this White Paper, we have provided an overview of some of the basic PM model components and signatures as well 
as those for a possible next step up the ladder towards a UV-completion based on an extended dark gauge group which leads to a highly enriched phenomenology and thus to many 
possible additional collider signatures.  This particular path upwards in energy scale is hardly unique and many other scenarios are possible. Not all of these many production processes have 
been sufficiently well studied and many require much more detailed and realistic simulations especially as the accessible model parameter space continues to grow at the 14 TeV LHC.

PM is a fundamental ingredient of the KM scenario and full implications of its possible existence are in need of further exploration.

\section*{Acknowledgements}
The author would like to particularly thank J.L. Hewett, D. Rueter and G. Wojcik for very valuable discussions and/or earlier collaboration related to this work. This work was supported by the 
Department of Energy, Contract DE-AC02-76SF00515.



\end{document}